\documentstyle[12pt,aaspp4]{article} 

\newcommand{\bq}{\begin{equation}} 
\newcommand{\eq}{\end{equation}}

\begin{document} 
\def\refitem{\par\parskip 0pt\noindent\hangindent 20pt} 

\normalsize 

\title{Type Ia Supernova Discoveries at $z>1$
From the {\it Hubble Space Telescope:}
Evidence for Past Deceleration and Constraints on Dark Energy Evolution\altaffilmark{1}}

\vspace*{0.3cm} 

{\it  \ \ \ \ \ \ \ \ \ \ \ \ \  \  \ \ \ \ \ \  \ To Appear in the Astrophysical Journal, June 2004}
\vspace*{0.3cm}

Adam G. Riess\altaffilmark{2}, 
Louis-Gregory Strolger\altaffilmark{2}, 
John Tonry\altaffilmark{3}, 
Stefano Casertano\altaffilmark{2}, 
Henry C. Ferguson\altaffilmark{2}, 
Bahram Mobasher\altaffilmark{2}, 
Peter Challis\altaffilmark{4}, 
Alexei V. Filippenko\altaffilmark{5}, 
Saurabh Jha\altaffilmark{5}, 
Weidong Li\altaffilmark{5}, 
Ryan Chornock\altaffilmark{5}, 
Robert P. Kirshner\altaffilmark{4}, 
Bruno Leibundgut\altaffilmark{6}, 
Mark Dickinson\altaffilmark{2}, 
Mario Livio\altaffilmark{2}, 
Mauro Giavalisco\altaffilmark{2},
Charles C. Steidel\altaffilmark{7}, 
Narciso Benitez\altaffilmark{8}
and
Zlatan Tsvetanov\altaffilmark{8}

\altaffiltext{1}{Based on observations with the NASA/ESA {\it Hubble Space 
Telescope}, obtained at the Space Telescope Science Institute, which is 
operated by AURA, Inc., under NASA contract NAS 5-26555.} 
\altaffiltext{2}{Space Telescope Science Institute, 3700 San Martin 
Drive, Baltimore, MD 21218.} 
\altaffiltext{3}{Institute for Astronomy, University of Hawaii, 
2680 Woodlawn Drive, Honolulu, HI 96822.} 
\altaffiltext{4}{Harvard-Smithsonian Center for Astrophysics, 60 Garden St., 
Cambridge, MA 02138.} 
\altaffiltext{5}{Department of Astronomy, 601 Campbell Hall, University of 
California, Berkeley, CA  94720-3411.} 
\altaffiltext{6}{European Southern Observatory, Karl-Schwarzschild-Strasse 
 2, Garching, D-85748, Germany.} 
\altaffiltext{7}{Department of Astronomy, 105-24, California Institute of
Technology, Pasadena, CA  91125.}
\altaffiltext{8}{Department of Physics and Astronomy, Johns Hopkins University, Baltimore, MD  21218.}
\begin{abstract} 

We have discovered 16 Type Ia supernovae (SNe~Ia) with the {\it Hubble Space
Telescope (HST)} and have used them to provide the first conclusive evidence
for cosmic deceleration that preceded the current epoch of cosmic acceleration.  These objects, discovered during the course of the
GOODS ACS Treasury program, include 6 of the 7 highest-redshift SNe~Ia known,
all at $z>1.25$, and populate the Hubble diagram in unexplored territory. The
luminosity distances to these objects, and to 170 previously reported SNe~Ia,
have been determined using empirical relations between light-curve shape and
luminosity.  A purely kinematic interpretation of the SN~Ia sample provides
evidence at the $>$  99\% confidence level for a transition from
deceleration to acceleration or similarly, strong evidence for a cosmic jerk. 
Using a simple model of the expansion
history, the transition between the two epochs is
constrained to be at $z=0.46 \pm 0.13$.  The data are consistent with the cosmic concordance model of
$\Omega_M \approx 0.3, \Omega_\Lambda \approx 0.7$ ($\chi^2_{dof}=1.06$), and are inconsistent with a
simple model of evolution or dust as an alternative to dark energy.  For a flat
Universe with a cosmological constant, we measure $\Omega_M=0.29 \pm
^{0.05}_{0.03}$ (equivalently, $\Omega_\Lambda=0.71$).  When combined with
external flat-Universe constraints including the cosmic microwave background
and large-scale structure, we find $w=-1.02 \pm ^{0.13}_{0.19}$ (and $w<-0.76$
at the 95\% confidence level) for an assumed static equation of state of dark
energy, $P = w\rho c^2$.  Joint constraints on both the recent equation of
state of dark energy, $w_0$, and its time evolution, $dw/dz$, are a factor of $\sim 8$ more
precise than its first estimate and twice as precise as those without the SNe~Ia
discovered with {\it HST}.  Our constraints are consistent with 
the static nature of 
and value of $w$ expected for a cosmological constant (i.e.,
$w_0 = -1.0$, $dw/dz = 0$), and are inconsistent with very rapid evolution of
dark energy.  We address consequences of evolving dark energy for the fate of the Universe.

\end{abstract} 
subject headings: galaxies: distances and redshifts ---
cosmology: observations --- cosmology: distance scale --- 
supernovae: general

\section{Introduction} 

  Observations of type Ia supernovae (SNe~Ia) at redshift $z < 1$
provide startling and puzzling evidence that the expansion of the Universe
at the present time appears to be {\it accelerating}, behavior attributed to ``dark energy'' with
negative pressure (Riess et al. 1998; Perlmutter et al. 1999; for reviews, see
Riess 2000; Filippenko 2001, 2004; Leibundgut 2001).  Direct evidence comes
from the apparent faintness of SNe~Ia at $z \approx 0.5.$.   Recently expanded
samples of SNe~Ia have reinforced the statistical significance of this result
(Knop et al. 2003) while others have also extended the SN~Ia sample to $z \approx 1$ (Tonry et
al. 2003; Barris et al. 2004).  Observations of large-scale structure (LSS),
when combined with measurements of the characteristic angular size of
fluctuations in the cosmic microwave background (CMB), provide independent
(though indirect) evidence for a dark-energy component (e.g., Spergel et al.
2003).  An independent, albeit more tentative investigation via the integrated Sachs-Wolfe (ISW) effect also provides evidence for dark energy (Scranton et al. 2003). The magnitude of the observed acceleration was not anticipated by
theory and continues to defy a {\it post facto} explanation.  Candidates for
the dark energy include Einstein's cosmological constant $\Lambda$ (with a phenomenally small value), evolving
scalar fields (modern cousins of the inflation field; Caldwell, Dav\'e, \&
Steinhardt 1998; Peebles \& Ratra 2002), and a weakening of gravity in our 3 +
1 dimensions by leaking into the higher dimensions required in string theories
(Deffayet, Dvali, \& Gabadadze 2002).  These explanations bear so greatly on
fundamental physics that observers have been stimulated to make extraordinary efforts to confirm the initial results on dark energy, test possible sources of error, and extend our empirical knowledge of this newly discovered component of the Universe.

Astrophysical effects could imitate the direct evidence from SNe Ia for an accelerating Universe.
A pervasive screen of grey dust could dim SNe~Ia with little telltale reddening
(Aguirre 1999a,b).  Luminosity evolution could corrupt the measurements if
SNe~Ia at $z \approx 0.5$ are intrinsically fainter than their low-redshift
counterparts.  To date, no evidence for an astrophysical origin of the apparent faintness of SNe Ia has been found (Riess
2000; Coil et al. 2001; Leibundgut 2001; Sullivan et al. 2003). However, given
the significance of the putative dark energy and the unique ability of SNe~Ia
to illuminate it, we need a more definitive test of the hypothesis that supernovae at $z\sim 0.5$ are intrinsically dimmer, or dimmed by absorption.

If cosmic acceleration is the reason why SNe Ia are dimmer at $z \sim 0.5$, then we expect cosmic deceleration at $z>1$ to reverse the sign of the observed effect.  The combination of recent acceleration and past deceleration is
a clear signature of a mixed dark-matter and dark-energy Universe and one which
is readily distinguishable from simple astrophysical dimming (Filippenko \& Riess 2001).

Furthermore, assuming SNe~Ia at $z > 1$ continue to trace the cosmological
world model, measurements of SNe~Ia in the next redshift octave provide the
unique ability to discriminate between a static and evolving dark-energy
equation of state.  This would provide a vital clue to distinguish a cosmological constant from other forms of dark energy that change with time.

Ground-based efforts to look for past deceleration with SNe~Ia have offered
hints of the effect, but ultimately they have suffered from insufficient signal-to-noise ratios
(Tonry et al. 2003; Barris et al. 2004).  Discovering, confirming, and then
monitoring transients at $I \approx 25$ mag on the bright sky is challenging
even with the largest telescopes and the best conditions.  A single SN~Ia at $z
\approx 1.7$, SN 1997ff, discovered with WFPC2 on the {\it Hubble Space
Telescope (HST)} (Gilliland, Nugent, \& Phillips 1999), provided a hint of past deceleration; however, inferences drawn from a single SN~Ia,
while plausible, are not robust (Riess et al. 2001; Ben\'\i tez et al. 2002;
Mortsell, Gunnarsson, \& Goobar 2001).

  To study the early expansion history of the Universe, we initiated the first systematic, space-based search and follow-up effort to
collect SNe~Ia at $z > 1$, carried out in conjunction with the Great
Observatories Origins Deep Survey (GOODS) Treasury program (Giavalisco et al.
2003) conducted with the Advanced Camera for Surveys (ACS) aboard {\it HST}.
(The ability to detect SNe at $z>1$ with the Space Telescope was an application first envisioned during its planning; Tammann 1977, Colgate 1979).   A
separate ``piggyback'' program was utilized to obtain target of opportunity
(ToO) follow-up {\it HST} observations of the SNe~Ia with ACS and NICMOS (the
Near-Infrared Camera and Multi-Object Spectrograph).  Elsewhere we present a
color-based method for discrimination of SNe~Ia at $z > 1$ from other
transients (Riess et al. 2003) and the full harvest of the SN survey (Strolger
et al. 2004).  We present the follow-up spectroscopy and photometry of 16
SNe~Ia in \S 2, light-curve analysis in \S 3, cosmological tests and
constraints in \S 4, and a discussion and summary in \S 5 and \S 6,
respectively.

\section{Target of Opportunity Follow-up: Light Curves and Spectra}

\subsection{Follow-up}

       The methods and criteria we used to search the GOODS ACS Treasury data
for SNe are described by Strolger et al. (2004) and are based on image
subtraction (Perlmutter et al. 1997; Schmidt et al. 1998).  Strolger et
al. (2004) provide the parameters of the search including search depth,
efficiency, timing, and false-positive discrimination, as well as a list of all
detected SNe.  Briefly, our search was conducted in the $F850LP$ (Z-band) to an effective limit of $\sim$26.0 (Vega) magnitude covering 0.1 square degree in 5 epochs (at intervals of $\sim$45 days).  Our limiting magnitude was 1 to 2 mag fainter than the expected peak of a SN Ia over the target range of $1<z<1.6$, therefore SNe Ia we collected (whose intrinsic dispersion is expected to be $<0.2$ mag)would not preferentially be selected from the bright tail of their intrinsic distribution.    In Table 1 we provide discovery data for the SNe~Ia reported
here.

Our ToO candidates were generally too faint to anticipate useful spectral
discrimination from the ground; it was therefore necessary to {\it initially}
identify SNe~Ia photometrically.  To discriminate SNe~Ia at $z > 1$ from SNe~II
and from SNe~I at lower redshifts, we used a combination of photometric
redshifts of the host galaxies (with 9 passbands) and rest-frame ultraviolet
(UV) colors; see Riess et al. (2003) for details. A comparison of the
photometric redshifts available at the start of the survey with spectroscopic
redshifts obtained thereafter yields an RMS of 0.05 for the quantity $(z_{phot} - z_{spec})/(1 + z_{spec})$ for 25 hosts (with a 4 $\sigma$ outlier; see Strolger et al. 2004). 

We selected 9 individual candidates for subsequent ToO observations (including
one pair of targets observed within the same ACS field).  In addition to these
primary targets, judicious positioning of the follow-up fields provided
serendipitous monitoring of 4 additional high-redshift SN~Ia candidates.  For 4
more SNe~Ia, the periodic imaging of the GOODS survey (sometimes augmented with
a few ground-based observations) provided sufficient characterization of their
light curves.  Subsections of the discovery and pre-discovery images, as well
as their difference (centered on each SN~Ia), are shown in Figure 1.

The great benefit of HST observations--dramatically reduced sky noise--is only fully realized if operational constraints can be overcome.  To get the full advantage of HST, the supernova search needs to be completed, the best targets selected, spectra taken, and the photometric follow-up set in motion in 2 weeks.  Otherwise, the peak of the light curve will be missed, the decline rate of the object will not be well measured, and, as the objects fade, the spectra will become too difficult to obtain.  Our challenge was to do all this without using  highly disruptive
and inefficient ``24 hour'' ToOs (for which the {\it HST} weekly schedule is
immediately interrupted, wasting precious observing time).
  To achieve
our goals, each GOODS epoch was scheduled during a 3--4 day interval
immediately preceding the weekly deadline for a ToO activation.  This allowed
the ToO observing to be scheduled and uploaded to the spacecraft for the
following week's observations without delay.  By adopting this prescription we
were able to achieve a time interval of typically 9 to 11 days between the
discovery and first follow-up observation, a span of less than 5 days in the
rest frame of a SN at $z \gtrsim 1$.

After discovery and photometric screening, we developed a follow-up plan which
was appropriate for our best estimate of the SN redshift. ACS and the $F850LP$
filter were typically used for 6--8 epochs to obtain a rest-frame $B$-band or
$U$-band light curve for SNe~Ia at $z > 1$ extending to $\sim$ 20 rest-frame
days past maximum brightness.  This strategy utilized results from Jha, Riess,
\& Kirshner (2004a), who demonstrated the utility of a large sample of $U$-band
light curves for constraining the light-curve shape parameter and phasing of
the optical light curves.  To provide rest-frame optical zero-points, where
SNe~Ia are best calibrated, we used NICMOS Camera 2.  SNe expected to be at
$1.0 < z < 1.3$ were imaged with NICMOS Camera 2 and $F110W$.  SNe expected to
be at $z > 1.3$ were imaged with both $F110W$ and $F160W$.  The one exception
was SN 2002ki, for which a ground-based spectrum yielded $z = 1.14$ from
narrow, host-galaxy [O~II] emission, and $F160W$ was used to sample the second
maximum in the rest-frame infrared (IR) to use as a secondary means of
classification.

ACS coupled with a grism filter (G800L) provides slitless spectroscopy over the
entire wide-field camera (WFC) field-of-view.  For obtaining spectra of
high-redshift SNe~Ia, this mode of observing has noteworthy advantages and
disadvantages as compared to spatial-blocking spectroscopy with STIS on {\it
HST} or from the ground with a large-aperture telescope.  The primary advantage
of ACS grism spectroscopy is its efficiency.  ACS is the most efficient camera
to fly on {\it HST} due to its rationed, silver-coated reflections.
Importantly, slitless spectroscopy with the grism retains the efficiency of
{\it HST} resolution along one spatial direction, dispersing the light of a
point-spread function (PSF) over few sky pixels.  Another advantage attuned to
identifying high-redshift SNe~Ia is the relatively low sky brightness between
0.8~$\mu$m and 1~$\mu$m from the vantage point of {\it HST}.

Using the {\it HST} grism has disadvantages as well.  {\it HST} is a factor of
3--4 smaller in size than the largest ground-based optical
telescopes. Moreover, the grism disperses a large spectral range of the sky
onto the position of the SN.  For G800L, wavelengths shorter than
$\sim$5500~\AA\ are blocked so the total sky counts are only $\sim$50\% greater
(hence $\sim$25\% more noise) than for the $I$-band ($F814W$).  A more
troublesome feature of grism observations is the superposition of multiple
sources from multiple diffraction orders on the same sky pixels.  Careful
consideration must be given to possible contamination of a SN spectrum by
nearby sources, especially the host galaxy (although this effect can be
mitigated by judicious choice of the telescope roll angle).  The ACS grism also
is limited in resolution to $R={\lambda / \Delta \lambda} = 200$.  However, this
resolution is well-matched to measure SNe~Ia whose blended
absorption features are broadened by ejecta dispersions of $10^4$ km s$^{-1}$.
Serendipitous spectra of SNe~Ia obtained during ACS commissioning and reported
by Blakeslee et al. (2003) provided the first in-orbit examples at high
redshift (specifically, SN 2002dc at $z=0.47$ and SN 2002dd at $z=0.95$). 

We employed the ACS grism for our ToOs when we expected to obtain a sufficient
signal-to-noise ratio (S/N) for classification and redshift determination in
fewer than 8 orbits of integration time and without detrimental contamination.
These included SNe 2002fw, 2003az, 2003dy, 2003es, and 2003eq as well as the
neighboring SN 2003eb. Spectra of other SNe or host galaxies were obtained with
ground-based telescopes (Table 3), including the VLT, Magellan, Keck-II with
NIRSPEC (McLean et al. 1998), and especially Keck-I with LRIS (Oke et
al. 1995).

\begin{deluxetable}{lllll} 
\footnotesize
\tablecaption{Discovery Data}
\tablehead{\colhead{SN}&\colhead{Nickname}&\colhead{UT
Date}&\colhead{SN $\alpha$(J2000)}&\colhead{SN $\delta$(J2000)}}
\startdata
\hline
\hline

2002fw &	Aphrodite & Sep. 19.86	& 03:32:37.52	 & $-$27:46:46.6 \nl
2002fx &	Athena & Sep. 20.84 &	03:32:06.80	 & $-$27:44:34.4 \nl
2002hp &	Thoth	 & Nov. 1.51 & 03:32:24.79	 & $-$27:46:17.8 \nl
2002hr &	Isis	 & Nov.  1.64 & 03:32:22.57	 & $-$27:41:52.2 \nl
2002kc &	Bilbo	 & Dec. 21.50 & 03:32:34.72	 & $-$27:39:58.3 \nl
2002kd &	Frodo	 & Dec. 21.64 & 03:32:22.34	 & $-$27:44.26.9 \nl
2002ki &	Nanna	 & Nov. 22.69 & 12:37:28.35	 & +62:20:40.0 \nl
2003aj &	Inanna &	Feb. 3.19  & 03:32:44.33 & $-$27:55:06.4 \nl
2003ak &	Gilgamesh  &  Feb. 3.19 & 03:32:46.90	 & $-$27:54:49.4 \nl
2003az &	Torngasek 	 & Feb. 20.91 & 12:37:19.67 & +62:18:37.5 \nl
2003bd &	Anguta	 & Feb. 21.95 & 12:37:25.06	 & +62:13:17.5 \nl
2003be &	Qiqirn	 & Feb. 22.08 & 12:36:25.97	 & +62:06:55.6 \nl
2003dy &	Borg	 & Apr. 4.67 & 12:37:09.16	 & +62:11:29.0 \nl
2003lv &	Vilas	 & Apr. 4.67 & 12:37:29.00	 & +62:11:27.8 \nl
2003eb &	McEnroe	 & Apr. 5.65 & 12:37:15.18	 & +62:13:34.6 \nl
2003eq &	Elvis	 & May  24.7 & 12:37:48.34	 & +62:13:35.3 \nl
2003es &	Ramone	 & May  25.5 & 12:36:55.39	 & +62:13:11.9 \nl
\enddata 
\end{deluxetable}

\subsection{Photometry}

After the search phase, all images were reprocessed using up-to-date reference
files and the CALACS pipeline in the STSDAS package in IRAF\footnote[8]{IRAF is
distributed by the National Optical Astronomy Observatories, which are operated
by the Association of Universities for Research in Astronomy, Inc., under
cooperative agreement with the National Science Foundation.}. This procedure
includes ``standard'' rectifications for the camera gain, overscan, spatial
bias, dark current, and flat fielding.  Due to the significant geometric
distortion of the ACS WFC (the cost of minimizing reflections), we applied the
drizzle algorithm (Fruchter \& Hook 1997) in the Multidrizzle software package
(Koekemoer et al. 2004).  Because ACS WFC images are undersampled at
wavelengths shortward of 11,000~\AA, a better sampled and more precise SN PSF
can be obtained by ``drizzling'' (i.e., resampling and combining) the images at
a pixel scale finer than the physical ACS WFC size of $0.05''$ pixel$^{-1}$.
However, such improvements can only be realized with well-dithered images.  The
relative size of the dither was measured for each frame using source catalogs.
Nearly all of the $F850LP$ images in the survey and its follow-up were obtained
at 4 independent dither positions and were subsequently resampled to $0.033''$
pixel$^{-1}$. Imaging in $F775W$ and $F606W$ utilized only 2 dither points and
the physical pixel scale was maintained.

  For NICMOS reductions the CALNICA pipeline in the STSDAS package in IRAF was
used to provide calibrated frames.  Then the well-dithered frames were drizzled
to half the physical pixel scale of Camera 2, i.e., $0.038''$ pixel$^{-1}$. The
size of each frame's dither was determined by cross-correlation of common
sources.

The phasing of GOODS (spanning more than 200 days) provided additional
observations of the host galaxies with negligible SN light to serve as
subtraction templates for the ACS data.  Subsequent light-curve fitting was
used to estimate the expected brightness of the SN at the time of each GOODS
epoch.  Epochs with negligible contamination from SN light were combined to
obtain deep subtraction templates.  By the last epoch of GOODS, SNe discovered
in the first search had faded sufficiently ($\sim$2 months past maximum and
3 to 4 mag below peak) to provide useful templates.  For SNe found in all
subsequent searches, the initial survey epochs provided the components of deep
templates.

Due to the remarkable stability of rectified ACS images it was generally not
necessary to ``blur'' (i.e., convolve) images to match the PSFs at different
epochs (a conclusion reached by fixed-aperture tests on field stars in
successive epochs).  Subtractions free from host contamination were obtained by
flux-conserving registration of the templates to the follow-up frames.  The two
exceptions were SN 2002hp and SN2003lv which resided $< 0.05''$ from the sharp nucleus of
bright elliptical hosts and for which the technique of matching
PSFs was used (Alard \& Lupton 1998).
 
For NICMOS imaging, the necessity of obtaining ``clean'' subtraction templates
was judged for each object based on the complexity of the SN site in the bluer,
$F850LP$ templates.  We judged late-time templates were needed and were
obtained for SN 2002hp, SN 2002ki, and SN 2003az. As with the ACS data, only SN
2002hp necessitated image convolution to match the PSF before subtraction.

The magnitudes of the SNe in the ACS images were calculated by fitting a PSF
produced using bright comparison stars scaled to match the infinite-aperture zero-points of
Sirianni et al. (2004).  All ACS passband magnitudes are given as
Vega-normalized magnitudes.  Residual ``sky'' flux was measured using annuli
centered on the SN and with an inner and outer radius of $0.66''$ and $1.00''$,
respectively.  The center of the PSF fit was determined from a centroid of a
stack of all SN images.

Gilliland \& Riess (2002) have shown that
the encircled energy of red stars in the IR is more dispersed
 than for blue stars,
likely due to backside scattering from the ACS WFC CCD mounting. 
This is an important effect for images in $F850LP$ with SNe~Ia at
$z > 0.8$ which are considerably red ($i-z \approx 1$ mag) and can be expected
to suffer the ``red halo'' effect.  To measure
the SNe~Ia in these images, we used a few bright, red
($i-z = 1.0$ mag) field stars as a PSF template. 

Statistical uncertainties for the SN magnitudes were determined by adding and
recovering artificial PSFs with the measured SN flux (Schmidt et al. 1998).
Additional uncertainty was included for the shot noise of the measured SN flux.
We found excellent agreement between the uncertainties calculated empirically
and synthetically from the STScI exposure-time calculator.

NICMOS measurements of supernovae were similar, except out model PSF and
zero-points were calculated from observations of standard stars P330E and
G191B2B in Cycle 11 (M. Dickinson et al. 2003, private communication).  For SN
2003es and SN 2003ak which have non-complex backgrounds, the underlying ``sky''
was determined from isophotal modeling of the host.  For SN2003dy and SN2003eq the mean background was estimated with local apertures.  Table 2 includes the measured magnitudes of
the SNe~Ia presented here.  Figure 2 shows their light curves.

\subsection{Spectra}

Supernovae are best classified by the presence and absence of diagnostic
features in their spectra (see Filippenko 1997 for review).  For 14 of the 17
SNe listed in Table 1, we obtained spectra of the SNe near maximum light.
For two of these (SN 2003lv and SN 2002hp), bright, elliptical hosts yielded redshifts but
overwhelmed the spectra of the SNe.  For the remaining three (SN 2003ak, SN
2003aj, and SN 2002fw), spectra of the hosts to identify their redshifts were
obtained when the SNe were no longer visible.  Spectra of the twelve visible SNe are
shown in Figure 3, and details of all the spectroscopy are provided in Table 3.

To classify the SNe, the detected SN spectra (shown in Figure 3) were
cross-correlated with template spectra (after removal of the continuum) to
identify their type and redshift using the ``SNID'' algorithm (Tonry et
al. 2003).  For the cases listed in Table 3 for which narrow-line host emission
was identified, the redshift was constrained to the value determined from the
host emission before cross-correlation.  For cases where the spectra were of
low S/N, the significance of a SN~Ia classification is much greater if the SN
redshift is fixed {\it a priori} by host emission.  For all 12 spectra shown in
Figure 3, SNID provided a significant classification for each as type Ia.
Although the diagnostic used by the SNID algorithm relies on the whole spectrum,
the majority of these SNe can also be classified as type Ia from the presence
of Si~II absorption at 4130~\AA\ (Coil et al. 2001).  Specifically, Si~II
absorption is detected in the two highest-redshift spectra presented here, SN
2003dy and SN 2002fw.  Broad Ca~II absorption near 3750~\AA\ is visible in all
the spectra as well, but this feature is much less secure than Si~II for SN~Ia
classification.
 
For two SNe (SN 2003lv and SN 2002hp) whose spectra were dominated by those of their hosts, the
nature of their red, elliptical hosts allowed us to classify them as highly
probable SNe~Ia (as was previously the case with SN 1997ff; Riess et al. 2001).

For the three SNe without any spectroscopy but with host redshifts,
classification requires greater consideration.  Based on the UV color selection
method described by Riess et al. (2003), SN 2002fx, SN 2003ak, and SN 2003aj
are likely to be SNe~Ia because of their red discovery far-UV colors.  However,
further follow-up of SN 2003aj yielded a rapid decline, uncharacteristic of
normal SNe~Ia (see Strolger et al. 2004).  The photometric records of the other
two are consistent with SNe~Ia.  A pre-discovery observation of SN 2003ak on
the rise is also consistent with the narrow range of SN~Ia rise behavior which rise in 3 weeks (Riess
et al. 1999b), but not with the majority of core-collapse SNe which typically rise in 2 weeks or less,
indicating it is very likely to be an SN~Ia.

For the 16 SNe~Ia (excluding SN 2003aj), we have sufficient quality of
photometry to yield robust luminosity distances.

\begin{deluxetable}{llll} 
\footnotesize
\tablecaption{SN~Ia Imaging}
\tablehead{\colhead{JD$^a$}&\colhead{Vega Mag}&\colhead{Epoch(rest)}&\colhead{K-Corr}}
\startdata
\hline
\multicolumn{4}{c}{SN 2002fx} \nl
\hline
\multicolumn{1}{c}{ } & \multicolumn{1}{l}{$F775W$} &
\multicolumn{2}{r}{$F775W\rightarrow U$} \nl
495.00&28.0(1.0)&-13.5&0.001(0.06)\nl
537.80&27.07(0.25)&  4.2&0.33(0.05)\nl
580.00&28.97(1.00)& 21.8&0.81(0.02)\nl
\multicolumn{1}{c}{ } & \multicolumn{1}{l}{$F850LP$} &
\multicolumn{2}{r}{$F850LP\rightarrow B$} \nl
490.39&27.48(0.45)&-15.5&-1.48(0.02)\nl
537.79&25.17(0.07)&  4.2&-1.16(0.03)\nl
580.49&27.07(0.27)& 22.0&-0.79(0.03)\nl
\hline
\multicolumn{4}{c}{SN 2003eq} \nl
\hline
\multicolumn{1}{c}{ } & \multicolumn{1}{l}{$F606W$} &
\multicolumn{2}{r}{$F606W\rightarrow U$} \nl
783.68&24.55(0.01)&  2.7&0.73(0.03)\nl
\multicolumn{1}{c}{ } & \multicolumn{1}{l}{$F775W$} &
\multicolumn{2}{r}{$F775W\rightarrow B$} \nl
783.68&23.19(0.01)&  2.7&-1.19(0.02)\nl
799.09&23.64(0.02)& 11.0&-1.15(0.02)\nl
807.27&24.12(0.04)& 15.4&-1.12(0.02)\nl
819.79&25.02(0.07)& 22.2&-0.99(0.02)\nl
838.30&26.18(0.30)& 32.2&-0.96(0.02)\nl
\multicolumn{1}{c}{ } & \multicolumn{1}{l}{$F850LP$} &
\multicolumn{2}{r}{$F850LP\rightarrow V$} \nl
783.68&23.03(0.01)&  2.7&-1.28(0.02)\nl
792.10&23.18(0.02)&  7.2&-1.22(0.02)\nl
799.09&23.37(0.02)& 11.0&-1.17(0.03)\nl
807.27&23.72(0.04)& 15.4&-1.12(0.06)\nl
819.79&24.19(0.06)& 22.2&-0.93(0.05)\nl
838.30&24.95(0.07)& 32.2&-0.69(0.02)\nl
\multicolumn{1}{c}{ } & \multicolumn{1}{l}{$F110W$} &
\multicolumn{2}{r}{$F110W\rightarrow R$} \nl
792.94&23.36(0.10)&  7.7&-1.24(0.02)\nl
\hline
\multicolumn{4}{c}{SN 2003es} \nl
\hline
\multicolumn{1}{c}{ } & \multicolumn{1}{l}{$F775W$} &
\multicolumn{2}{r}{$F775W\rightarrow U$} \nl
784.50&24.09(0.03)&  7.7&-0.87(0.02)\nl
\multicolumn{1}{c}{ } & \multicolumn{1}{l}{$F850LP$} &
\multicolumn{2}{r}{$F850LP\rightarrow B$} \nl
784.50&23.67(0.02)&  7.7&-1.35(0.02)\nl
792.37&24.05(0.05)& 11.7&-1.37(0.02)\nl
801.27&24.64(0.06)& 16.2&-1.43(0.03)\nl
807.81&24.90(0.07)& 19.6&-1.49(0.03)\nl
820.91&25.76(0.08)& 26.2&-1.55(0.02)\nl
838.10&26.07(0.15)& 35.0&-1.56(0.02)\nl
\multicolumn{1}{c}{ } & \multicolumn{1}{l}{$F110W$} &
\multicolumn{2}{r}{$F110W\rightarrow V$} \nl
792.76&24.20(0.08)& 11.9&-1.32(0.02)\nl
\hline
\multicolumn{4}{c}{SN 2003az} \nl
\hline
\multicolumn{1}{c}{ } & \multicolumn{1}{l}{$F775W$} &
\multicolumn{2}{r}{$F775W\rightarrow U$} \nl
690.89&25.06(0.05)&  0.0&-0.22(0.03)\nl
701.16&25.42(0.05)&  4.5&-0.15(0.02)\nl
\multicolumn{1}{c}{ } & \multicolumn{1}{l}{$F850LP$} &
\multicolumn{2}{r}{$F850LP\rightarrow B$} \nl
690.89&24.31(0.04)&  0.0&-1.43(0.03)\nl
701.16&24.44(0.04)&  4.5&-1.39(0.02)\nl
709.09&24.60(0.05)&  8.0&-1.38(0.02)\nl
716.92&25.04(0.06)& 11.5&-1.37(0.02)\nl
726.52&25.47(0.08)& 15.7&-1.33(0.03)\nl
733.25&25.71(0.09)& 18.7&-1.27(0.04)\nl
\multicolumn{1}{c}{ } & \multicolumn{1}{l}{$F110W$} &
\multicolumn{2}{r}{$F110W\rightarrow V$} \nl
703.62&24.22(0.06)&  5.6&-1.59(0.03)\nl
710.58&24.37(0.06)&  8.7&-1.54(0.03)\nl
\hline
\multicolumn{4}{c}{SN 2002kc} \nl
\hline
\multicolumn{1}{c}{ } & \multicolumn{1}{l}{$F606W$} &
\multicolumn{2}{r}{$F606W\rightarrow V$} \nl
629.62&22.27(0.01)& -7.5&-0.23(0.02)\nl
672.33&23.25(0.01)& 27.5&0.23(0.02)\nl
\multicolumn{1}{c}{ } & \multicolumn{1}{l}{$F775W$} &
\multicolumn{2}{r}{$F775W\rightarrow R$} \nl
629.62&21.78(0.02)& -7.5&-0.43(0.02)\nl
642.50&21.32(0.10)&  3.0&-0.47(0.02)\nl
672.33&22.08(0.01)& 27.5&-0.44(0.02)\nl
\multicolumn{1}{c}{ } & \multicolumn{1}{l}{$F850LP$} &
\multicolumn{2}{r}{$F850LP\rightarrow I$} \nl
629.62&21.66(0.01)& -7.5&-0.46(0.02)\nl
672.33&21.89(0.01)& 27.5&-0.19(0.04)\nl
\hline
\multicolumn{4}{c}{SN 2003eb} \nl
\hline
\multicolumn{1}{c}{ } & \multicolumn{1}{l}{$F606W$} &
\multicolumn{2}{r}{$F606W\rightarrow U$} \nl
734.58&24.23(0.02)& -1.3&0.84(0.04)\nl
783.47&27.02(0.15)& 24.0&1.36(0.02)\nl
\multicolumn{1}{c}{ } & \multicolumn{1}{l}{$F775W$} &
\multicolumn{2}{r}{$F775W\rightarrow B$} \nl
734.65&23.02(0.02)& -1.3&-1.15(0.02)\nl
745.65&23.12(0.02)&  4.3&-1.11(0.02)\nl
783.54&25.18(0.06)& 24.1&-0.83(0.02)\nl
799.10&25.87(0.10)& 32.2&-0.79(0.02)\nl
\multicolumn{1}{c}{ } & \multicolumn{1}{l}{$F850LP$} &
\multicolumn{2}{r}{$F850LP\rightarrow V$} \nl
734.65&22.79(0.02)& -1.3&-1.34(0.02)\nl
745.65&22.81(0.02)&  4.3&-1.29(0.05)\nl
751.12&22.94(0.02)&  7.2&-1.22(0.04)\nl
763.61&23.45(0.02)& 13.7&-1.11(0.08)\nl
773.89&23.98(0.04)& 19.1&-0.91(0.08)\nl
783.54&24.42(0.04)& 24.1&-0.72(0.06)\nl
792.37&24.72(0.05)& 28.7&-0.59(0.02)\nl
792.10&24.79(0.06)& 28.5&-0.59(0.02)\nl
799.09&25.05(0.08)& 32.2&-0.52(0.02)\nl
801.28&24.93(0.06)& 33.3&-0.52(0.02)\nl
807.81&25.07(0.07)& 36.7&-0.56(0.02)\nl
820.92&25.33(0.07)& 43.5&-0.60(0.02)\nl
838.10&25.37(0.08)& 52.5&-0.66(0.02)\nl
\hline
\multicolumn{4}{c}{SN 2003lv} \nl
\hline
\multicolumn{1}{c}{ } & \multicolumn{1}{l}{$F606W$} &
\multicolumn{2}{r}{$F606W\rightarrow U$} \nl
733.65&25.14(0.05)&  6.0&0.857(0.02)\nl
\multicolumn{1}{c}{ } & \multicolumn{1}{l}{$F775W$} &
\multicolumn{2}{r}{$F775W\rightarrow B$} \nl
733.65&23.54(0.05)&  6.0&-1.12(0.02)\nl
745.65&24.09(0.05)& 12.5&-1.09(0.02)\nl
783.54&26.36(0.25)& 32.8&-0.91(0.02)\nl
\multicolumn{1}{c}{ } & \multicolumn{1}{l}{$F850LP$} &
\multicolumn{2}{r}{$F850LP\rightarrow V$} \nl
692.42&26.39(0.30)&-16.1&-1.17(0.02)\nl
733.65&23.25(0.04)&  6.0&-1.23(0.03)\nl
745.65&23.57(0.05)& 12.5&-1.13(0.04)\nl
751.12&23.86(0.06)& 15.4&-1.05(0.07)\nl
763.61&24.60(0.08)& 22.1&-0.80(0.05)\nl
773.89&24.92(0.12)& 27.6&-0.68(0.02)\nl
783.54&25.63(0.15)& 32.8&-0.65(0.02)\nl
799.09&26.05(0.25)& 41.2&-0.71(0.02)\nl
807.27&26.28(0.30)& 45.6&-0.73(0.02)\nl
\hline
\multicolumn{4}{c}{SN 2002hr} \nl
\hline
\multicolumn{1}{c}{ } & \multicolumn{1}{l}{$F606W$} &
\multicolumn{2}{r}{$F606W\rightarrow U$} \nl
579.64&24.01(0.03)& -6.3&0.09(0.04)\nl
629.47&25.78(0.15)& 26.2&-0.38(0.02)\nl
674.11&27.38(0.30)& 55.5&-0.37(0.04)\nl
\multicolumn{1}{c}{ } & \multicolumn{1}{l}{$F775W$} &
\multicolumn{2}{r}{$F775W\rightarrow B$} \nl
579.64&23.53(0.03)& -6.3&-0.86(0.02)\nl
589.68&23.17(0.12)&  0.2&-0.89(0.02)\nl
590.01&23.25(0.11)&  0.4&-0.89(0.02)\nl
596.65&23.41(0.14)&  4.7&-0.95(0.05)\nl
614.50&23.88(0.08)& 16.4&-1.29(0.11)\nl
629.47&24.39(0.05)& 26.2&-1.68(0.02)\nl
674.11&25.62(0.10)& 55.5&-1.60(0.02)\nl
\multicolumn{1}{c}{ } & \multicolumn{1}{l}{$F850LP$} &
\multicolumn{2}{r}{$F850LP\rightarrow V$} \nl
579.64&23.34(0.03)& -6.3&-0.97(0.02)\nl
629.47&23.75(0.04)& 26.2&-1.19(0.02)\nl
674.11&24.97(0.07)& 55.5&-1.15(0.02)\nl
\hline
\multicolumn{4}{c}{SN 2003bd} \nl
\hline
\multicolumn{1}{c}{ } & \multicolumn{1}{l}{$F606W$} &
\multicolumn{2}{r}{$F606W\rightarrow U$} \nl
691.94&24.54(0.04)& 11.6&0.161(0.02)\nl
735.44&27.45(0.12)& 37.6&0.180(0.02)\nl
\multicolumn{1}{c}{ } & \multicolumn{1}{l}{$F775W$} &
\multicolumn{2}{r}{$F775W\rightarrow B$} \nl
691.94&23.38(0.03)& 11.6&-1.05(0.02)\nl
735.44&25.72(0.10)& 37.6&-1.26(0.02)\nl
745.65&26.00(0.10)& 43.7&-1.25(0.02)\nl
\multicolumn{1}{c}{ } & \multicolumn{1}{l}{$F850LP$} &
\multicolumn{2}{r}{$F850LP\rightarrow V$} \nl
642.50&26.98(0.40)&-17.9&-1.10(0.02)\nl
691.94&23.14(0.03)& 11.6&-1.08(0.02)\nl
735.44&24.75(0.05)& 37.6&-1.03(0.02)\nl
745.65&24.95(0.06)& 43.7&-1.03(0.02)\nl
751.18&25.04(0.07)& 47.0&-1.03(0.02)\nl
763.60&25.30(0.08)& 54.5&-1.04(0.02)\nl
773.95&25.40(0.08)& 60.7&-1.04(0.02)\nl
792.09&25.72(0.10)& 71.5&-1.05(0.02)\nl
807.27&25.67(0.25)& 80.6&-1.05(0.02)\nl
819.79&25.72(0.25)& 88.1&-1.05(0.02)\nl
\hline
\multicolumn{4}{c}{SN 2002kd} \nl
\hline
\multicolumn{1}{c}{ } & \multicolumn{1}{l}{$F606W$} &
\multicolumn{2}{r}{$F606W\rightarrow U$} \nl
629.42&24.93(0.03)& -8.9&0.326(0.02)\nl
673.52&25.78(0.05)& 16.4&0.330(0.02)\nl
\multicolumn{1}{c}{ } & \multicolumn{1}{l}{$F775W$} &
\multicolumn{2}{r}{$F775W\rightarrow B$} \nl
629.43&23.75(0.02)& -8.9&-1.07(0.02)\nl
644.50&22.93(0.17)& -0.2&-1.11(0.02)\nl
645.50&22.73(0.10)&  0.3&-1.11(0.02)\nl
673.53&24.15(0.05)& 16.4&-1.12(0.02)\nl
\multicolumn{1}{c}{ } & \multicolumn{1}{l}{$F850LP$} &
\multicolumn{2}{r}{$F850LP\rightarrow V$} \nl
629.47&23.75(0.01)& -8.9&-1.11(0.02)\nl
639.38&23.04(0.02)& -3.1&-1.13(0.02)\nl
673.58&23.73(0.01)& 16.5&-1.01(0.02)\nl
\hline
\multicolumn{4}{c}{SN 2003be} \nl
\hline
\multicolumn{1}{c}{ } & \multicolumn{1}{l}{$F606W$} &
\multicolumn{2}{r}{$F606W\rightarrow U$} \nl
692.04&24.44(0.03)& 12.1&0.087(0.02)\nl
732.46&27.21(0.20)& 36.8&0.091(0.02)\nl
\multicolumn{1}{c}{ } & \multicolumn{1}{l}{$F775W$} &
\multicolumn{2}{r}{$F775W\rightarrow B$} \nl
641.30&28.0(0.8)&-18.7&-0.72(0.06)\nl
692.04&23.42(0.03)& 12.1&-1.05(0.03)\nl
732.47&25.46(0.10)& 36.8&-1.29(0.02)\nl
784.28&26.04(0.10)& 68.4&-1.21(0.02)\nl
\multicolumn{1}{c}{ } & \multicolumn{1}{l}{$F850LP$} &
\multicolumn{2}{r}{$F850LP\rightarrow V$} \nl
641.30&28.0(0.8)&-18.7&-1.05(0.02)\nl
692.04&23.04(0.02)& 12.1&-1.07(0.02)\nl
732.46&24.52(0.03)& 36.8&-1.09(0.02)\nl
784.28&25.53(0.10)& 68.4&-1.07(0.02)\nl
\hline
\multicolumn{4}{c}{SN 2003dy} \nl
\hline
\multicolumn{1}{c}{ } & \multicolumn{1}{l}{$F850LP$} &
\multicolumn{2}{r}{$F850LP\rightarrow U$} \nl
692.42&27.47(0.75)&-17.1&-0.86(0.04)\nl
733.65&24.43(0.05)&  0.3&-0.96(0.04)\nl
745.65&24.62(0.07)&  5.4&-1.05(0.02)\nl
751.11&24.75(0.07)&  7.7&-1.06(0.02)\nl
763.60&25.22(0.08)& 12.9&-1.12(0.02)\nl
773.89&25.53(0.10)& 17.3&-1.16(0.02)\nl
783.54&26.87(0.34)& 21.4&-1.17(0.02)\nl
801.28&27.00(0.43)& 28.9&-1.21(0.02)\nl
807.81&27.21(0.64)& 31.6&-1.19(0.02)\nl
\multicolumn{1}{c}{ } & \multicolumn{1}{l}{$F110W$} &
\multicolumn{2}{r}{$F110W\rightarrow B$} \nl
751.59&24.50(0.07)&  7.9&-1.71(0.02)\nl
754.59&24.60(0.08)&  9.1&-1.73(0.02)\nl
\multicolumn{1}{c}{ } & \multicolumn{1}{l}{$F160W$} &
\multicolumn{2}{r}{$F160W\rightarrow R$} \nl
751.72&23.90(0.08)&  7.9&-1.95(0.02)\nl
\hline
\multicolumn{4}{c}{SN 2002ki} \nl
\hline
\multicolumn{1}{c}{ } & \multicolumn{1}{l}{$F775W$} &
\multicolumn{2}{r}{$F775W\rightarrow U$} \nl
600.76&24.76(0.10)& -0.3&-0.47(0.02)\nl
643.57&26.78(0.17)& 19.6&-0.48(0.02)\nl
\multicolumn{1}{c}{ } & \multicolumn{1}{l}{$F850LP$} &
\multicolumn{2}{r}{$F850LP\rightarrow B$} \nl
600.76&23.85(0.04)& -0.3&-1.50(0.02)\nl
643.57&25.78(0.10)& 19.6&-1.38(0.02)\nl
652.21&25.98(0.10)& 23.7&-1.34(0.02)\nl
663.68&26.88(0.20)& 29.0&-1.34(0.02)\nl
\multicolumn{1}{c}{ } & \multicolumn{1}{l}{$F160W$} &
\multicolumn{2}{r}{$F160W\rightarrow I$} \nl
652.44&24.77(0.25)& 23.8&-1.65(0.02)\nl
664.47&24.96(0.25)& 29.4&-1.55(0.03)\nl
\hline
\multicolumn{4}{c}{SN 2003ak} \nl
\hline
\multicolumn{1}{c}{ } & \multicolumn{1}{l}{$F850LP$} &
\multicolumn{2}{r}{$F850LP\rightarrow U$} \nl
627.35&27.33(0.40)&-14.0&-0.72(0.02)\nl
673.13&25.63(0.08)&  3.8&-0.70(0.02)\nl
680.12&25.82(0.08)&  6.6&-0.69(0.02)\nl
694.48&26.52(0.12)& 12.2&-0.65(0.02)\nl
\multicolumn{1}{c}{ } & \multicolumn{1}{l}{$F110W$} &
\multicolumn{2}{r}{$F110W\rightarrow B$} \nl
681.06&25.15(0.10)&  7.0&-1.72(0.02)\nl
693.01&25.25(0.10)& 11.6&-1.75(0.02)\nl
708.48&25.60(0.15)& 17.7&-1.82(0.04)\nl
715.35&26.02(0.15)& 20.4&-1.87(0.02)\nl
\multicolumn{1}{c}{ } & \multicolumn{1}{l}{$F160W$} &
\multicolumn{2}{r}{$F160W\rightarrow V$} \nl
681.26&24.10(0.05)&  7.0&-2.32(0.02)\nl
693.41&24.30(0.07)& 11.8&-2.28(0.02)\nl
701.61&24.70(0.08)& 15.0&-2.25(0.07)\nl
\hline
\multicolumn{4}{c}{SN 2002hp} \nl
\hline
\multicolumn{1}{c}{ } & \multicolumn{1}{l}{$F775W$} &
\multicolumn{2}{r}{$F775W\rightarrow U$} \nl
579.49&25.49(0.10)&  3.2&-0.05(0.04)\nl
\multicolumn{1}{c}{ } & \multicolumn{1}{l}{$F850LP$} &
\multicolumn{2}{r}{$F850LP\rightarrow B$} \nl
537.31&26.80(0.39)&-15.0&-1.50(0.03)\nl
579.49&24.29(0.04)&  3.2&-1.30(0.02)\nl
589.05&24.86(0.06)&  7.3&-1.26(0.02)\nl
595.38&25.04(0.08)& 10.1&-1.25(0.02)\nl
603.81&25.65(0.08)& 13.7&-1.22(0.02)\nl
613.80&26.07(0.14)& 18.1&-1.13(0.05)\nl
629.30&26.59(0.17)& 24.8&-1.02(0.02)\nl
639.38&27.29(0.35)& 29.2&-1.01(0.02)\nl
\multicolumn{1}{c}{ } & \multicolumn{1}{l}{$F110W$} &
\multicolumn{2}{r}{$F110W\rightarrow V$} \nl
589.17&24.29(0.07)&  7.4&-1.54(0.03)\nl
595.50&24.44(0.07)& 10.1&-1.49(0.03)\nl
\hline
\multicolumn{4}{c}{SN 2002fw} \nl
\hline
\multicolumn{1}{c}{ } & \multicolumn{1}{l}{$F775W$} &
\multicolumn{2}{r}{$F775W\rightarrow U$} \nl
536.80&25.34(0.05)& -5.8&-0.24(0.02)\nl
548.30&24.98(0.08)& -0.8&-0.16(0.04)\nl
578.40&26.41(0.10)& 12.2&0.09(0.04)\nl
\multicolumn{1}{c}{ } & \multicolumn{1}{l}{$F850LP$} &
\multicolumn{2}{r}{$F850LP\rightarrow B$} \nl
536.80&24.57(0.05)& -5.8&-1.43(0.02)\nl
548.30&24.25(0.06)& -0.8&-1.43(0.03)\nl
552.30&24.47(0.06)&  0.8&-1.43(0.03)\nl
557.40&24.44(0.06)&  3.0&-1.39(0.02)\nl
567.80&24.56(0.08)&  7.6&-1.36(0.02)\nl
577.50&24.99(0.06)& 11.8&-1.34(0.02)\nl
578.50&25.13(0.07)& 12.2&-1.33(0.02)\nl
595.40&25.97(0.16)& 19.6&-1.19(0.04)\nl
603.40&26.88(0.25)& 23.0&-1.09(0.02)\nl
628.00&27.31(0.30)& 33.7&-1.06(0.02)\nl
\multicolumn{1}{c}{ } & \multicolumn{1}{l}{$F110W$} &
\multicolumn{2}{r}{$F110W\rightarrow V$} \nl
549.51&24.15(0.08)& -0.3&-1.69(0.02)\nl
557.57&24.25(0.08)&  3.1&-1.66(0.03)\nl
\multicolumn{1}{c}{ } & \multicolumn{1}{l}{$F160W$} &
\multicolumn{2}{r}{$F160W\rightarrow R$} \nl
549.75&23.92(0.08)& -0.2&-1.92(0.05)\nl
557.76&23.85(0.09)&  3.2&-1.86(0.03)\nl
\enddata 
\tablenotetext{a}{Actually JD$-$2,450,000. \\ Uncertainties in magnitudes are listed
in parentheses.}
\end{deluxetable}

\begin{deluxetable}{lllll} 
\footnotesize
\tablecaption{Spectroscopic Data}
\tablehead{\colhead{SN}&\colhead{UT
Date}&\colhead{Instrument}&\colhead{exposure(sec)}&\colhead{$z$}}
\startdata
\hline
\hline

2002fw & Sep. 31, 2002 & {\it HST} ACS	 & 15000 & 1.30$^{a,1}$  \nl
2002fx & Sep.14, 2003 & Keck-II NIRSPEC &  2000 &  	1.40$^{b,3}$ \nl
2002hp & Nov. 7, 2002 & Keck-I LRIS & 7800 & 1.305$^{b,2}$ \nl
          & Nov. 7, 2002 & VLT FORS & 14000 & 1.305$^{b,2}$ \nl
2002hr & Nov. 8, 2002 & Keck-I  LRIS & 7800 & 0.526$^{c,1}$   \nl
2002kc & Jan. 7, 2003 & Keck-I  LRIS & 1500 & 0.216$^{c,1}$  \nl
2002kd & Jan. 1, 2003 & Magellan LDSS & 7200 & 	 0.735$^{c,1}$  \nl
2002ki & 	Jan. 7, 2003 & Keck-I LRIS & 2700 & 1.141$^{c,1}$  \nl
2003aj &  Oct 1-3, 2003 & 	VLT FORS2 &  16800 & 1.307$^{b,4}$  \nl
2003ak & Sep. 11, 2003 & Keck-II NIRSPEC, VLT FORS2 &  14000 & 1.551$^{b,3}$  \nl
2003az &	Mar. 3, 2003 & {\it HST} ACS & 6500 & 1.27$^{a,1}$ \nl
2003bd &	Feb. 27/28, 2003 & Keck-I LRIS & 16500 &   0.67$^{a,1}$  \nl
2003be &	Feb. 28, 2003 & Keck-I LRIS & 5400 & 0.64$^{c,1}$  \nl
2003dy &	Apr. 16, 2003& {\it HST} ACS	 & 15000 & 1.34$^{d,1}$  \nl
2003lv &	Apr. 16, 2003& {\it HST} ACS	 & 15000 & 0.935$^{d,2}$  \nl
2003eb &	Apr. 16, 2003& {\it HST} ACS	 & 15000 & 0.899$^{d,1}$  \nl
2003eq &	Jun. 2, 2003& {\it HST} ACS	 & 6000 & 0.839$^{a,1}$  \nl
2003es &	Jun. 2, 2003& {\it HST} ACS	 &  6000 & 0.954$^{d,1}$  \nl
\enddata 
\tablenotetext{a}{From cross-correlation with broad SN features.}
\tablenotetext{b}{From narrow features in the host-galaxy spectrum.}
\tablenotetext{c}{From both (a) and (b).}
\tablenotetext{d}{From (a) and Cowie et al. (2004), Wirth et al (2004).}
\tablenotetext{1}{Classified as SN~Ia with high confidence from
spectrum.}
\tablenotetext{2}{Classified as SN~Ia with high confidence from early-type, red host.}
\tablenotetext{3}{Photometric properties indicate likely SN~Ia.}
\tablenotetext{4}{Uncertain type.}
\tablenotetext{ }{For VLT host spectra of 2003ak,aj see Nonino et al 2004}
\end{deluxetable}

\section{Light-Curve Fitting}

   Distances to SNe~Ia with individual precision approaching 7\% can be measured utilizing
empirical relationships between light-curve shape and luminosity as well as
color-curve shape and extinction.  Our primary analysis uses a revision of the
multi-color light-curve shape (MLCS) fitting method (Riess, Press, \& Kirshner
1995, 1996a; Riess et al. 1998) described by Jha (2002) and Jha et al. (2004a),
and hereafter referred to as ``MLCS2k2.''  Previous versions of MLCS fit rest-frame $BVRI$, but MLCS2k2 includes $U$ band templates based on a new set of 25 well-observed SNe~Ia in
the $U$ band from Jha (2002) and Jha et al. (2004b).  This UV extension of MLCS
allows us to extend the Hubble diagram of SNe~Ia
to $z > 1$ using light curves observed in the reddest available band on ACS.
The results from Jha (2002) demonstrate that rest-frame $U$-band light curves
provide similar information on the epoch of maximum, relative luminosity,
reddening, and distance as optical light curves, albeit with lower
precision (by a factor of 1.5).  Because the MLCS2k2 covariance matrix contains the
bandpass-dependent variance of SNe~Ia (primarily in the form of autocorrelation
along the diagonals as well as in the form of two-point covariance in
off-diagonal terms), the $U$-band light curves can be used together with
optical light-curve data with the appropriate weight and propagation of
uncertainty in the multi-color fit.

Additional improvements to MLCS2k2 include a more self-consistent treatment of
the allowed range of extinction and extinction laws, as well as an improved
determination of the unreddened SN~Ia color (see Jha et al. 2004a for details).
This method still empirically models a light curve as the sum of a fiducial
template and a set of phase-dependent vectors (linear and quadratic) whose
contribution scales with the luminosity offset from the fiducial curve at peak.
Luminosity corrections are not extrapolated beyond the range observed in the
local sample.  Simultaneous fitting in multiple colors constrains the
line-of-sight reddening and distance.

Extinction priors are used together with the observed reddening to estimate the expected dimming of the SN
magnitudes resulting from dust.  Recent analyses by the
high-z supernova search team (HZT; Schmidt et al. 1998) and the supernova
cosmology project (SCP; Perlmutter et al. 1997), published respectively by
Tonry et al. (2003) and Knop et al. (2003), make use of a Galactic prior for
the extinction law (which is corroborated at low redshifts; Riess, Press,
Kirshner 1996b) and flag SNe with large reddening to reduce systematic errors
from non-Galactic-type dust.  The HZT has employed an exponential extinction prior whose functional form derives from modeled lines-of-sight (Hatano,
Branch, \& Deaton 1998) and from the {\it a posteriori} distribution of
extinction values.  Here we
 utilize the exponential prior on extinction and the gaussian prior on the intrinsic color derived from the {\it a
posteriori} distribution of Jha et al. (2004a) and flag SNe with large measured
reddening as unreliable.  Each SN Ia is corrected for galactic reddening as estimated in the direction of each SN by Schlegel, Finkbeiner, \& Davis (1999) before the observed colors are used to estimate the host reddening.  

K-corrections are used to account for the SN redshift and provide a
transformation between an observed-frame magnitude and a rest-frame magnitude
(Oke \& Sandage 1968).  We use composite spectra of SNe Ia from Nugent et
al. (2003) to calculate Vega-normalized ``cross-band'' K-corrections.
Individual K-corrections were calculated to the best-matching passbands for the
appropriate phase and redshift of the SN.  Each SN is then fit using MLCS2k2 to
provide a custom model of the phase-dependent colors.  The colors of the Nugent
et al. (2003) spectral energy distributions are then matched to the model by
multiplication of a spline-interpolation.  Next, the K-corrections are
recalculated, and the process continues until convergence (usually after two to
three iterations).  The final K-corrections are given for each measured
magnitude in Table 2.  The MLCS2k2 fits to each of the 16 new SNe~Ia are shown
in Figure 2 and the fit parameters are given in Table 4.

We have also used an additional light-curve fitting method, ``Bayesian Template
Method'' (BATM; Tonry et al. 2003), to provide independent estimates of the
luminosity distances of the new SN~Ia data presented here.  BATM is a
``template-fitting'' method which seeks to identify close matches between an
individual SN~Ia and a well-observed, local
 counterpart and then calculates their
distance ratio by synthetically redshifting the nearby template to the observed
frame.  The current realization of BATM has far fewer $U$-band examples than
MLCS2k2, because BATM was developed without the Jha (2002) data.  Consequently, BATM did
not converge on a solution for three sparsely observed SNe~Ia (SNe 2002kc,
2003be, and 2002fx), whose MLCS2k2 fits relied on sampling in the rest-frame
$U$ band.

\subsection{The Ground-Based Discovery Set}

We construct the expansion history of the Universe by using this new set of HST-discovered objects together with published observations of supernovae over a wide range in redshift.  Tonry et al. (2003) have recently compiled the distances and redshifts for 172 SNe Ia.  Using available results from different light-curve
fitting methods (including BATM, MLCS, $\Delta m_{15}$, snapshot, and stretch)
for each SN Ia, Tonry et al. (2003) corrected for zeropoint differences
between methods and provided best estimates of the distance to each SN~Ia from
a median of the distance estimates from individual methods.

Although the Tonry et al (2003) data set represented the state of the art in February, 2003, when it was submitted, there have been some significant developments since then that need to be included to build the most relaible data set for analyzing the HST-discovered objects.  For the SCP, Knop et al. (2003) report on a new set of 11 SNe~Ia
at $0.4 < z < 0.85$ as well as on a reanalysis of the original high-redshift
SNe~Ia from the SCP (Perlmutter et al. 1999).  In this reanalysis they now
exclude 15 of the 42 high-redshift SNe from Perlmutter et al. (1999) due to
inaccurate color measurements and uncertain classification.  Knop et al. (2003)
flag an additional 6 of the original 42 SNe, as well as 5 of the new 11 SNe, as
likely SNe~Ia but failing a ``strict SN~Ia'' sample cut.  For the HZT, Barris
et al. (2004) report on a large set of new high-redshift SNe (22 in all), with
widely varying degrees of completeness of the spectroscopic and photometric records.
Finally, Blakeslee et al. (2003) report on 2 new SNe~Ia discovered with ACS on
{\it HST}.  The development of the MLCS2k2 method, which includes $U$-band observations when they are available, makes it worthwhile to revisit the previously published data.

We recompiled a set of previously observed SNe~Ia relying on large, published
samples, whenever possible, to reduce systematic errors from differences in
calibration.  To compile SNe~Ia at $0.01 < z < 0.15$ we used the 3 largest,
modern data sets of such SNe~Ia published to date: the Cal\'{a}n-Tololo Survey
(29 SNe~Ia; Hamuy et al. 1996), the CfA Survey I (22 SNe~Ia; Riess et
al. 1999a), and the CfA Survey II (44 SNe~Ia; Jha et
al. 2004b).\footnote[9]{Note that the CfA SNe were generally discovered by
other searches, especially the Lick Observatory SN Search with the Katzman
Automatic Imaging Telescope (Filippenko et al. 2001; Filippenko 2003).}  At
higher redshifts we used SN~Ia photometry published by Riess et al. (1998,
2001), Tonry et al.  (2003), Knop et al. (2003), and Barris et al. (2004), as
well as SNe~Ia tabulated by Tonry et al. (2003) from Suntzeff et al. (2004),
Leibundgut et al. (2004), Clochiatti et al. (2004), and Jha et al. (2004c).
Despite the apparently large number of sources of data, the majority of the
data at $z>0.1$ comes from the HZT (Schmidt et al. 1998; Riess et al. 1998),
and thus the methods used to calibrate all of these SNe~Ia are extremely
similar (and familiar to the authors of the present paper).

In order to reduce systematic errors which arise from differences in
light-curve fitting (and K-correcting) methods, we have made every effort to
consistently refit all the past data with a single method, MLCS2k2 (Jha et
al. 2004a). An exception was made for the high-redshift SNe~Ia from the SCP
(Perlmutter et al. 1999) because photometry of these SNe~Ia remain
unpublished.  For these we utilized the reanalyzed distances as given by Knop
et al. (2003).  We transformed them to the MLCS2k2 distance scale by solving
for the SN~Ia luminosity zero-point required to match the mean distances to
low-redshift objects fit by the two methods.

  Unfortunately, a large variation exists in the quality and breadth of the
photometric and spectroscopic records of individual SNe.  Ideally, each SN~Ia
would have the same well-defined spectroscopic features (Filippenko 1997), a
spectroscopic redshift, and well-sampled light curves and color curves.
However, this is often not the case.  As with the photometry,
 most of the extant
high-redshift spectra from the SCP have not been published (e.g., Perlmutter et
al. 1999; Knop et al. 2003).  This makes it difficult to apply a uniform set
of criteria to SNe~Ia in constructing a cosmological sample.

  To reflect the differences in the quality of the spectroscopic and photometric record for individual supernovae, we divide the objects into ``high-confidence'' (hereafter ``gold") and ``likely but not certain'' (hereafter
``silver") subsets.  Ideally, we would assign each supernova a weight in any overall fit that reflected its individual uncertainty.  However, distance errors resulting from
spurious problems more common to lower-confidence SNe~Ia such as SN
misclassification, large extinction (amplifying uncertain extragalactic
extinction laws), and poorly constrained colors are difficult to quantify.  So we use the coarser approach of separating the high-confidence gold events from the larger set.

   We adopted in our gold set all SNe~Ia which were included by the original
sources in their most stringent subsets (when such discriminations were made).
Any SN~Ia flagged as having a cause for a specific concern was not included in this set.  The two primary reasons for
rejecting a SN~Ia from this set are that (1) the classification, though
plausible, was not compelling (see discussion), and (2) the photometric record
is too incomplete to yield a robust distance (i.e., the number of model
parameters is roughly equal to the number of effective samplings of the light curve).

SNe~Ia included in previous cosmological samples but rejected from our gold
sample include SN 1999fh (poorly constrained light curve; Tonry et al. 2003),
SN 1997ck (poor color information; Garnavich et al. 1998), all ``snapshot''
SNe~Ia from Riess et al. (1998), 15 SNe~Ia from Perlmutter et al. (1999) later
discarded by Knop et al. (2003) as well as 11 additional SNe flagged by Knop et
al. (2003; flag values $1,2,3$), SNe~Ia from Barris et al. (2003) without SNID
classification, and any SN~Ia with more than 1 mag of extinction or whose light
curve begins more than 10 rest-frame days after maximum as determined from the
MLCS2k2 fit.

 The same criteria were applied to the GOODS SNe~Ia whose individual
classifications were described in \S 2.3.  As a result, two of these SNe were
rejected from the gold sample: SN 2002fx, whose classification was not certain
enough, and SN 2002kc, whose fit indicated $> 1$ mag extinction.  The gold set
contains a total of 157 SNe~Ia.

The silver set contains the objects identified by the above sources as likely
SNe~Ia but failing one criterion for inclusion in the gold category.  The
silver set includes a total of 29 SNe~Ia.  SNe failing more than one criterion
were excluded from the analyses. The final membership rosters of the subsets
are tabulated in the Appendix.

For most of our cosmological analyses we focus on results derived from the gold
set, but for a few analyses with the largest number of free parameters (and
thus the most limited in statistical inference) we include the silver set (with the caveats arising from its reduced reliability). 

\section{Cosmological Constraints}

   Distance estimates from SN~Ia light curves are derived from the luminosity
distance, \bq d_{L} = \left(\frac{{\cal L}}{4 \pi {\cal
F}}\right)^{\frac{1}{2}}, \eq where ${\cal L}$ and ${\cal F}$ are the intrinsic
luminosity and observed flux of the SN within a given passband, respectively.  Equivalently,
logarithmic measures of the flux (apparent magnitude, $m$) and luminosity
(absolute magnitude, $M$) were used to derive extinction-corrected distance
moduli, $\mu_0=m-M=5\log d_L +25 $ ($d_L$ in units of megaparsecs).  In this context, the
luminosity is a ``nuisance parameter'' whose value is unimportant for kinematic
(and most cosmological) studies.  We have used the MLCS2k2 method and the data
described in \S 2 to derive accurate and individual {\it relative} distance
moduli for the sets of SNe described in \S 3.

In Figure 4 we show the Hubble diagram of distance moduli and redshifts for the
new HST-discovered SNe~Ia in the gold and silver sets.  Although these new SNe~Ia span a
wide range of redshift ($0.21 < z < 1.55$), their most valuable contribution to
the SN~Ia Hubble diagram is in the highest-redshift region where they
effectively delineate the range at $0.85 < z < 1.55$ with 11 new SNe~Ia,
including 6 of the 7 highest-redshift SNe known (the seventh being SN 1997ff;
Riess et al. 2001).

The relationship between distance and redshift over a significant fraction of
the Hubble time can be considered either empirically as a record of the
(integrated) expansion history of the Universe, or theoretically as constraints
on the mass-energy terms contained in the Friedman equation and affecting the expansion.  In the
next subsections we consider both approaches.

\subsection{Expansion History: A Kinematic Description}

It is valuable to consider the distance-redshift relation of SNe~Ia as a purely
{\it kinematic} record of the expansion history of the Universe, without regard
to its cause.  An {\it empirical} description of the time variation of the
scale factor, $a(t)$, can provide answers to basic questions (e.g., ``When was
the Universe (if ever) accelerating or decelerating?'') and model-independent constraints with which to
test cosmological models.

Following Turner \& Riess (2002), we empirically define the luminosity distance
in Euclidean space (i.e., $\Omega_{\rm total} = 1.0$; as motivated by inflation) as the integral of the inverse of the preceding expansion
rate,

\begin{equation}
d_L =c(1+z)\int_0^z {du\over H(u)} =
c(1+z)H_0^{-1} \int_0^z\, \exp \left[-\int_0^u\, [1+q(u)]d\ln (1+u) \right]\, du,
\label{eq:dLq}
\end{equation}

\noindent
where

\begin{equation}
H(z) = {{\dot a} \over a},
\end{equation}

\begin{equation}
q(z) \equiv (-\ddot a /a)/H^2(z) = {d H^{-1}(z) \over dt} -1.
\label{eq:q(z)}
\end{equation}

\noindent 
Note that equation (2) is not an approximation but is an exact expression for
the luminosity distance in a geometrically flat Universe (though generalizable
for non-zero curvature), given an expression for the epoch-dependent
deceleration parameter, $q(z)$, and the present Hubble constant, $H_0$.  Here
we employ equation (2) as a kinematic model of the SN~Ia data with parametric
representations for $q(z)$.

Given evidence that the Universe has recently been
 accelerating [i.e., $q(z\sim 0) <
0$], hints that it may have once been decelerating [i.e., $q(z>1) > 0$; Riess
et al. 2001; Turner \& Riess 2002], and the large leverage in redshift
of the current SN sample, we consider resolving $q(z)$ into two distinct
components or epochs.  A linear two-parameter expansion for $q(z)$ which is
continuous and smooth is $q(z)=q_0+z{dq/dz}$, where ${dq/dz}$ is
defined to be evaluated at $z=0$.

The likelihood for the parameters $q_0$ and ${dq/dz}$ can be determined
from a $\chi^2$ statistic, where \bq \chi^2(H_0,q_0,{dq/dz})=\sum_i 
{ (\mu_{p,i}(z_i; H_0, q_0, {dq/dz})-\mu_{0,i})^2 \over
\sigma_{\mu_{0,i}}^2+\sigma_v^2} \eq, $\sigma_v$ is the dispersion
in supernova redshift (transformed to units of distance moduli) due to peculiar
velocities and $\sigma_{\mu_{0,i}}$ is the uncertainty in the individual distance moduli. This term also includes the uncertainty in galaxy
redshift.  Due to the extreme redshift of our distant sample and the abundance
of objects in the nearby sample, our analysis is insensitive to the value
we assume for $\sigma_v$ within its likely range of 200 km s$^{-1}$ $\leq 
\sigma_v \leq$ 500 km s$^{-1}$.  For our
analysis we adopt $\sigma_v=400$ km s$^{-1}$.  For high-redshift SNe~Ia 
whose redshifts were determined from the broad features in the SN
spectrum, we add 2500 km s$^{-1}$ in quadrature to $\sigma_v$.

Marginalizing our likelihood functions over the nuisance parameter, $H_0$ (by
integrating the probability density $P \propto e^{- \chi^2 / 2}$ for all values
of $H_0$), yields the confidence intervals shown in Figure 5.  As shown, both
the gold set or the gold and silver sets strongly favor a
Universe with recent acceleration ($q_0 < 0$) and previous deceleration
(${dq/dz} > 0$) with 99.2\% and 99.8\% likelihood (summed within this
quadrant), respectively. With this same model we can also derive the likelihood
function for the transition redshift, $z_t$, defined as $q(z_t)=0$.  Summing
the probability density in the $q_0$ vs. ${dq/dz}$ plane along lines of
constant transition redshift, $z_t = -q_0/(dq/dz)$, yields the likelihood
function in Figure 5.  We find a transition redshift of $z_t = 0.46 \pm 0.13$.
In Figure 6 we show the Hubble diagram for the SNe~Ia compared to a
discrete set of kinematic models.

An alternate, kinematic model is derived using the first three time derivatives
of the scale factor. Following Visser (2003), the Hubble, deceleration, and
jerk parameters are defined as

\begin{equation}
H(t) = +{\dot a} /a 
\; ,
\end{equation}

\begin{equation}
q(t) = -({\ddot a}/a)({\dot a}/a)^{-2}
\; ,~{\rm and}
\end{equation}

\begin{equation}
j(t) = +({\dot {\ddot a}}/a)({\dot a}/a)^{-3}
\; .
\end{equation}

The deceleration and jerk parameters are dimensionless, and 
a Taylor expansion of the scale factor around $t_0$ provides
 
\begin{eqnarray} 
a(t)= a_0 \;
\left\{ 1 + H_0 \; (t-t_0) - {1\over2} \; q_0 \; H_0^2 \;(t-t_0)^2 
+{1\over3!}\;  j_0\; H_0^3 \;(t-t_0)^3 + O([t-t_0]^4) \right\},
\nonumber\\
\end{eqnarray}

\noindent
and hence for the luminosity distance (in Euclidean space),

\begin{equation}
d_L(z) =  {c\; z\over H_0}
\left\{ 1 + {1\over2}\left[1-q_0\right] {z} 
-{1\over6}\left[1-q_0-3q_0^2+j_0 \right] z^2
+ O(z^3) \right\}
\end{equation}
(cf. Visser 2003).

Though related, the $j_0$ parameter as defined here and by Visser (2003) is not
precisely equivalent to our previous ${dq/dz}$ parameter, providing an
alternative parameterization.  The SN subsets constrain the $j_0$ parameter to
the positive domain at the 92\% to the 95\% confidence level.  That is, the expansion history over the range of the SN data is equally well-described by recent acceleration and a constant jerk.  Models with
discrete values of $j_0$ are shown in Figure 6.

In summary, we find strong evidence for a change in the sign of cosmic
acceleration in the past. 

\subsection{Cosmological Constant or Astrophysical Dimming?}

SNe~Ia at $z \approx 0.5$ appear fainter by $\sim$0.25 mag relative to a
Universe with $\Omega_M=0.3$ and $\Omega_\Lambda = 0$, a result readily
accommodated by a cosmological constant with $\Omega_\Lambda=0.7$ (Riess et
al. 1998; Perlmutter et al. 1999).  Despite the lack of any independent evidence, an alternative explanation for this dimming could lie in the astrophysics of supernovae or in the propogation of their light to us.  Speculative models for astrophysical contamination of the SN~Ia signal
have been posited; these include extragalactic gray dust with negligible
tell-tale reddening or added dispersion (Aguirre 1999a,b; Rana 1979, 1980), and
a pure luminosity evolution (Drell, Loredo, \& Wasserman 2000).  Here we limit
our consideration to the observable differences between these hypotheses.

The luminosity distance expected in a Friedmann-Robertson-Walker (FRW) cosmology
with mass density $\Omega_M$ and vacuum energy density (i.e., the cosmological
constant) $\Omega_\Lambda$ is
\bq
 d_L= c H_0^{-1}(1+z)\left | \Omega_k \right |^{-1/2}{\rm sinn}\lbrace\left | 
\Omega_k \right |^{1/2}
\int_0^zdz[(1+z)^2(1+\Omega_Mz)-z(2+z)\Omega_\Lambda ]^{-1/2}\rbrace,
\eq
where $\Omega_k=1-\Omega_M-\Omega_\Lambda$, and ``sinn'' is $\sinh$ for
$\Omega_k > 0$ and $\sin$ for $\Omega_k < 0$
(Carroll, Press, \& Turner 1992). For $\Omega_k=0$, 
equation (11) reduces to $c H_0^{-1}(1+z)$ times the integral.  
With $d_L$ in units of megaparsecs, the predicted 
distance modulus is \bq \mu_p=5\log d_L +25
.\eq

Following Goobar, Bergstrom, \& Mortsell (2002) we consider two models of gray
extinction by a homogeneous component of dust:
 $\rho_{\rm dust}(z) = \rho^\circ_{\rm dust}(1+z)^\alpha$, where 
\begin{eqnarray*}
 \alpha(z) =  
      \left\{ 
        \begin{array}{lc}
          3 {\rm \ for \ all \ }  z &  {\rm ``high}-z  \ {\rm dust,''} \\
          0 {\rm \ for \ }  z > 0.5 {\rm \ (3 \ for \ lower \ } z ).  &
      {\rm ``replenishing \ dust.''} 
       \end{array}
        \right.
\end{eqnarray*}
\noindent

The ``high-z dust'' model represents a smooth background of dust present (presumably ejected from
galaxies) at a redshift which is greater than the SN sample (i.e., $z > 2$) and
diluting as the Universe expands.  
The total extinction is then calculated as the attenuation 
integrated along the photon path, $-2.5\, {\rm log}\,
({\rm exp}({\int_0^z \rho_{\rm
dust}(z) r(z) dz})),$ where $r(z)$ is the coordinate distance traversed
by the SN photons.  A single free (opacity) parameter is fixed by requiring the total
extinction at $z \approx 0.5$ to match the observed peak brightness of SNe~Ia
in a cosmology with $\Omega_\Lambda=0$. 

 The ``replenishing dust'' represents a constant density
of dust which is continually replenished at precisely the same rate in which it is
diluted by the expanding Universe (i.e., $\alpha$ = 0).  This model is also
tuned to match the extinction implied by SNe Ia at $z \approx 0.5$ in the
absence of a cosmological constant, hence it requires the tuning of two
parameters (as well as fast-moving dust which quickly provides a homogeneous
background without added dispersion from uneven lines-of-sight).  We also
consider a third model (following Filippenko \& Riess 2001, Riess et al. 2001,
and Blakeslee et al. 2003) to mimic simple evolution which scales as $z$ in
percent dimming.  Our set of models for astrophysical dimming is not an
exhaustive set of all possibilities, but rather is drawn from {\it physically}
motivated hypotheses (in contrast to purely parametric models of astrophysical
dimming, e.g., Drell et al. 2000).

\begin{deluxetable}{ll} 
\tablecaption{$\chi^2$ Comparison of Gold Set Data to Models }
\tablehead{\colhead{Model}&\colhead{$\chi^2$(for 157 SNe Ia)}}
\startdata
$\Omega_M=0.27, \Omega_\Lambda=0.73$ & 178 \nl
$\Omega_M=1.00, \Omega_\Lambda=0.00$ & 325 \nl
$\Omega_M=0.00, \Omega_\Lambda=0.00$ & 192 \nl
High-redshift gray dust (with $\Omega_M=1.00, \Omega_\Lambda=0.00$) & 307 \nl
Replenishing dust (with $\Omega_M=1.00, \Omega_\Lambda=0.00$) & 175 \nl
Dimming $\propto z$ (with $\Omega_M=1.00, \Omega_\Lambda=0.00$)  & 253 \nl 
\enddata 
\end{deluxetable}

In Figure 7 we show the Hubble diagram of SNe~Ia relative to the cosmological
and astrophysical hypotheses.  As seen in Table 4, the SN dataset is consistent
with an $\Omega_M=0.27, \Omega_\Lambda=0.73$ cosmology, yielding $\chi^2=178$
for 157 SNe~Ia (degrees of freedom, dof; $\chi_{dof}^2=1.13$) in the gold set.
The total $\chi^2 $ is significantly worse for the high-redshift gray dust
model ($\Delta \chi^2=122$; 11$\sigma$ for 1 dof) as well as for the simple
model of evolution with dimming $\propto z$ ($\Delta \chi^2=70$; 8$\sigma$ for
1 dof), allowing us to reject both hypotheses with high confidence.
Interestingly, the ``replenishing dust'' model is nearly indistinguishable from
an $\Omega_\Lambda$ model because the dimming is directly proportional to
distance traveled and thus mathematically quite similar to the effects of a
cosmological constant.  Consequently, we cannot discriminate this model from an
$\Omega_\Lambda$-dominated model strictly from its behavior in the
magnitude-redshift plane (and probably never will be able to, given the small
magnitude differences).  However, the fine tuning required of this dust's
opacity, replenishing rate, and velocity ($> 1000$ km s$^{-1}$ for it to fill
space uniformly without adding detectable dispersion) makes it unattractive as
a simpler alternative to a cosmological constant.

\subsection{Exploring Dark Energy}

Despite the results of the last section which favor the dark-energy
interpretation of SNe~Ia, we avoid using this conclusion as a starting point
for exploring the nature of dark energy.  To do so would be to engage in ``circular
reasoning'' or to incur more free parameters than our limited dataset can
usefully constrain.  Instead, we embark on a parallel study from the previous
section.  Here we use distance-independent information to justify the
cosmological interpretation of SNe~Ia and combine with other experiments to
study dark energy.

  The potential for luminosity evolution of corrected SN~Ia distances has been
studied using a wide range of local host environments.  No dependence of the
distance measures on the host morphology, mean stellar age, radial distance
from the center, dust content, or mean metallicity has been seen (Riess et
al. 1998; Perlmutter et al. 1999; Hamuy et al 2000).  No differences in the
inferred cosmology were seen by Sullivan et al. (2003) for SNe~Ia in early-type
hosts or late-type hosts at high redshifts. These studies limit morphology
dependence of SN~Ia distances to the 5\% level.  Detailed studies of
distance-independent observables of SNe~Ia such as their spectral energy
distribution and temporal progression have also been employed as probes of
evolution; see Riess (2000), Leibundgut (2001), and Perlmutter \& Schmidt
(2003) for reviews.  The consensus interpretation is that there is no evidence
for evolution with limits at or below the statistical constraints on the
average high-redshift apparent brightness of SNe~Ia.  The observed nominal
dispersion of high-redshift SNe~Ia substantially limits the patchiness of
uncorrected extinction, and near-IR observations of a high-redshift SN~Ia
demonstrate that a large opacity from grayish dust is unlikely (Riess et
al. 2001).  Non-SN constraints on gray dust from QSOs observed in X-rays
(Paerels et al. 2002; Ninomiya, Yaqoob, \& Khan 2003) and a partial or complete
resolution of the far-IR background by SCUBA (Chapman et al. 2003) place
stringent limits of less than a few percent of dimming at $z=0.5$ from gray dust.

Based on this evidence, we will adopt in the following analysis an {\it a
priori} constraint that the net {\it astrophysical} contamination of SN~Ia
distance measures does not exceed their statistical uncertainty in their mean
brightness.  Quantitatively, our adopted limit on systematics is defined to be
5\% per $\Delta z$ at $z>0.1$.

First we consider the SN data within an FRW cosmology of unknown curvature and
mass density (with a flat prior on all parameters), with the simplest description of a dark-energy component (i.e., a
cosmological constant) using equation (11).  Joint confidence intervals in the
$\Omega_M - \Omega_\Lambda$ plane were derived after numerical integration of
the probability density $P(H_0) \propto {\rm exp}(-{\chi^2(H_0)/2})$ over all
values of $H_0$ and are shown in Figure 8.  Compared to the same analysis in
Riess et al. (1998), the gold sample presented here reduces the area of the $1
\sigma$ contour by a factor of 6 (a factor of 7 including the silver
sample).  With the current sample, the $4 \sigma$ confidence intervals (i.e.,
$> 99.99\%$ confidence) are now fully contained within the region where
$\Omega_\Lambda > 0$.  The ``concordance'' model,  of $\Omega_M=0.27,
\Omega_\Lambda=0.73$ lies within the 1$\sigma$ contour (though just outside of
it with the addition of the silver set).  For a flat geometry prior, we measure
$\Omega_M=0.29 \pm ^{0.05}_{0.03}$ (equivalently $\Omega_\Lambda=0.71$).  The
{\it HST}-discovered SNe~Ia alone decrease the area of the $1 \sigma$ contour
by a factor of 1.5 (in the gold sample) due to their high mean redshift.

An alternative approach with good precedent (Garnavich et al. 1998; Perlmutter
et al. 1999) is to consider a flat Universe and a generalized dark-energy
component parameterized by its (assumed) constant equation of state, $w={P/\rho
c^2}$.  Flatness is assumed either on theoretical grounds (i.e., as a
consequence of inflation) or on observational grounds from the
characteristic angular size scale of the CMB fluctuations (Spergel et al. 2003, and
references therein).  In this case the luminosity distance is given by

\bq
 d_L= c H_0^{-1}(1+z)
\int_0^z dz[(1+z)^3(\Omega_{M})+(1-\Omega_{M})(1+z)^{3(1+w)} ]^{-1/2}. \eq

We determined the probability density in the $\Omega_M-w$ plane in the same
manner as above and the results are shown in the left panel of Figure 9.  The
SN~Ia data alone require $w < -0.5$ for any value of $\Omega_M$ at the 95\%
confidence level and are consistent with $w=-1$ (i.e., dark energy resembling a
cosmological constant) at the 68\% confidence level for $0.20 < \Omega_M <
0.35$.

Utilizing SN-independent constraints in this plane (primarily to constrain
$\Omega_M$) yields far more precise constraints for $w$ due to the strong
degeneracy between $w$ and $\Omega_M$ for SNe~Ia.  In the left panel of Figure
9 we use $\Omega_M=0.27 \pm 0.04$ (at $z=0$) as a simple approximation to the constraints
derived from numerous SN-independent experiments (see Freedman \& Turner 2003
for a review).  Alternatively (right panel, Figure 9), we used the WMAPext
(Bennett et al. 2003; Spergel et al. 2003) measurement of the reduced distance
to the surface of last scattering at $z=1089$ {\it and} the Two-Degree Field
Galaxy Redshift Survey (2dFGRS) measurement of the growth parameter,
$f=({\Omega_M/ (\Omega_M+(1-\Omega_M)(1+z)^{3w})})^{0.6}$, to derive
independent constraints in the $\Omega_M-w$ plane.  The results from either
approach are similar.  We find $w=-1.02 \pm ^{0.13}_{0.19}$ and $w=-1.08 \pm
^{0.20}_{0.18}$ with the use of an $\Omega_M$ prior and from WMAPext+2dFGRS,
respectively.  Using the prior on $\Omega_M$, the 95\% confidence interval is
$-0.78 > w > -1.46$, or $w < -0.76$ if $w \geq -1$.  The 99\% level constrains
$w < -0.72$.  These results are somewhat more constraining than those from
Tonry et al. (2003), Barris et al. (2004), Knop et al. (2003).

The high relative redshifts of the {\it HST}-discovered SNe~Ia provide little
additional power to constrain a $w$ parameter which is fixed {\it a priori} to
be redshift-independent.  The precision of such a constrained study of dark
energy is most sensitive to the sheer number of SNe~Ia with a relatively weak
dependence on their redshifts (a useful approximation is $\Delta m \approx
[1.8z/(1+e^z)] \Delta w$ for small $\Delta w$, $w \approx -1$, and $z<2$) until
the systematic error limit is reached.

A more ambitious and potentially more revealing approach to studying dark
energy is to allow for both an unconstrained value of the equation of state (at
some fiducial redshift, e.g., $z=0$) and its time evolution, i.e.,
$w(z)=w_0+w'z$, where $w' \equiv {dw \over dz}|_{z=0}$.  This parameterization
provides the minimum possible resolving power to distinguish a cosmological
constant and a rolling scalar field from their time variation (or lack
thereof).  Indeed, rejection of the hypothesis that $w'=0$ would rule out a
cosmological constant as the dark energy (as would the determination that
$w\neq -1$).  The measured value of $w'$ would provide an estimate of the
scale length of a dark-energy potential.  The only previous estimate of $w'$,
by Di Pietro \& Claeskens (2003), used the set of SNe~Ia from Perlmutter et
al. (1999) and the constraints $\Omega_{total} \equiv 1$ and $\Omega_M \equiv 0.3$, and
concluded $-12 < w' < 12$ at the 95\% confidence level (best fit: $w_0=-1.4,
w'=2.3$).

For $w(z)=w_0+w'z$, we employ (following Linder 2003)

\bq
 d_L= c H_0^{-1}(1+z)
\int_0^z dz[(1+z)^3(\Omega_{M})+(1-\Omega_{M})(1+z)^{3(1+w_0-w')}e^{3w'z}
 ]^{-1/2} . \eq
  A strong degeneracy exists among the
three free parameters $w_0$, $w'$, and $\Omega_M$, requiring the use of
independent experimental constraints in this space to make progress.  To this
end we use the previous prior, $\Omega_M=0.27 \pm 0.04$ (at $z=0$), which has the
advantage of being independent of redshift while providing a good 
approximation to all non-SN cosmological constraints in this space.

 We have avoided using the CMB measurements {\it directly} as an additional
constraint on the time evolution of the equation of state of dark energy due to
difficulties which arise in the analysis of the CMB at $z>>1$ where the linear
expansion of the epoch-dependent equation of state diverges.  Because
constraints on $w(z)$ from the CMB are derived from an integration between
$z=0$ and $z=1089$, diverging formulations of $w(z)$ are unsuitable.  Linder
(2003) has proposed a more stable parameterization of $w(z)=w_0+w_a({z/
(1+z)})$; however, for large values of $w_0+w_a$, which are not rejected by the
Di Pietro \& Claeskens (2003) analysis, the CMB integral remains ill-behaved.  In
addition, the CMB measurements provide little direct leverage on $w_a$ or $w'$
compared to SNe~Ia at $z \approx 1$ (because $\Omega_{Dark Energy} \approx 0$ at
$z>>1$).  Therefore, we used the $w'$ parameterization which is well-suited to
our SN sample, and a simple prior on $\Omega_M$ which avoids problems
evaluating functions involving $w(z)$ at $z >>1$.

In Figure 10 we show constraints in the $w_0-w'$ plane (after marginalizing
over $\Omega_M$).  Using the gold subset, we find $w_0=-1.31 \pm
^{0.22}_{0.28}$ and $w'=1.48 \pm ^{0.81}_{0.90}$ with the uncertainties in both parameters strongly correlated.  A cosmological constant
(i.e., $w_0=-1$, $w'=0$) is separated from the best fit along the direction of the major axis of the error ellipse, lying at the boundary of the {\it joint} 68\% confidence level.  If
we constrain the recent behavior of dark energy to be like a cosmological
constant (i.e., $w_0=-1$) we find $w'=0.60 \pm 0.47$. Models with $w_0 < -1$ or
``phantom energy'' (Caldwell et al 2003) violate the Dominant-Energy condition ($\rho+p > 0$) and are extremely
speculative at this point (and may rip apart the Universe in the future), so if
we constrain the analysis to $w_0 > -1$ we find $w'=0.6 \pm 0.5$ and $w_0 <
-0.72$ with 95\% confidence.  Unfortunately, few theoretical predictions about
the size of $w'$ exist for {\it dynamic} models of dark energy.  However, we
can reject the possibility that dark-energy evolution is currently {\it very} rapid (i.e., $\vert w' \vert > $ a few).  This conclusion alone limits the rate at
which simple rolling scalar fields could reach their true minima.  The
consequences of this statement for predicting the future fate of the Universe
are discussed in \S 5.

Our new constraints in the $w' - w$ plane provide a substantial factor of $\sim
8$ improvement over the analysis by Di Pietro \& Claeskens (2003) of
the SCP (Perlmutter et al. 1999) data.  Still, greater precision for the
measurement of $w'$ is needed before the proximity (or separation) from $w'=0$
would provide a compelling, empirical case for (or against) a static dark energy (i.e., a
cosmological constant).  The addition of the silver sample has only a
modest impact on this analysis, as seen in Figure 10 (although a cosmological
constant crosses to just outside the nominal 68\% confidence interval).

The {\it HST}-discovered SNe~Ia provide significant leverage in the $w_0-w'$
plane due to their high mean redshift.  Figure 10 (upper left panel) shows the $w'-w$ plane without including the HST-discovered objects.   The {\it HST}-discovered SNe~Ia alone
increase the precision (i.e., reduce the area of the confidence intervals) by an impressive
factor of 1.9, although they account for only 10\% of the sample.  Previous studies in support of a dedicated, space-based mission
to measure $w_0$ and $w'$ have concluded that a SN~Ia sample must extend to
$z>1.5$ to adequately break degeneracies in this parameter space (Linder \&
Huterer 2003), a conclusion supported by our analysis. The current relative
dearth of SNe~Ia at $z>1$ compared to their number at $z<1$ indicates that
significant progress can still be made in the constraints on $w'$.  Proposals
for a Supernova Acceleration Probe (SNAP) or a Joint Dark Energy Mission (JDEM) predict an improved constraint for
$w'$ over our current analysis by a factor of 3 to 4, assuming a similar-sized
improvement in our knowledge of $\Omega_M$ from the Planck Satellite and a continued ability
to reduce systematic errors (Linder \& Huterer 2003).  However, the current sample is rapidly growing in size and we may expect progress in our constraints on the nature of dark energy in the next few years.

\section{Discussion}

\subsection{Cosmological Constraints}

SNe~Ia at $z>1$ provide valuable and unique contributions
to our current understanding of the cosmological model.  The current sample of
such SNe~Ia, though greatly expanded here, remains small (i.e., $< 10$).  Our
capacity to constrain {\it simultaneously} the full debated range of
cosmological and environmental parameters is therefore limited.  Consequently,
we have chosen to test specific and narrow questions in the context of
well-defined assumptions or in conjunction with independent information.  It is
important to recognize that the conclusions garnered from any analysis cannot
furnish {\it a priori} information for a subsequent analysis.  Readers should
carefully consider which {\it priors} they are using and where they came from  before selecting which analysis presented here provides a relevant
incremental gain.

The two most extreme analyses (in the sense of the breadth of their priors)
presented here also realize the most significant gains from the addition of our
highest-redshift SNe~Ia discovered with {\it HST}.  The kinematic (i.e., cause-independent) interpretation of SN~Ia distances and redshifts (independent of
all other experiments) is most consistent with two distinct epochs of
expansion: a recent accelerated expansion and a previous decelerated expansion
with a transition between the two at $z \approx 0.5$.  This is a generic
requirement of a mixed dark matter and dark energy universe, and it may even be
a feature of unrelated cosmological paradigms (which are beyond our scope to
consider here).  The data are not consistent with many astrophysical
interpretations posited in lieu of dark energy.  Notable examples include the
attenuation produced in a universe filled with gray dust at $z>1$, or a
luminosity evolution which is a simple, monotonic function of redshift. These
interpretations are robust against the exclusion of any individual SN~Ia used
in the analyses and therefore represent an improvement over the results of
Riess et al. (2001).

A vacuum-driven metamorphosis model (VCDM) has been proposed by Parker \& Raval
(1999) to explain the cause of accelerated expansion.  In this model the
Universe makes a transition to a constant-scalar-curvature (induced by a
quantized non-interacting scalar field of very small mass in its vacuum state)
at $z \approx 1$.  This model differs from a quintessence model in that the
scalar field is free and thus interacts only with the gravitational field.  The
transition as described by Parker, Komp, \& Vanzella (2003) is far more abrupt
than for a cosmological constant and is characterized by $w_0=-1.3$ and
$w'=-0.8$ for values of $\Omega_M$ within its likely range.  These values lay just outside the 99.5\% confidence level.  However, other future
variations of the central idea of this model may yield better fits to the data.

The ``replenishing dust'' model addressed in \S 4.2 is one example of a variety
of astrophysical dimming models for which most of the apparent dimming occurs
by $z \approx 0.5$, with little additional dimming at $z > 0.5$.  Such models
require an additional parameter or a logarithmic parameterization to dampen the
dimming and fit the new SNe~Ia at higher redshifts.  Another model with this
behavior would be evolution which is proportional to look-back time (Wright
2002).  While possible, such dimming behavior, especially if in the form of
luminosity evolution, would seem implausible.  We may expect evolution (or dust
production) to be coupled to the observed evolution of stellar populations,
galaxy morphologies, sizes, large-scale structure, or even chemical enrichment.
None of these known varieties of evolution are largely completed by $z=0.5$ starting from their properties at $z=0$; quite the contrary, most of them have hardly begun,
looking back to $z=0.5$.  A strong empirical argument against recent luminosity
evolution is the independence of SN~Ia distance measures on gross host
morphology (Riess et al. 1998; Sullivan et al. 2003).  The range of progenitor
formation environments spanned by SNe~Ia in early-type and late-type hosts
greatly exceeds any evolution in the mean host properties between $z=0$ and
$z=0.5$.  In the end, however, the only ``proof'' against astrophysical
contamination of the cosmological signal from SNe~Ia is to test the results
against other experiments, independent of SNe~Ia.

SNe~Ia are no longer unique in their requirement of a dominant dark-energy
component.  The bevy of SN-independent cosmological experiments, most notably
the CMB, LSS and the ISW effect, provide a prior constraint of $\Omega_{total}=1$ and $\Omega_M
\approx 0.3$.  These priors preserve the limited discriminating power of SN~Ia
data for resolving the nature of dark energy.  What is its equation of state
and has it been evolving?  The constraints obtained here provide a substantial
improvement in our ability to answer the latter (i.e., not rapidly), but the
results are far from a compelling, empirical case for a cosmological constant
or evolving dark energy.

A parametric reconstruction of $w(z)$ by Alam et al (2003) of the full set of SN Ia distances from Tonry et al (2003) and  Barris et al (2003) concluded that dark energy ``evolves rapidly'' and has ``metamorphosized'' from $w \sim 0$ at $z \sim 1$ to $w < -1$ at $z \sim 0$.  Their analysis differs from ours in a few important ways.  Alam et al (2003) use a significant number of SNe that would fail our selection criteria of our gold sample including 26 SNe from the SCP (now excluded or flagged by Knop et al 2003) and another 13 from Barris et al (2003) of dubious reliability.  Alam et al (2003) also use a different parameterization for $w(z)$ and slightly different external cosmological constraints.   These differences do account for a small part of the difference in our conclusions.  However, despite these differences our results are fairly similar to theirs {\it without the inclusion of the new HST-discovered SNe Ia presented here} as seen in the upper left panel of Figure 10.  Without the HST-discovered SNe, the 95\% confidence region resides in the quadrant in which $w_0 < -1$ and $w(z)$ was less negative in the past.  Yet, with the addition of the HST-discovered SNe Ia, the contours shrink in area by a factor of 2 and shift in the direction of a cosmological constant, a position in the $w_0-w'$ plane which is now at the boundary of the $\sim$ 1 $\sigma$ contour.  Such large changes in the size and position of the error contours with the addition of new data indicates the crucial need for even more data before firmer conclusions can be reached.

Knowledge of the equation of state of dark energy {\it and} its time evolution
has profound implications for determining the fate of the Universe.  For
example, a joint constraint of $w_0 > -1$ and $w' < 0$ could provide the
signature of a future recollapse (i.e., ``Big Crunch'') for a linear potential
field which becomes negative in the future (Kallosh \& Linde 2003).
Interestingly, the data indicate that this quadrant of the $w_0-w'$ plane is
the least favored.  Values (in this quadrant) farthest from the origin ($w_0=-1$
and $w'=0$) would forecast the earliest possible recollapse and are the least favored by our
data.  Taking the Kallosh \& Linde (2003) toy model at face value, our
constraint would rule out such a future recollapse of the Universe in less than
$\sim$ 30 or 15 Gyr at the 95\% or 99\% confidence levels, respectively. This
is somewhat more reassuring than the previous minimal remaining time to
recollapse of 11 Gyr allowed by the observation that the dark-energy density
has had time to evolve to $\Omega_\Lambda \leq$ 0.72 (Kallosh et al. 2003).

A qualitatively different and more speculative fate, a hierarchical ``ripping'' of progressively
smaller bound systems (i.e., ``Big Rip'') might occur if $w(z)$ evolves to or
remains at values less than $-1$ (Caldwell, Kamionkowski, \& Weinberg 2003).
In this case the dark-energy density within any bound system would increase
without limit as $\rho \propto a^{-3(1+w)}$, overcoming its binding energy.  This eventuality may be foretold
from joint constraints of $w_0 < -1$ and $w' > 0$ and (an assumed) linear
extrapolation of the dark-energy scalar field.  The current data are not
inconsistent with this quadrant of the $w_0 - w'$ plane (and will remain
consistent if dark energy is a cosmological constant and measurements have
finite precision).  For an assumed constant equation of state, our 95\%
confidence interval would limit a ``Big Rip'' to be no sooner than 25 Gyr from
now, only slightly more reassuring than the previous estimate of minimal
remaining time to the Rip of 22 Gyr estimated by Caldwell, Kamionkowski, \&
Weinberg (2003).  However, if $w(z)$ is evolving to progressively more negative
values (a result empirically consistent with our data), a Rip could occur far sooner.  We are limited by our understanding of how the future equation of state of dark energy would track the scale factor, time, or some other global parameter of the Universe.  Therefore, if dark energy is evolving in this direction, we cannot place any meaningful empirical
limit on the minimum time to the Rip.

We caution against naively extrapolating our empirical dynamical constraint
(fit over a very small range in redshift) far from the time at which it was
derived and is likely to be meaningful.  The empirical approach to determining
the fate of the Universe must rely on large extrapolations until dark
energy is better understood.

\subsection{Tests of the Utility of Supernovae}

Many more SNe~Ia at $z>1$ may yield a more precise probe of dark energy,
provided the statistical power of a larger sample can be realized.  It is
therefore of interest to use the current set of SNe~Ia at $z > 1$ to
consider whether there are any reasons to believe this goal could not be
accomplished.

The dispersion of the SNe~Ia at $z > 1.0$ around the best fit model is 0.29 mag, similar to
the 0.27 mag dispersion of the sample at $0.1 < z<1.0$.  (We note that these values
are higher than the nominal intrinsic dispersion of $\sim$0.15 mag due to the
sparse sampling and noisy photometry of SNe~Ia observed at high redshifts.)  In
addition, the average $\chi^2$ per SN at $z > 1.0$ (9 objects in the gold
sample) is 0.95 for the concordance model.  We conclude that there is no
evidence of excess dispersion for SNe~Ia at $z > 1$.

Another test of evolutionary effects is to compare the distribution of
light-curve shapes for the {\it HST}-discovered SNe~Ia with those at low
redshift.  In Figure 11 we compare the individually normalized (i.e., after
subtraction of the individual peak day and apparent magnitude), rest-frame
(i.e., after K-corrections and correcting for $1+z$ time dilation) light curves
of these samples.  Statistically, the measured magnitudes of {\it
HST}-discovered SN light curves (with median redshift 1) are consistent with
having been drawn from a parent population at low redshift.  The fitted
distributions of light-curve shape parameters from the MLCS2k2
parameterization, $\Delta$ (Riess et al.  1996a; Jha et al. 2004a), are also
consistent for the two populations as shown in Figure 12.

Some evidence for anomalously blue colors of the high-redshift SNe~Ia in the
objects from Riess et al. (1998) was noted by Falco et al. (1999) and Leibundgut (2001).  The indicated mean difference was a few
hundredths of a magnitude in data transformed to rest-frame $B-V$ with a
significance approaching 2$\sigma$.  No such difference was seen in the data
from Perlmutter et al. (1999), although the individual precision of their
measured colors may have precluded the detection of such an anomaly.  Here we
have used the expanded data set and its redshift leverage to look for evidence
of color evolution in both rest-frame $B-V$ and $U-B$.  The ultraviolet colors of SNe Ia are predicted by theory to be the most sensitive to chemical abundances  (H\"{o}flich, Wheeler, \& Thielemann 1998).  Figure 13 shows both of these
colors as a function of redshift for the sample in a modest range around the
unreddened colors of SNe~Ia.  In Figure 14 we have constructed histograms for
the colors in three redshift ranges: $z<0.1$, $0.1 < z < 0.6$, and $0.6 < z <
1.6$.  For $B-V$, the middle bin is bluer in the mean by 0.02 mag than the
lowest bin, and despite the increased sample size the significance of this
difference remains 2$\sigma$.  This result is not surprising because it makes
use of similar data as the previous analysis.  Interestingly, the size of the
mean differences {\it decreases} for the highest-redshift bin to 0.01 mag with
an even lower significance of 1$\sigma$.  For the $U-B$ data, the lowest and
highest-redshift bins are consistent in their means of $-0.45 \pm 0.02$ mag and
$-0.47 \pm 0.02$ mag, respectively.  (Negligible data for $U-B$ in the middle
redshift bin is available.)  Of the three comparisons presented here, only the
previously noted $B-V$ comparison between low and intermediate redshifts is
suspect.  If the colors of SNe~Ia were evolving, we would expect the comparison
at higher redshifts, and additionally with more chemically sensitive colors, to
provide amplified evidence for the effect. In addition, the increased sample
size in the low and intermediate-redshift bins has not produced an increase in
the significance of the anomaly as would be expected for a real effect.  We
therefore conclude there is insufficient evidence for color evolution of SNe~Ia
with the current sample, but we encourage future investigators to remain aware
of (and test for) this possibility.

Overall, the data presented here suggest that SNe~Ia at $z > 1$ remain {\it
empirically} well-behaved and show promise for providing robust cosmological
measurements.

To test the sensitivity of our results to the light-curve fitting method used,
we compared the MLCS2k2 distances of the {\it HST}-discovered SNe~Ia to those
derived from BATM (Tonry et al. 2003).  For this comparison, the distance
scales of the two methods were normalized by matching the mean distances
calculated for the same SNe~Ia from Tonry et al (2003).  In Figure
15 we show a comparison for the SNe~Ia fit by both methods over a wide range in
redshift.  The mean difference (weighted by the quadrature sum of the
individual uncertainties) for the {\it HST}-discovered SNe~Ia is 0.047 mag (in
the sense that the BATM distances are larger). This difference is consistent
(at the $1 \sigma$ confidence level) with an independent mean uncertainty of
0.18 mag for each method.  If we limit the comparison to the 9 {\it
HST}-discovered SNe~Ia at $z > 0.9$, the mean difference diminishes to $< 0.01$
mag.  Because this higher-redshift range provides the leverage utilized for the
updated cosmological conclusions presented here, we conclude that our results
are insensitive to the light-curve fitting method.

   The problem of heterogeneity of SN~Ia data may be addressed by
future, massive SN surveys such as ESSENCE (Smith et al. 2002; Garnavich et
al. 2002) or the CFHT Legacy Survey (Pain et al. 2002), which will obtain large
sets of SNe~Ia over the full range of explored redshifts (obviating the need
for older, heterogeneous data).  Until then, we must rely on a careful and
judicious compilation of the available data and a clear description of how they
were compiled.

Farrah et al. have used the statistical measurement (most individual hosts were not detected) of sub-millimetre emission from high-redshift SN Ia hosts to infer the average internal extinction of these galaxies.  They then assume that the line-of-sight extinction of supernovae matches the inferred mean extinction of the hosts.  They conclude that the SNe Ia from hosts at higher redshifts may suffer more extinction than their lower redshift counterparts and this effect may bias the inferred cosmology.  However, it has not been demonstrated that radio emission correlates well (if at all) with SN extinction.  In contrast, apparent reddening has been shown to provide a significant reduction in distance dispersion and should account for any change in the relative proportions of host extinction along the line-of-sight.  More relevant to this work, a trend of increased and uncorrected extinction with redshift does not match the apparent magnitudes of the new SNe Ia presented here in the highest redshift bin.

     \subsection{Lensing}

   Until now, we have evaluated the SN data in the context of a homogeneous
model for the matter distribution, also known as the ``filled-beam'' model.
The inhomogeneity of matter along the SN line-of-sight distorts space-time,
directing more or less photon-filled lines-of-sight to our detectors depending
on the matter content along the line of sight.  Averaged over enough
lines-of-sight, SN flux is conserved, and the inferred cosmology is unaffected.
However, further investigation of possible lensing is warranted for small
sample sizes and in regard to possible selection biases.

We have estimated the expected lensing along the line-of-sight of each of the
{\it HST}-discovered SNe~Ia as well as for 100 randomly selected positions in
each of the GOODS fields.  The lensing is estimated using the same multiple
lens-plane methodology employed by Ben\'\i tez et al. (2002) for SN 1997ff.
Photometric and spectroscopic redshifts of all foreground lenses were derived
from the GOODS catalogs.  Their masses were estimated from the rest-frame
$B$-band luminosities and from the Tully-Fisher and Faber-Jackson relations
corrected for evolution (Ziegler et al. 2001; Boehm et al. 2003; Kochanek 1996; Treu et al. 2002).  Because SN luminosity distances
here and elsewhere have been modeled using ``filled-beam" cosmological models,
we have calculated the amplifications relative to a filled-beam model and for
$\Omega_M=0.3$, $\Omega_\Lambda=0.7$.  The results are shown in Figure
16.

As expected, the distribution of net amplifications for random 3-dimensional
positions in the fields scatters about unity with a width steadily increasing
with redshift (simulations predict $\sigma=0.03$ mag per unit redshift as
indicated; D. Holz 2003, private communication).  Relatively underdense and
frequent lines-of-sight resulting in net deamplification are compensated by an
extended tail of strong magnifications, resulting in an increasingly skewed
distribution at higher redshifts.  These results are in good accord with Monte
Carlo simulations of this effect (Holz 1998; Metcalf \& Silk 1999).

We find that the magnification estimates for the SNe~Ia discovered in the GOODS
survey are consistent with having been drawn from those in the random positions
sample.  We conclude that there is no evidence of a lensing-induced selection
bias of our sample.  This is not unexpected since most of our SNe~Ia were
discovered at $> 2$ mag brighter than the survey limit and lensing would
contribute an insignificant $\leq 0.2$ mag of amplification for 98\% of the
population at $z< 1.5$.

Interestingly, there is an indication that the discovery of SNe~Ia in the HDF-N
found by other surveys could have been favored by amplification.  The net
amplification of SN 1997ff of $\sim 0.3$ mag (Benitze et al. 2002) is unusually large and the SN was
within $0.5$ mag of its survey limit (Gilliland, Nugent, \& Phillips 1999).  It
is less clear whether the discovery of SN 2002dd ($z=0.95$) could have been
favored by its $0.2$ mag amplification due to the serendipity of its discovery
(Blakeslee et al. 2003).

In the absence of an apparent selection bias (or bad luck) in our SN survey, it
is unnecessary (and even undesirable) to correct individual SNe for the
estimated net amplification, as the sample will approach a mean amplification of
unity when considered in a filled-beam model.  An important exception would be to correct (or flag) SNe whose predicted amplification places them on the strong lensing tail, such as SN 1997ff as done by Benitez et al (2001), Tonry et al (2003) and here.  
However, an independent test of
this conclusion is available by comparing the predicted amplifications with the
observed residuals from a good-fit cosmological model.  In Figure 17 we show
this comparison and their derived correlation.  Empirically we find the
residuals to be $1.6 \pm 0.9$ times the predicted magnifications, consistent
with the theoretical relation (i.e., unity) and inconsistent with no
correlation at the 1.7$\sigma$ (80\%) confidence level.  Much of the leverage
comes from SN 1997ff; without it, the correlation is $1.0 \pm 1.2$.  As the
sample expands, it may soon be possible to provide independent evidence of the
correlation of dark matter and light from the high-redshift SN sample.

    Mortsell, Gunnarsson, \& Goobar (2001) (hereafter MGG2001) and more
recently Gunnarsson (2004) have made predictions of the expected magnification
of two of the SNe~Ia in our sample, both contained in the original HDF-N.
For SN 1997ff ($z=1.7$) these authors concluded the magnification was uncertain
and potentially large (i.e., as high as a factor of a few).  In contrast,
Ben\'\i tez et al. (2002) estimated a much more modest magnification of $0.34
\pm 0.12$ mag, in good agreement with the values estimated by Lewis \& Ibata
(2001) and Riess et al. (2001).  The source of the potentially higher
magnification can be traced to differences in the treatment of the mass scaling
of the foreground lenses.  MGG2001 treated the lensing galaxies as unevolved
with an unknown but possibly high mass scaling (e.g., velocity dispersions of
200 to 300 km s$^{-1}$ for $M_*$ galaxies with $M_B=-19.5+5\, {\rm log}\, h$).
Ben\'\i tez et al. (2002) used the $B$-band Tully-Fisher relation observed by
Ziegler et al. (2002) for late-type galaxies at redshifts 0.1--1.0 to provide a
slope and normalization which accounts for evolution (which yields a velocity dispersion $M_*\sim 150$ km s$^{-1}$).  The result is a much
smaller mass scale than that considered by MGG2001 for the foreground lenses of
high-redshift SNe.

For SN 2003es, Gunnarsson (2004) estimated a magnification factor of 1.15 (with
a velocity dispersion of 170 km s$^{-1}$) for 13 apparent foreground galaxies
(at $z<0.968$) based primarily on optical photometric redshifts.  To span the
rest-frame optical breaks in the spectral energy distributions of potential
high-redshift foreground galaxies and reliably estimate their redshifts, it is
important to use near-IR observations with good precision.  We reanalyzed the
photometric redshifts of these 13 galaxies using the high-resolution near-IR
data from NICMOS (Budavari et al. 2000).  In the majority of cases we found
that the data from Fernandez-Soto, Lanzetta \& Yahil (1999) and Gwynn \&
Hartwick (1996) contained detections in only 3 or 4 bands whereas the NICMOS
data provided a total of 5 to 7 bands and with greater leverage.  We found 7 of
the 13 galaxies to lie in the {\it background} of the SN.  Repeating the
analysis of Gunnarsson (2004) using their {\it Q-LET} algorithm for the 7
remaining foreground galaxies reduced the predicted magnification factor to
1.10.  However, calculations from the {\it Q-LET} algorithm are made relative
to a non-filled-beam cosmology and hence all lines-of-sight are amplified (in
the absence of any foreground lenses the {\it Q-LET} amplification would be
1.0).  At the redshift of SN 2003es, the average strong lensing amplification
would be $\sim$0.03 mag so the remaining excess would be a factor of 1.07.  The
remaining difference between this value and our estimate of 1.03 results from
the previously discussed difference in the treatment of the size and evolution
of the mass scale.  Our analysis, comparing the predicted magnification and
observed cosmological residuals, is consistent with the mass scaling from
Ziegler et al. (2002) as utilized here (and would be inconsistent with a scale
approximately twice as large).

It is tempting to consider that we have reached the end of the beginning in the
exploration of dark energy.  Two reliable and independent routes require it in
addition to a third more tentative investigation via the integrated Sachs-Wolfe
effect (Scranton et al. 2003).  SNe~Ia continue to provide the most direct
route to illuminating dark energy because their light
 can be measured propagating from
within its era of dominance.  Two clues about dark energy, its equation of
state and its recent time evolution, would be invaluable aids to test and
provoke theories.  We suggest that the most efficient way forward in the near
term is by simultaneously mining both ends of the observable redshift range: at
$z<1$ generally from the ground, and at $z>1$ generally from space.  The
constraints presented here in the $w_0-w'$ plane have reduced the allowable
range of $w'$ from a factor of $\sim 10$ to $\lesssim 1$ while retaining the
constraints on $w_0$ within $-1.4 < w_0 < -0.7$.  With continued determination,
an improvement in precision by a factor of a few in this plane is expected.
   
\section{Summary and Conclusions}

     We have conducted the first space-based SN search and follow-up campaign
using the ACS on board {\it HST}.  The search parameters and the full list of
42 new SNe are provided elsewhere (Strolger et al. 2004).  We reviewed the
sample of SNe~Ia harvested from the survey and examined its cosmological
significance.  The key results can be summarized as follows.

(1) We obtained multi-color light curves and spectroscopic redshifts for 16 new
SNe~Ia which uniformly sample the redshift range $0.2 < z < 1.6$.  Twelve of
these are classified by their spectra, 2 from their red, early-type host
galaxies, and 2 by photometric diagnostics.  Three of the SN spectra are at the
highest redshifts yet observed for SNe.  Six of the SNe~Ia are among the seven
highest-redshift known; all are at $z>1.25$. These data provide a robust
extension of the Hubble diagram to $1 < z < 1.6$.

(2) Utilizing a simple kinematic description of the magnitude-redshift data, we
find that the SNe~Ia favor recent acceleration and past deceleration at the
99.2\% confidence level.  An alternate kinematic parameterization requires a positive jerk (third derivative of the scale factor).  The best-fit redshift of the transition between these
kinematic phases is $z=0.46 \pm 0.13$, although the precise value depends on the
kinematic model employed.

(3) We have compared the goodness-of-fit of cosmological models and simple
models of astrophysical dimming. The ``gold'' sample of 157 SNe~Ia is
consistent with the ``cosmic concordance'' model ($\Omega_M=0.3,
\Omega_\Lambda=0.7$) with $\chi^2_{dof}=1.06$.  The data reject at
high-confidence simple, monotonic models of astrophysical dimming which are
tuned to mimic the evidence for acceleration at $z \approx 0.5$.  These
models include either a universe filled with gray dust at high redshift, or
luminosity evolution $\propto z$.  More complex parameterizations of
astrophysical dimming which peak at $z \approx 0.5$ and dissipate at $z>1$
remain consistent with the SN data (but appear unattractive on other grounds).

(4) For a flat Universe with a cosmological constant, we measure $\Omega_M=0.29
\pm ^{0.05}_{0.03}$ (equivalently, $\Omega_\Lambda=0.71$).  When combined with
external flat-Universe constraints including the CMB and LSS, we find for the
dark-energy equation of state $w=-1.02 \pm ^{0.13}_{0.19}$ (and $w<-0.76$ at
the 95\% confidence level) for an assumed static equation of state of dark
energy, $P = w\rho c^2$.

(5) Joint constraints on both the recent equation of state of dark energy and
its time evolution are a factor of $\sim 8$ more precise than its first estimate
and twice more precise than those derived without the SNe~Ia discovered by {\it
HST}.  Both of these dark energy properties are consistent with a cosmological
constant (i.e., with $w_0=-1.0$, $w'=0$) and are inconsistent with very rapid
evolution of dark energy (i.e., $\vert w' \vert > $ a few).  The absence of
rapid evolution places constraints on the time in which a simple scalar field
could evolve to recollapse the Universe.  Specifically, the
timescale to a potential recollapse is larger than $\sim$30 Gyr.
If dark energy is evolving towards more negative $w$, we cannot place any meaningful limit on the minimum time to a (speculative) Big Rip.
 
\bigskip 
\medskip 

We wish to thank Richard Hook, Swara Ravindranath, Tomas Dahlen, Peter
Garnavich, Duilia de Mello, Ed Taylor, Soo Kim, Rafal Idzi, Carl Biagetti, Lexi
Moustakas, Marin Richardson, Vicki Laidler, Ann Hornschmeier, Ray Lucas, Norman
Grogin, Claudia Kretchmer, Brian Schmidt, Stephane Blondin, and Anton Koekemoer for their help in
the supernova search.  We are grateful to Dorothy Fraquelli, Sid Parsons, Al
Holm, Tracy Ellis, Richard Arquilla, and Mark Kochte for their help in assuring
rapid delivery of the data.  We thank Tom Matheson, Dan Stern, Hy Spinrad, Piero Rosati,
Mario Nonino, Alice Shapley, Max Pettini and Dawn Erb for their efforts to obtain redshifts
of some SN host galaxies.  We appreciate the guidance of Anton Koekemoer and
Eddie Bergeron. Partly based on observations collected at the European Southern Observatory, Chile (Prog. Nr. 70.A-0497).
Financial support for this work was provided by NASA through
programs GO-9352 and GO-9583 from the Space Telescope Science Institute, which
is operated by AURA, Inc., under NASA contract NAS 5-26555. Some of the data
presented herein were obtained at the W. M. Keck Observatory, which is operated
as a scientific partnership among the California Institute of Technology, the
University of California, and NASA; the Observatory was made possible by the
generous financial support of the W. M. Keck Foundation.

\appendix 

\section {Appendix: The Full Sample}

Distance measurements to individual SNe depend on the algorithms used to
estimate their K-corrections, fit their light curves, and infer their
extinction.  There is currently no single set of algorithms which are
considered by consensus to be the optimal ones.  Rather, different methods may
have advantages depending on the breadth and quality of the observational
record of any individual SN~Ia.  In addition, algorithms improve as their
training samples grow.  Here we present the full cosmological sample of SNe~Ia
used in this work in Table 5.  Their virtue is that all distance estimates were derived
from a single set of algorithms, MLCS2k2 (Jha et al. 2004a), with the broadest set of training data available at this time including $U$-band data.
  Additional
advantages include a consistent and thorough reanalysis of quality criteria for
all currently published SNe~Ia, an exercise resulting in the rejection of many
SNe from our ``gold'' sample whose observational records have one or more
shortcomings (see \S 3.1 for discussion).  

The zeropoint, distance scale, absolute magnitude of the fiducial SN Ia or Hubble constant derived from Table 5 are all closely related (or even equivalent) quantities which were arbitrarily set for the sample presented here.  Their correct value is not relevant for the analyses presented which only make use of differences between SN Ia magnitudes.  Thus the analysis are independent of the aforementioned normalization parameters.     

\begin{deluxetable}{llllll} 
\footnotesize
\tablecaption{MLCS2k2 Full Sample}
\tablehead{\colhead{SN}&\colhead{$z$}&\colhead{$\mu_0^a$}&\colhead{$\sigma^*$}&\colhead{host $A_V$}&\colhead{sample}}

\startdata
\hline
\hline
SN 1990T &   0.0400 &     36.38 &    0.19 &    0.37 & gold \nl
SN 1990af &   0.050 &     36.84 &    0.21 &   -0.04 & gold \nl
SN 1990O  &  0.0307 &     35.90 &    0.20 &    0.11 & gold \nl
SN 1991S  &  0.0560 &     37.31 &    0.18 &    0.20 & gold \nl
SN 1991U  &  0.0331 &     35.54 &    0.20 &    0.37 & gold \nl
SN 1991ag &  0.0141 &  34.13 &  0.25 &  0.12  & gold \nl
SN 1992J &  0.0460 &  36.35 &  0.21 & 0.25  & gold \nl
SN 1992P &  0.0265 &  35.64 &  0.20 & 0.17  & gold \nl
SN 1992aq &  0.101 &  38.73 &  0.20 & -0.03  & gold \nl
SN 1992ae &  0.075 &  37.77 &  0.19 & 0.16  & gold \nl
SN 1992au &  0.061 &  37.30 &  0.22 & 0.09  & gold \nl
SN 1992al &  0.0141 &  34.12 &  0.25 & 0.05  & gold \nl
SN 1992ag &  0.0262 &  35.06 &  0.24 & 0.54  & gold \nl
SN 1992bl &  0.0430 &  36.53 &  0.19 & -0.04  & gold \nl
SN 1992bh &  0.0450 &  36.97 &  0.18 & 0.35  & gold \nl
SN 1992bg &  0.036 &  36.17 &  0.19 & 0.21  & gold \nl
SN 1992bk &  0.058 &  37.13 &  0.19 & 0.03  & gold \nl
SN 1992bs &  0.063 &  37.67 &  0.19 & 0.26  & gold \nl
SN 1992bc &  0.0186 &  34.96 &  0.22 & -0.04  & gold \nl
SN 1992bp &  0.079 &  37.94 &  0.18 & 0.03  & gold \nl
SN 1992br &  0.088 &  38.07 &  0.28 & -0.04  & gold \nl
SN 1992bo &  0.0178 &  34.70 &  0.23 & -0.01  & gold \nl
SN 1993B &  0.071 &  37.78 &  0.19 & 0.36  & gold \nl
SN 1993H &  0.0251 &  35.09 &  0.21 & 0.05  & gold \nl
SN 1993O &  0.052 &  37.16 &  0.18 & 0.13  & gold \nl
SN 1993ah &  0.0286 &  35.53 &  0.21 & 0.26  & gold \nl
SN 1993ac &  0.0490 &  36.90 &  0.20 & 0.54  & gold \nl
SN 1993ag &  0.050 &  37.08 &  0.19 & 0.28  & gold \nl
SN 1993ae &  0.0180 &  34.29 &  0.23 & 0.00  & gold \nl
SN 1994B &  0.089 &  38.50 &  0.17 & 0.00  & silver \nl
SN 1994C &  0.051 &  36.67 &  0.16 & 0.00  & silver \nl
SN 1994M &  0.0244 &  35.09 &  0.20 & 0.23  & gold \nl
SN 1994Q &  0.0290 &  35.70 &  0.19 & 0.33  & gold \nl
SN 1994S &  0.0161 &  34.50 &  0.24 & 0.06  & gold \nl
SN 1994T &  0.0360 &  36.01 &  0.20 & 0.09  & gold \nl
SN 1995E &  0.0116 &  32.96 &  0.29 & 2.48  & silver \nl
SN 1995K &  0.478 &  42.48 &  0.23 & 0.04  & gold \nl
SN 1995M &  0.053 &  37.17 &  0.15 & 0.00  & silver \nl
SN 1995ap &  0.230 &  40.44 &  0.46 & 0.00  & silver \nl
SN 1995ao &  0.300 &  40.76 &  0.60 & 0.00  & silver \nl
SN 1995ae &  0.067 &  37.54 &  0.34 & 0.00  & silver \nl
SN 1995az &  0.450 &  42.13 &  0.21 & ---  & gold \nl
SN 1995ay &  0.480 &  42.37 &  0.20 & ---  & gold \nl
SN 1995ax &  0.615 &  42.85 &  0.23 & ---  & gold \nl
SN 1995aw &  0.400 &  42.04 &  0.19 & ---  & gold \nl
SN 1995as &  0.498 &  43.21 &  0.24 & ---  & silver \nl
SN 1995ar &  0.465 &  42.81 &  0.22 & ---  & silver \nl
SN 1995ac &  0.0490 &  36.52 &  0.20 & 0.40  & gold \nl
SN 1995ak &  0.0219 &  34.70 &  0.22 & 0.56  & gold \nl
SN 1995ba &  0.3880 &  42.07 &  0.19 & ---  & gold \nl
SN 1995bd &  0.0152 &  34.11 &  0.25 & 0.70  & gold \nl
SN 1996C &  0.0276 &  35.90 &  0.20 & 0.34  & gold \nl
SN 1996E &  0.425 &  41.70 &  0.40 & 0.35  & gold \nl
SN 1996H &  0.620 &  43.11 &  0.30 & 0.09  & gold \nl
SN 1996I &  0.570 &  42.81 &  0.25 & 0.14  & gold \nl
SN 1996J &  0.300 &  41.01 &  0.25 & 0.23  & gold \nl
SN 1996K &  0.380 &  42.02 &  0.22 & 0.02  & gold \nl
SN 1996R &  0.160 &  39.08 &  0.40 & 0.00  & silver \nl
SN 1996T &  0.240 &  40.68 &  0.43 & 0.00  & silver \nl
SN 1996U &  0.430 &  42.33 &  0.34 & 0.08  & gold \nl
SN 1996V &  0.0247 &  35.33 &  0.25 & 0.00  & silver \nl
SN 1996ab &  0.124 &  39.20 &  0.22 & 0.00  & gold \nl
SN 1996bo &  0.0165 &  33.82 &  0.27 & 0.77  & gold \nl
SN 1996bv &  0.0167 &  34.21 &  0.23 & 0.71  & gold \nl
SN 1996bl &  0.0348 &  36.17 &  0.19 & 0.33  & gold \nl
SN 1996cg &  0.490 &  42.58 &  0.19 & 0.63  & silver \nl
SN 1996cm &  0.450 &  42.58 &  0.19 & ---  & silver \nl
SN 1996cl &  0.828 &  43.96 &  0.46 & ---  & gold \nl
SN 1996ci &  0.495 &  42.25 &  0.19 & ---  & gold \nl
SN 1996cf &  0.570 &  42.77 &  0.19 & ---  & silver \nl
SN 1997E &  0.0132 &  34.02 &  0.26 & 0.12  & gold \nl
SN 1997F &  0.580 &  43.04 &  0.21 & ---  & gold \nl
SN 1997H &  0.526 &  42.56 &  0.18 & 0.45  & gold \nl
SN 1997I &  0.172 &  39.79 &  0.18 & ---  & gold \nl
SN 1997N &  0.180 &  39.98 &  0.18 & ---  & gold \nl
SN 1997P &  0.472 &  42.46 &  0.19 & ---  & gold \nl
SN 1997Q &  0.430 &  41.99 &  0.18 & ---  & gold \nl
SN 1997R &  0.657 &  43.27 &  0.20 & ---  & gold \nl
SN 1997Y &  0.0166 &  34.54 &  0.23 & 0.25  & gold \nl
SN 1997ai &  0.450 &  42.10 &  0.23 & ---  & gold \nl
SN 1997ac &  0.320 &  41.45 &  0.18 & ---  & gold \nl
SN 1997aj &  0.581 &  42.63 &  0.19 & ---  & gold \nl
SN 1997aw &  0.440 &  42.57 &  0.40 & 0.80  & gold \nl
SN 1997as &  0.508 &  41.64 &  0.35 & 0.85  & gold \nl
SN 1997am &  0.416 &  42.10 &  0.19 & 0.00  & gold \nl
SN 1997ap &  0.830 &  43.85 &  0.19 & ---  & gold \nl
SN 1997af &  0.579 &  42.86 &  0.19 & ---  & gold \nl
SN 1997bh &  0.420 &  41.76 &  0.23 & 0.60  & gold \nl
SN 1997bb &  0.518 &  42.83 &  0.30 & 0.11  & gold \nl
SN 1997bj &  0.334 &  40.92 &  0.30 & 0.34  & gold \nl
SN 1997ck &  0.970 &  44.13 &  0.38 & 0.17  & silver \nl
SN 1997cn &  0.0175 &  34.52 &  0.25 & 0.01  & gold \nl
SN 1997cj &  0.500 &  42.74 &  0.20 & 0.15  & gold \nl
SN 1997ce &  0.440 &  42.08 &  0.19 & 0.08  & gold \nl
SN 1997dg &  0.0297 &  36.12 &  0.20 & 0.28  & gold \nl
SN 1997do &  0.0104 &  33.73 &  0.33 & 0.44  & gold \nl
SN 1997ez &  0.778 &  43.81 &  0.35 & ---  & gold \nl
SN 1997ek &  0.860 &  44.03 &  0.30 & ---  & gold \nl
SN 1997eq &  0.538 &  42.66 &  0.18 & ---  & gold \nl
SN 1997ff &  1.755 &  45.53 &  0.35 & 0.00  & gold \nl
SN 1998I &  0.886 &  42.91 &  0.81 & 0.95  & gold \nl
SN 1998J &  0.828 &  43.61 &  0.61 & 0.49  & gold \nl
SN 1998M &  0.630 &  42.62 &  0.24 & 0.75  & gold \nl
SN 1998V &  0.0170 &  34.47 &  0.23 & 0.28  & gold \nl
SN 1998ac &  0.460 &  41.83 &  0.40 & 0.48  & gold \nl
SN 1998ay &  0.638 &  43.30 &  0.36 & ---  & silver \nl
SN 1998bi &  0.740 &  43.35 &  0.30 & ---  & gold \nl
SN 1998be &  0.644 &  42.78 &  0.26 & ---  & silver \nl
SN 1998ba &  0.430 &  42.36 &  0.25 & ---  & gold \nl
SN 1998bp &  0.0104 &  33.21 &  0.32 & 0.19  & gold \nl
SN 1998co &  0.0171 &  34.68 &  0.24 & 0.20  & gold \nl
SN 1998cs &  0.0327 &  36.08 &  0.19 & -0.03  & gold \nl
SN 1998dx &  0.053 &  36.97 &  0.18 & 0.04  & gold \nl
SN 1998ef &  0.0170 &  34.18 &  0.23 & 0.07  & gold \nl
SN 1998eg &  0.0234 &  35.36 &  0.20 & 0.29  & gold \nl
SN 1999Q &  0.460 &  42.56 &  0.27 & 0.23  & gold \nl
SN 1999U &  0.500 &  42.75 &  0.19 & 0.04  & gold \nl
SN 1999X &  0.0257 &  35.41 &  0.20 & 0.31  & gold \nl
SN 1999aa &  0.0157 &  34.58 &  0.24 & 0.02  & gold \nl
SN 1999cc &  0.0316 &  35.85 &  0.19 & 0.09  & gold \nl
SN 1999cp &  0.0104 &  33.56 &  0.31 & 0.08  & gold \nl
SN 1999da &  0.0121 &  34.05 &  0.32 & 0.58  & silver \nl
SN 1999dk &  0.0141 &  34.43 &  0.26 & 0.20  & gold \nl
SN 1999dq &  0.0136 &  33.73 &  0.26 & 0.43  & gold \nl
SN 1999ef &  0.0380 &  36.67 &  0.18 & 0.05  & gold \nl
SN 1999fw &  0.278 &  41.00 &  0.41 & 0.26  & gold \nl
SN 1999fk &  1.056 &  44.25 &  0.23 & 0.19  & gold \nl
SN 1999fm &  0.949 &  43.99 &  0.25 & 0.11  & gold \nl
SN 1999fj &  0.815 &  43.76 &  0.33 & 0.23  & gold \nl
SN 1999ff &  0.455 &  42.29 &  0.28 & 0.19  & gold \nl
SN 1999fv &  1.19 &  44.19 &  0.34 & 0.24  & gold \nl
SN 1999fh &  0.369 &  41.62 &  0.31 & 0.70  & silver \nl
SN 1999fn &  0.477 &  42.38 &  0.21 & 0.15  & gold \nl
SN 1999gp &  0.0260 &  35.62 &  0.20 & 0.18  & gold \nl
SN 2000B &  0.0193 &  34.59 &  0.23 & 0.28  & gold \nl
SN 2000bk &  0.0266 &  35.36 &  0.21 & 0.19  & gold \nl
SN 2000cf &  0.0360 &  36.39 &  0.18 & 0.21  & gold \nl
SN 2000cn &  0.0233 &  35.14 &  0.21 & 0.08  & gold \nl
SN 2000ce &  0.0164 &  34.47 &  0.23 & 1.02  & silver \nl
SN 2000dk &  0.0164 &  34.41 &  0.24 & -0.05  & gold \nl
SN 2000dz &  0.500 &  42.75 &  0.24 & 0.09  & gold \nl
SN 2000eh &  0.490 &  42.41 &  0.25 & 0.20  & gold \nl
SN 2000ee &  0.470 &  42.74 &  0.23 & 0.13  & gold \nl
SN 2000eg &  0.540 &  41.96 &  0.41 & 0.12  & gold \nl
SN 2000ea &  0.420 &  40.79 &  0.32 & 1.05  & silver \nl
SN 2000ec &  0.470 &  42.77 &  0.21 & 0.13  & gold \nl
SN 2000fr &  0.543 &  42.68 &  0.19 & ---  & gold \nl
SN 2000fa &  0.0218 &  35.06 &  0.21 & 0.44  & gold \nl
SN 2001V &  0.0162 &  34.13 &  0.23 & 0.28  & gold \nl
SN 2001fs &  0.873 &  43.75 &  0.38 & 0.64  & gold \nl
SN 2001fo &  0.771 &  43.12 &  0.17 & 0.05  & gold \nl
SN 2001hy &  0.811 &  43.97 &  0.35 & 0.03  & gold \nl
SN 2001hx &  0.798 &  43.88 &  0.31 & 0.31  & gold \nl
SN 2001hs &  0.832 &  43.55 &  0.29 & 0.10  & gold \nl
SN 2001hu &  0.882 &  43.90 &  0.30 & 0.12  & gold \nl
SN 2001iw &  0.340 &  40.71 &  0.27 & 0.73  & gold \nl
SN 2001iv &  0.397 &  40.89 &  0.30 & 0.91  & gold \nl
SN 2001iy &  0.570 &  42.88 &  0.31 & -0.04  & gold \nl
SN 2001ix &  0.710 &  43.05 &  0.32 & 0.53  & gold \nl
SN 2001jp &  0.528 &  42.77 &  0.25 & 0.10  & gold \nl
SN 2001jh &  0.884 &  44.23 &  0.19 & -0.01  & gold \nl
SN 2001jb &  0.698 &  43.33 &  0.32 & 0.15  & silver \nl
SN 2001jf &  0.815 &  44.09 &  0.28 & 0.23  & gold \nl
SN 2001jm &  0.977 &  43.91 &  0.26 & 0.18  & gold \nl
SN 2001kd &  0.935 &  43.99 &  0.38 & 0.14  & silver \nl
SN 2002P &  0.719 &  43.22 &  0.26 & 0.11  & silver \nl
SN 2002ab &  0.422 &  42.02 &  0.17 & 0.10  & silver \nl
SN 2002ad &  0.514 &  42.39 &  0.27 & 0.09  & silver \nl
SN 2002dc &  0.475 &  42.14 &  0.19 & 0.23  & gold \nl
SN 2002dd &  0.95 &  44.06 &  0.26 & 0.24  & gold \nl
SN 2002fw &  1.30 &  45.27 &  0.19 & 0.21  & gold \nl
SN 2002fx &  1.40 &  45.09 &  0.45 & 0.49  & silver \nl
SN 2002hr &  0.526 &  43.01 &  0.27 & 0.74  & gold \nl
SN 2002hp &  1.305 &  44.70 &  0.22 & 0.19  & gold \nl
SN 2002kc &  0.216 &  40.33 &  0.18 & 1.29  & silver \nl
SN 2002kd &  0.735 &  43.09 &  0.19 & 0.21  & gold \nl
SN 2002ki &  1.140 &  44.84 &  0.30 & 0.09  & gold \nl
SN 2003az &  1.265 &  45.20 &  0.20 & 0.25  & gold \nl
SN 2003ak &  1.551 &  45.30 &  0.22 & 0.86  & gold \nl
SN 2003bd &  0.67 &  43.19 &  0.28 & 0.27  & gold \nl
SN 2003be &  0.64 &  43.07 &  0.21 & 0.23  & gold \nl
SN 2003dy &  1.340 &  45.05 &  0.25 &  0.54  & gold \nl
SN 2003es &  0.954 &  44.28 &  0.31 &  0.07  & gold \nl
SN 2003eq &  0.839 &  43.86 &  0.22 &  0.22  & gold \nl
SN 2003eb &  0.899 &  43.64 &  0.25 &  0.26  & gold \nl
SN 2003lv &  0.94 &  43.87 &  0.20 & 0.15 & gold \nl
\enddata 
\tablenotetext{*}{Peculiar velocity of 400 km s$^{-1}$ included for all SNe and 2500 km s$^{-1}$ if $z$ from SN.} 
\tablenotetext{a}{Distance normalization is arbitrary; see Appendix}
\end{deluxetable}

\vfill
\eject

Figure Captions

Figure 1: Discovery-image sections from ACS $F850LP$ images around each SN.
Panels on the left and middle show the discovery epoch and the preceding
(template) epoch, respectively. The panels on the right show the results of the
subtraction (discovery epoch minus template). Arrows indicate position of the
SNe.  Image scales and orientations are given.

Figure 2: Multi-color light curves of SNe~Ia.  For each SN~Ia,
multi-color photometry transferred to rest-frame passbands is plotted.
The individual, best-fit MLCS2k2 model is shown as a solid line, with a
$\pm 1 \sigma$ model uncertainty, derived from the model covariance
matrix, above and below the best fit.

Figure 3: Identification spectra (in f$_{\lambda}$) of 12 of the new {\it
HST}-discovered high-redshift SNe~Ia, shown in the rest frame.  Classification
features are analyzed in \S 2.3.  The data are compared to nearby SN~Ia spectra
of the same age as determined by the light curves (see Table 3).
Classification of the 5 SNe without spectra (SN 2003lv, $z=0.94$; SN 2002fx,
$z=1.40$; SN 2002hp, $z=1.31$; SN 2003ak, $z=1.55$; and SN 2003aj, $z=1.31$)
are discussed in section \S 2.3 and \S 3.1.

Figure 4: MLCS2k2 SN~Ia Hubble diagram.  SNe~Ia from ground-based discoveries
in the gold sample are shown as diamonds, {\it HST}-discovered SNe~Ia are shown
as filled symbols.  Overplotted is the best fit for a flat cosmology:
$\Omega_M=0.29$, $\Omega_\Lambda=0.71$.

Figure 5: Left panels: Joint confidence intervals for a two-parameter model of
the expansion history, $q(z)=q_0 + z{dq/dz}$, from SNe~Ia.  The upper left shows
the constraints derived from the gold sample, the lower left includes both gold
and silver samples.  For either set, the data favor the quadrant with recent
acceleration ($q_0<0$) and past deceleration (${dq/dz} > 0$) with
high confidence.  Lines of fixed transition redshift ($q(z_t)=0$) are shown.
Panels on the right illustrate the likelihood function for the transition redshift
derived from the same samples.

Figure 6: Kinematic SN~Ia residual Hubble diagram.  Upper panel: SNe~Ia from
ground-based discoveries in the gold sample are shown as diamonds, {\it
HST}-discovered SNe~Ia are shown as filled symbols.  Bottom panel: weighted
averages in fixed redshift bins are given for illustrative purposes only.  Data
and kinematic models of the expansion history are shown relative to an
eternally coasting model, $q(z)=0$.  Models representing specific kinematic
scenarios (e.g., ``constant acceleration'') are illustrated.

Figure 7: SN~Ia residual Hubble diagram comparing cosmological models and
models for astrophysical dimming. Upper panel: SNe~Ia from ground-based
discoveries in the gold sample are shown as diamonds, {\it HST}-discovered
SNe~Ia are shown as filled symbols.  Bottom panel: weighted averages in fixed
redshift bins are given for illustrative purposes only.  Data and models are
shown relative to an empty Universe model $(\Omega=0)$. The $\chi^2$ fit
statistics for each model are listed in Table 4.

Figure 8: Joint confidence intervals for ($\Omega_M$,$\Omega_\Lambda$) from
SNe~Ia.  The solid contours are results from the gold sample of 157 SNe~Ia
presented here.  The dotted contours are the results from Riess et al. (1998)
illustrating the earlier evidence for $\Omega_\Lambda > 0$.  Regions
representing specific cosmological scenarios are illustrated.  Contours are
closed by their intersection with the line $\Omega_M=0$.

Figure 9: Joint confidence intervals for $\Omega_M$ and a static equation of
state for dark energy, $w$.  In the left-hand panel, constraints from the gold
SN~Ia sample (dotted contours) are combined with a prior of $\Omega_M=0.27 \pm
0.04$ to yield the solid contours.  In the right-hand panel, the same SN
constraints are combined with those from WMAPext and 2dfGRS to yield the solid
contours.

Figure 10: Joint confidence intervals derived from SN samples for a
two-parameter model of the equation of state of dark energy, $w(z)=w_0+w'z$.
For each panel, constraints from a SN sample is combined with the prior,
$\Omega_M=0.27 \pm 0.04$, to yield the indicated confidence intervals.  The
position of a cosmological constant $(-1,0)$ is indicated as a filled symbol.
The lower-right panel shows the impact of adding or subtracting a systematic
error in distance modulus of $0.05z$ mag to the gold sample.

Figure 11: Comparison of composite rest-frame light curves of SNe~Ia.  Each
SN~Ia is individually transformed to the rest frame (K-corrected; corrected for
$1+z$ time dilation).  Each is normalized by subtraction of the peak magnitude
and its date.  Data from SNe~Ia with $z<0.1$ are shown as open symbols; data
from the {\it HST}-discovered SNe~Ia are shown as filled symbols.

Figure 12: Light-curve shape distributions of low-redshift and high-redshift
SNe~Ia.  Histograms of the MLCS2k2 parameter $\Delta$ (relative peak visual
luminosity) are shown for SNe~Ia with $z<0.1$ and the {\it HST}-discovered SNe~Ia.

Figure 13: Rest-frame $B_{max}-V_{max}$ and $U_{max}-B_{max}$ colors of SNe~Ia
versus redshift.  Expected, unreddened colors of SNe~Ia are shown as dotted
lines.

Figure 14: Histograms of the color data shown in Figure 13.  The distributions
from three different redshift bins are shown as indicated.

Figure 15: Comparison of individual distance difference estimated by the
MLCS2k2 and BATM methods for the {\it HST} and ground-discovered SNe~Ia.  The zero-points of both methods are normalized by using the same set of SNe Ia.

Figure 16: Predicted lensing magnifications of SNe~Ia and of random positions in
the CDF-S and HDF-N.  For the SNe~Ia discovered in the GOODS fields, the
expected magnification was calculated using a multiple-lens plane formalism,
with estimates of foreground lens redshifts and masses derived from the GOODS
catalog.  Expected magnifications were also calculated for 100 randomly
selected positions (redshift and angular position).  The solid and dotted lines
show redshift bin averages and dispersion, respectively.  SNe~Ia found in the
GOODS survey and in other {\it HST} searches are indicated.

Figure 17: Correlation of the predicted magnification and the best-fit
cosmological model residual for individual SNe~Ia.  The predicted
magnifications are as described in Figure 16.  The residuals are the
difference in distance modulus as predicted from the best-fit model
($\Omega_M=0.3$, $\Omega_\Lambda=0.7$) and as observed.  The empirical
correlation is expected to be unity (if the lens light traces their
mass) and is shown for the whole sample and without SN 1997ff, the SN
with the largest residual and predicted magnification.

\vfill

\eject

{\bf References}

\refitem
Aguirre, A. N. 1999a, ApJ, 512, L19

\refitem
------. 1999b, ApJ, 525, 583

\refitem
Alard, C., \& Lupton, R. H. 1998, ApJ, 503, 325

\refitem
Barris, B., et al. 2004, ApJ, in press

\refitem
Ben\'\i tez, N., Riess, A., Nugent, P., Dickinson, M., Chornock, R.,
   Filippenko, A. V. 2002, ApJ, 577, L1

\refitem
Bennett, C., et al., 2003, ApJS, 148, 1

\refitem
Blakeslee, J. P., et al. 2003, ApJ, in press (astro-ph/0302402)

\refitem
Boehm, A. et al.  2003, A\&A in press, (astro-ph/0309263)

\refitem
Budav\'ari, T., Szalay, A. S., Connolly, A. J., Csabai, I., \& 
  Dickinson, M. 2000, AJ, 120, 1588

\refitem
Caldwell, R. R., Dav\'e, R., \& Steinhardt, P. J. 1998, Ap\&SS, 261, 303

\refitem
Caldwell, R. R., Kamionkowski, M., \& Weinberg, N. N. 2003, 
   Phys. Rev. Let., 91, 71301

\refitem
Carroll, S. M., Press, W. H., \& Turner, E. L. 1992, ARA\&A, 30, 499

\refitem 
Chapman, S. C., Blain, A. W., Ivison, R. J., \& Smail, I. 2003,
   Nature, 422, 695

\refitem
Clochiatti, A., et al. 2004, in preparation

\refitem
Coil, A. L., et al. 2000, ApJ, 544, L111

\refitem
Colgate, S. 1979, ApJ, 232, 404

\refitem Cowie, L.L., Barger, A. J., Hu, E. M., Capak, P., \& Songaila, A., 2004, AJ, submitted (astro-ph/0401354)

\refitem
Deffayet, C., Dvali, G., \& Gabadadze, G. 2002, Phys. Rev. D, 65044023

\refitem 
Di Pietro, E., \& Claeskens, J. 2003, MNRAS, 341, 1299 

\refitem
Drell, P. S., Loredo, T. J., \& Wasserman, I. 2000, ApJ, 530, 593

\refitem 
Falco, E., et al. 1999, ApJ, 523, 617

\refitem
Farrah, D., Fox, M., Rowan-Robinson, M., \& Clements, D., 2004, ApJ, in press

\refitem
Fernandez-Soto, A., Lanzetta, K. M., \& Yahil, A. 1999, ApJ, 513, 34

\refitem
Filippenko, A. V. 1997, ARA\&A, 35, 309 

\refitem
------. 2001, PASP, 113, 1441

\refitem
------. 2003, in From Twilight to Highlight: The Physics of Supernovae, ed.
   W. Hillebrandt \& B. Leibundgut (Berlin: Springer-Verlag), 171.
   
\refitem
------. 2004, in Carnegie Observatories Astrophysics Series,
  Vol. 2: Measuring and Modeling the Universe, ed. W. L. Freedman (Cambridge:
  Cambridge Univ. Press), in press (astro-ph/0307139).

\refitem
Filippenko, A. V., Li, W. , Treffers, R. R., \& Modjaz, M. 2001, in 
   Small-Telescope Astronomy on Global Scales, ed.  W. P. Chen, C. Lemme, \& B. 
   Paczy\'{n}ski (San Francisco: ASP), 121

\refitem
Filippenko, A. V., \& Riess, A. G. 2001, in Particle Physics and
   Cosmology: Second Tropical Workshop, ed. J. F. Nieves (New York: 
   AIP), 227

\refitem
Freedman, W., \& Turner, M. 2003, Rev. Mod. Phys. Colloquia, in press 
   (astro-ph /0308418)

\refitem
Fruchter, A., \& Hook, R. 1997, SPIE, 3164, 120F

\refitem
Garnavich, P. M., et al. 1998, ApJ, 509, 74 

\refitem
------. 2002, BAAS, 78.09

\refitem
Giavalisco, M., et al. 2003, ApJL, in press (astro-ph/0309105)

\refitem
Gilliland, R. L., Nugent, P. E., \& Phillips, M. M. 1999,
   ApJ, 521, 30

\refitem
Gilliland, R. L., \& Riess, A. G. 2002, in {\it HST Calibration 
   Workshop}, ed. S. Arribas, A. Koekemoer, \& B. 

\refitem
Goobar, A., Bergstrom, L., \& Mortsell, E. 2002, A\&A, 384, 1

\refitem
Gunnarsson, C. 2004, JCAP, submitted (astro-ph/0311380)

\refitem
Gwynn, S. D. J., \& Hartwick, F. D. A. 1996, ApJ, 468, L77

\refitem
------. 1996a, AJ, 112, 2391 

\refitem
------. 1996b, AJ, 112, 2398

\refitem
Hamuy, M., Trager, S. C., Pinto, P. A., Phillips, M. M.,
  Schommer, R. A., Ivanov, V., \& Suntzeff, N. B. 2000, AJ, 120, 1479

\refitem
Hatano, K., Branch, D., \& Deaton, J. 1998, ApJ, 502, 177

\refitem
H\"{o}flich, P., Wheeler, J. C., \& Thielemann, F. K. 1998, ApJ, 495, 617

\refitem
Holz, D. E. 1998, ApJ, 506, L1

\refitem

\refitem
Jha, S. 2002, Ph.D. thesis, Harvard University

\refitem
Jha, S., Riess, A. G., \& Kirshner, R. P. 2004a, in preparation

\refitem
Jha, S., et al. 2004b, in preparation

\refitem
Jha, S., et al. 2004c, in preparation

\refitem
Kallosh, R., \& Linde, A. 2003, JCAP, 302, 2

\refitem
Kallosh, R., Kratochvil, J., Linde, A., Linder, E. V., \& 
   Shmakova, M. 2003, JCAP, 310, 15

\refitem
Kantowski, R. 1998, ApJ, 507, 483

\refitem
Knop, R., et al. 2003, ApJ, in press (astro-ph/0309368)

\refitem Kochanek, C. S., 1996, ApJ, 473, 595

\refitem
Koekemoer, A., et al. 2004, in preparation

\refitem
Leibundgut, B. et al. 2004, in preparation

\refitem
Leibundgut, B. 2001, ARAA, 39, 67

\refitem
Lewis, G. F., \& Ibata, R. A. 2001, MNRAS, submitted
  (astro-ph/0104254)

\refitem
Linder, E. V. 2003, Phys. Rev. Lett., 90, 91301

\refitem
Linder, E. V., \& Huterer, D. 2003, Phys. Rev., D67, 81303

\refitem
McLean, I. S., et al. 1998, Proc. SPIE, 3354, 566

\refitem
Metcalf, R. B., \& Silk, J. 1999, ApJ, 519, 1

\refitem
Mortsell, E., Gunnarsson, C., \& Goobar, A, 2001, ApJ, 561, 106 (MGG2001)

\refitem
Ninomiya, Yaqoob, T. \& Khan 2003, private communication

\refitem
Nonino, M. et al., 2004, in preparation

\refitem
Nugent, P., Kim, A., \& Perlmutter, S. 2002, PASP, 114, 803

\refitem
Oke, J. B., \& Sandage, A. 1968, ApJ, 154, 21

\refitem
Oke, J. B., et al. 1995, PASP, 107, 375

\refitem
Paerels, F., Petric, A., Telis, G., \& Helfand, D. J. 2002, 
   BAAS, 201, 9703

\refitem
Pain, R., et al., 2002, BAAS, 43.02

\refitem 
Parker, L., Komp, W., \& Vanzella, D. 2003, ApJ, in press

\refitem 
Parker, L., \& Raval, A. 1999, Phys. Rev., D60, 123502

\refitem
Peebles, P. J., \& Ratra, B. 2003, Rev. Mod. Phys., 75, 559

\refitem
Perlmutter, S., \& Schmidt, B. P. 2003, in Supernovae \& Gamma-Ray 
   Bursts, ed. K. Weiler (New York: Springer, Lecture Notes in 
   Physics), in press

\refitem
Perlmutter, S., et al. 1997, ApJ, 483, 565

\refitem
------. 1999, ApJ, 517, 565

\refitem
Poznanski, D., Gal-Yam, A., Maoz, D., Filippenko, A. V.,
  Leonard, D. C., \& Matheson, T. 2002, PASP, 114, 833

\refitem 
Rana, N. C. 1979, Ap\&SS, 66, 173

\refitem 
Rana, N. C. 1980, Ap\&SS, 71, 123

\refitem
Riess, A. G. 2000, PASP, 112, 1284

\refitem
Riess, A. G., et al. 1998b, AJ, 116, 1009 

\refitem
------. 1999a, AJ, 117, 707  

\refitem
------. 1999b, AJ, 118, 2675   

\refitem
------. 2001, ApJ, 560, 49

\refitem
------. 2003, ApJ, 600, in press

\refitem
Riess, A. G., Nugent, P. E., Filippenko, A. V., Kirshner,
  R. P., \& Perlmutter, S. 1998a, ApJ, 504, 935

\refitem
Riess, A. G., Press, W. H., \& Kirshner, R. P. 1995, ApJ, 438, L17

\refitem
------. 1996a, ApJ, 473, 88  

\refitem
------. 1996b ApJ, 473, 588. 

\refitem
Schlegel, D. J., Finkbeiner, D. P., \& Davis, M. 1998, ApJ, 500, 525

\refitem
Schmidt, B. P., et al. 1998, ApJ, 507, 46

\refitem
Scranton, R., et al. 2003, Phys. Rev. Lett., submitted (astro-ph/0307335)

\refitem
Sirianni, M. et al. 2004, in preparation

\refitem
Smith, R. C., et al., 2002, BAAS, 78.08

\refitem
Spergel, D. N., et al. 2003, ApJS, 148, 175

\refitem
Strolger, L.-G., et al. 2004, in preparation

\refitem
Sullivan, M., et al. 2003, MNRAS, 340, 1057

\refitem
Suntzeff, N., et al. 2004, in preparation.

\refitem
Tammann, G. A. 1979, in ESA/ESO Workshop on Astronomical Uses of the Space Telescope, ed. D. Macchetto, F. Pacini, \& M. Tarenghi (Geneva: ESO), 329

\refitem
Tonry, J. T., et al. 2003, ApJ, 594, 1

\refitem Treu, T., Stavelli, M., Casertano, S., Moller, P., \& Bertin, G., 2002, ApJ, 564, 13

\refitem
Turner, M., \& Riess, A. G. 2001, ApJ, 569, 18

\refitem
Visser, M. 2003, gr-qc/0309109

\refitem
Wambsganss, J., Cen, R., \& Ostriker, J. P. 1998, ApJ, 494, 29

\refitem Wirth, G. D., et al., 2004, AJ, submitted (astro-ph/0401353)

\refitem
Wright, E. L. 2002, BAAS, 161.17 (astro-ph/0201196)

\refitem
Ziegler, et al. 2002, ApJ, 564, 69

\vfill \eject 

\begin{figure}[h]
\vspace*{120mm}
\caption {Figure 1.  See attached jpeg image of SN hosts.}
\end{figure}

\vfill \eject 

\begin{figure}[h]
\vspace*{80mm}
\includegraphics{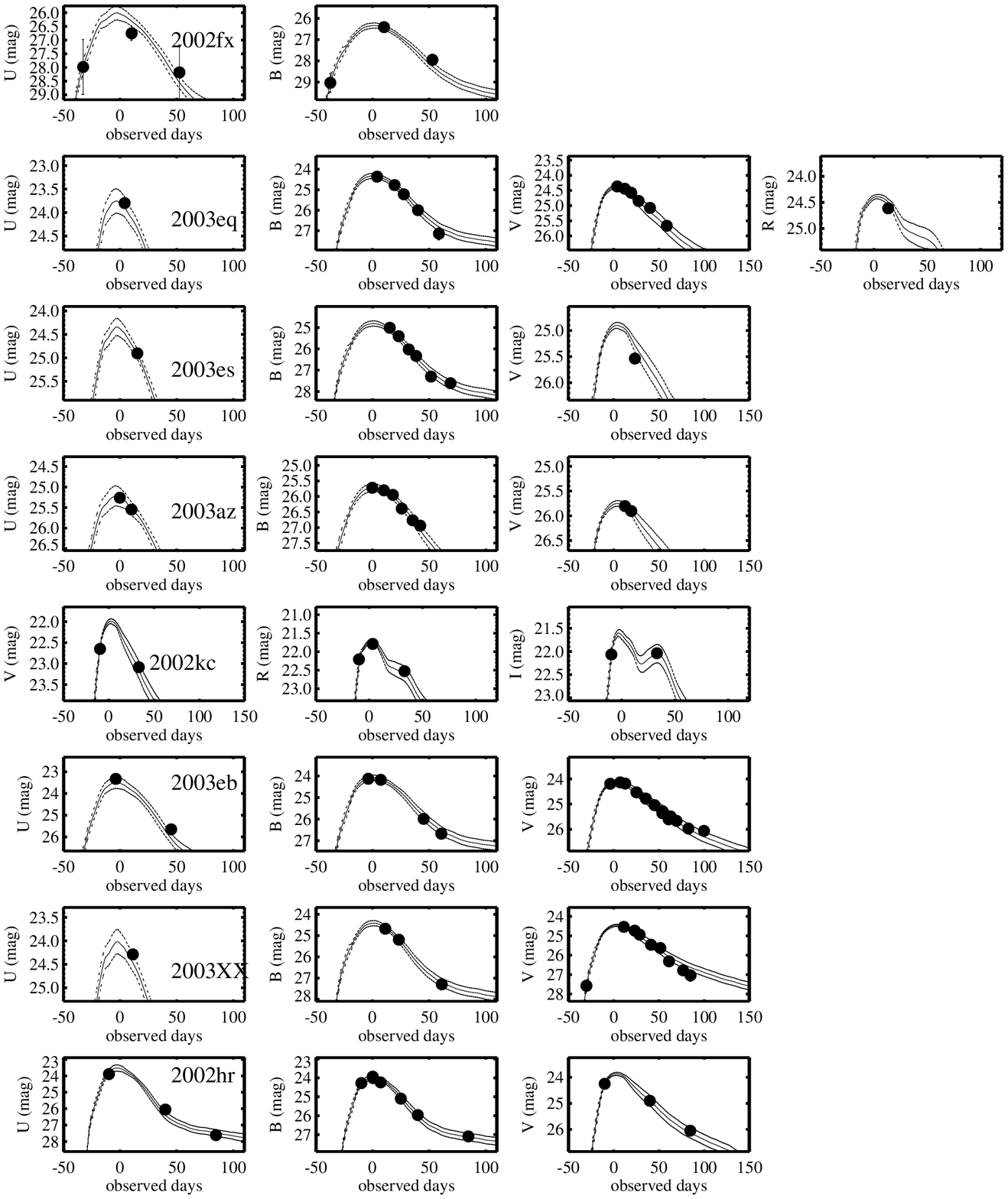}
\end{figure}

\vfill \eject

\begin{figure}[h]
\vspace*{80mm}
\includegraphics{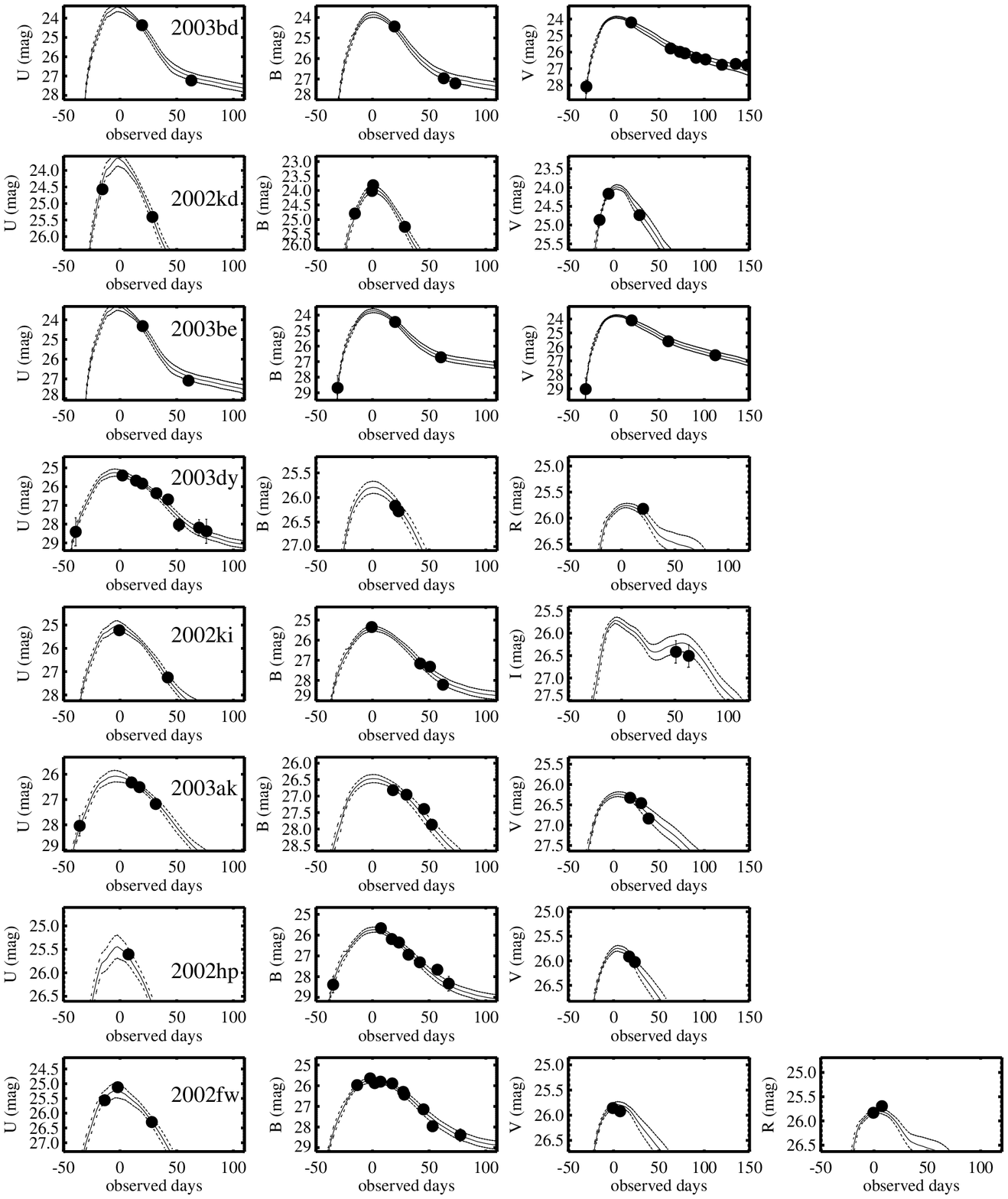}
\end{figure}

\vfill \eject

\begin{figure}[h]
\vspace*{150mm}
\includegraphics{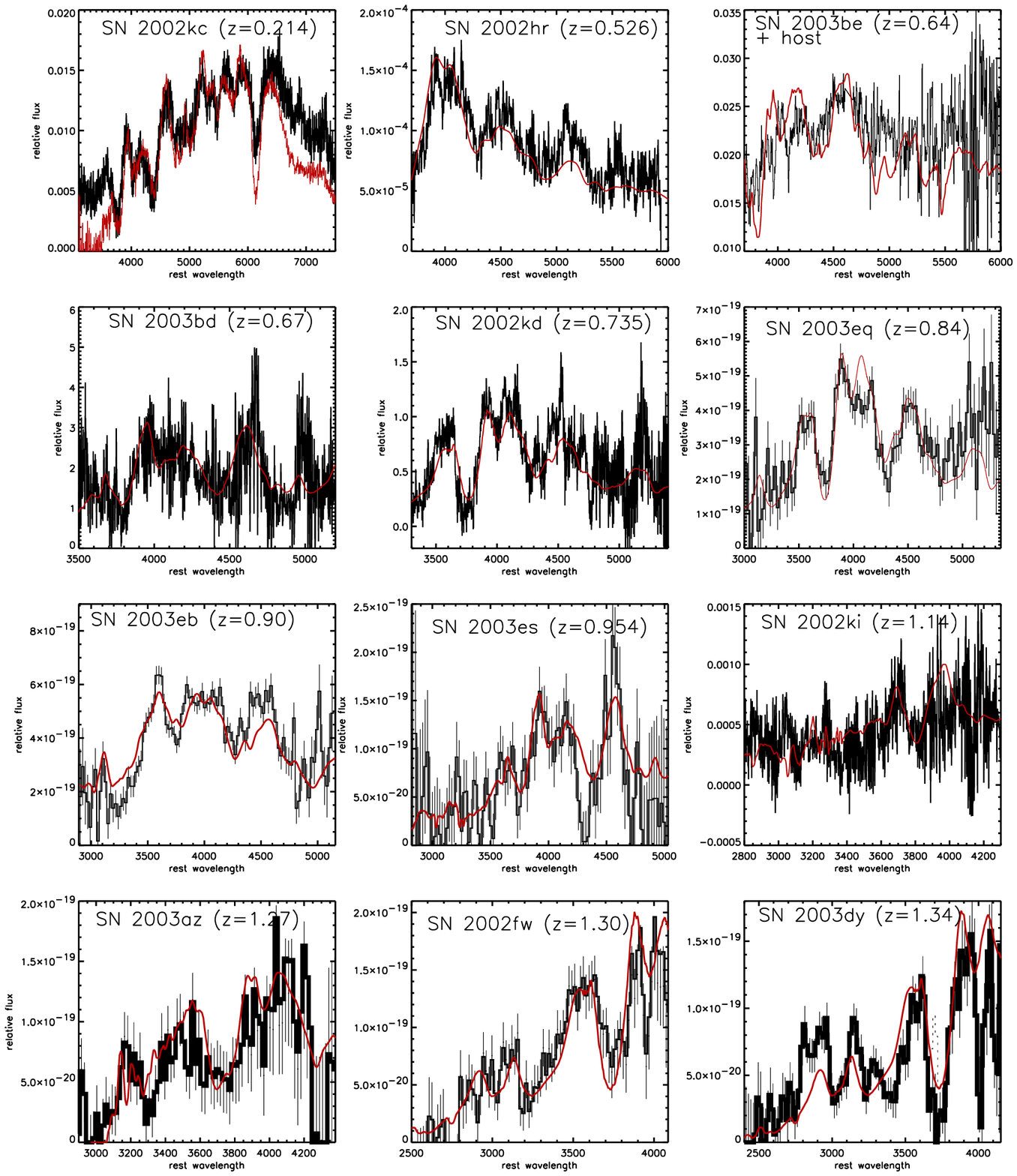}
\end{figure}

\vfill \eject

\begin{figure}[h]
\vspace*{150mm}
\includegraphics{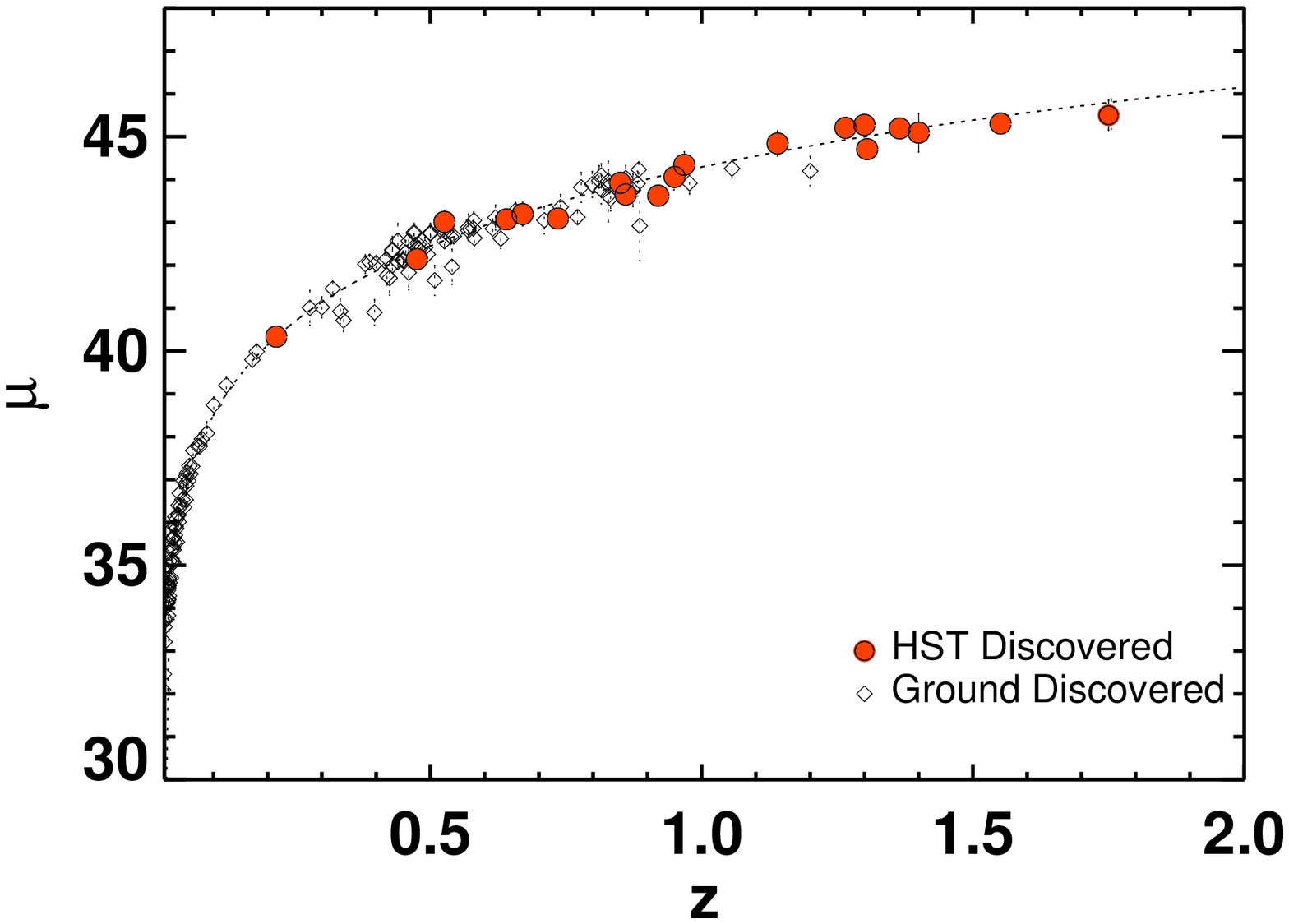}
\end{figure}

\vfill \eject

\begin{figure}[h]
\vspace*{150mm}
\includegraphics{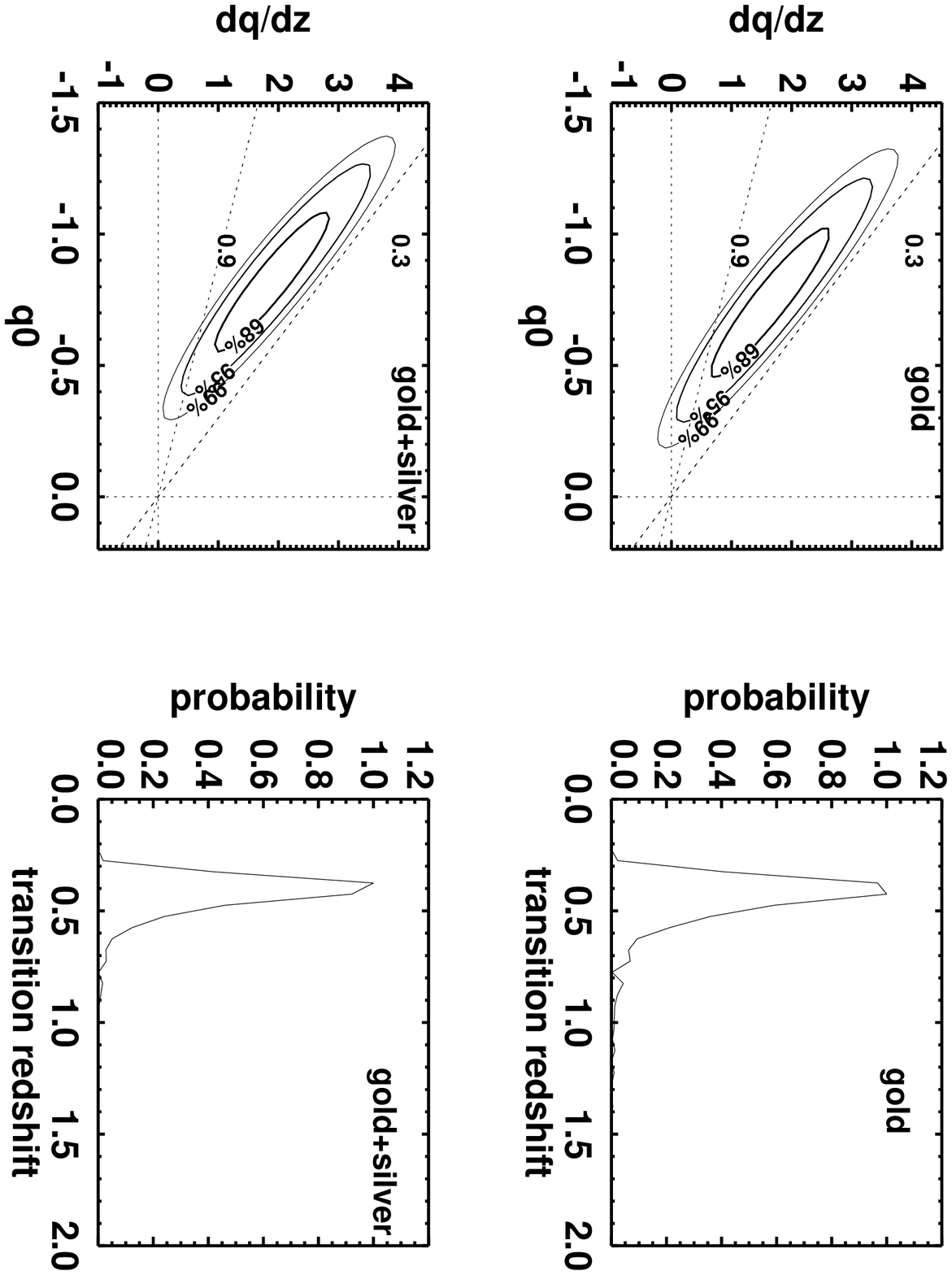}
\end{figure}

\vfill \eject

\begin{figure}[h]
\vspace*{150mm}
\includegraphics{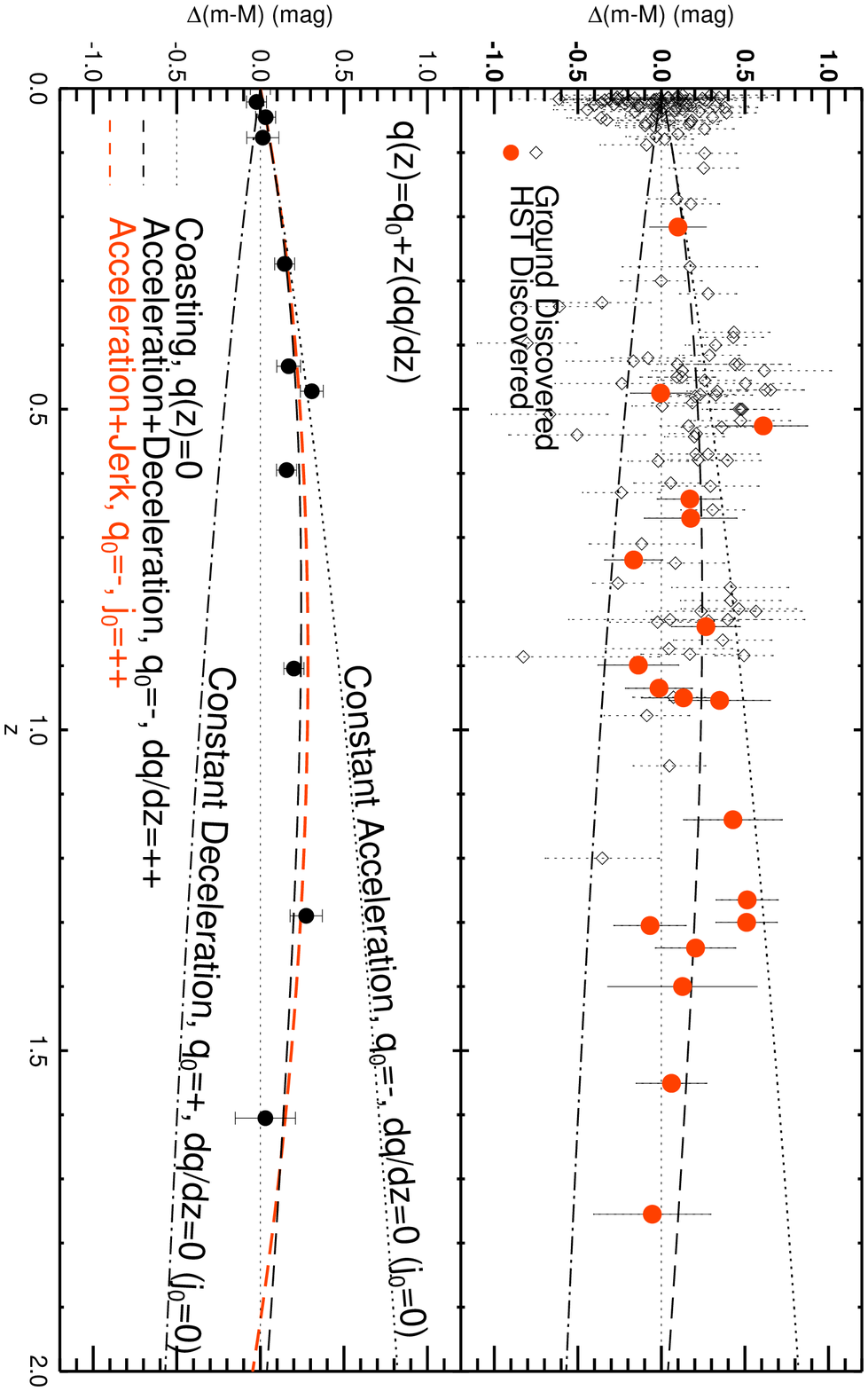}
\end{figure}

\vfill \eject

\begin{figure}[h]
\vspace*{150mm}
\includegraphics{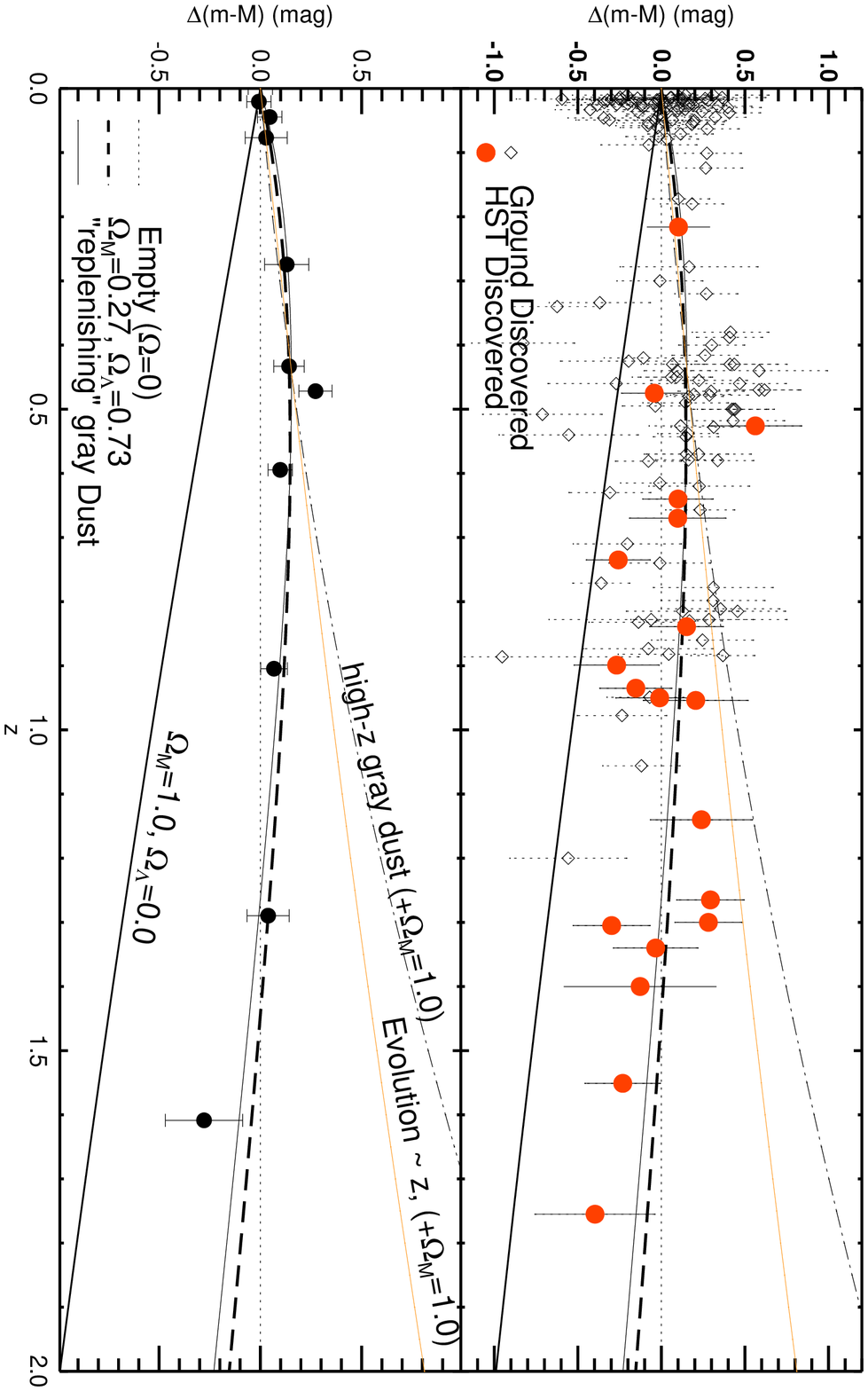}
\end{figure}

\vfill \eject

\begin{figure}[h]
\vspace*{150mm}
\includegraphics{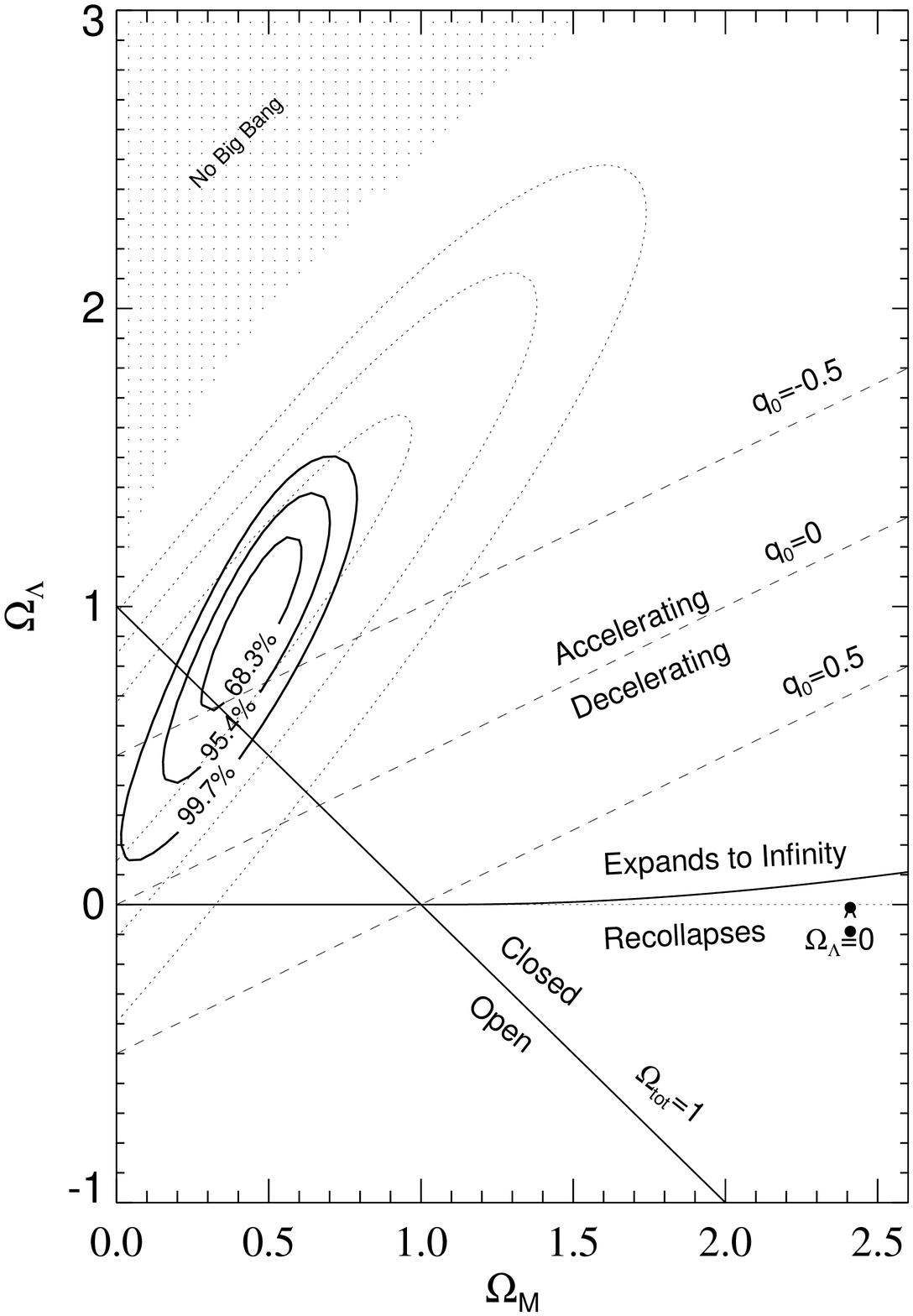}
\end{figure}

\vfill \eject

\begin{figure}[h]
\vspace*{150mm}
\includegraphics{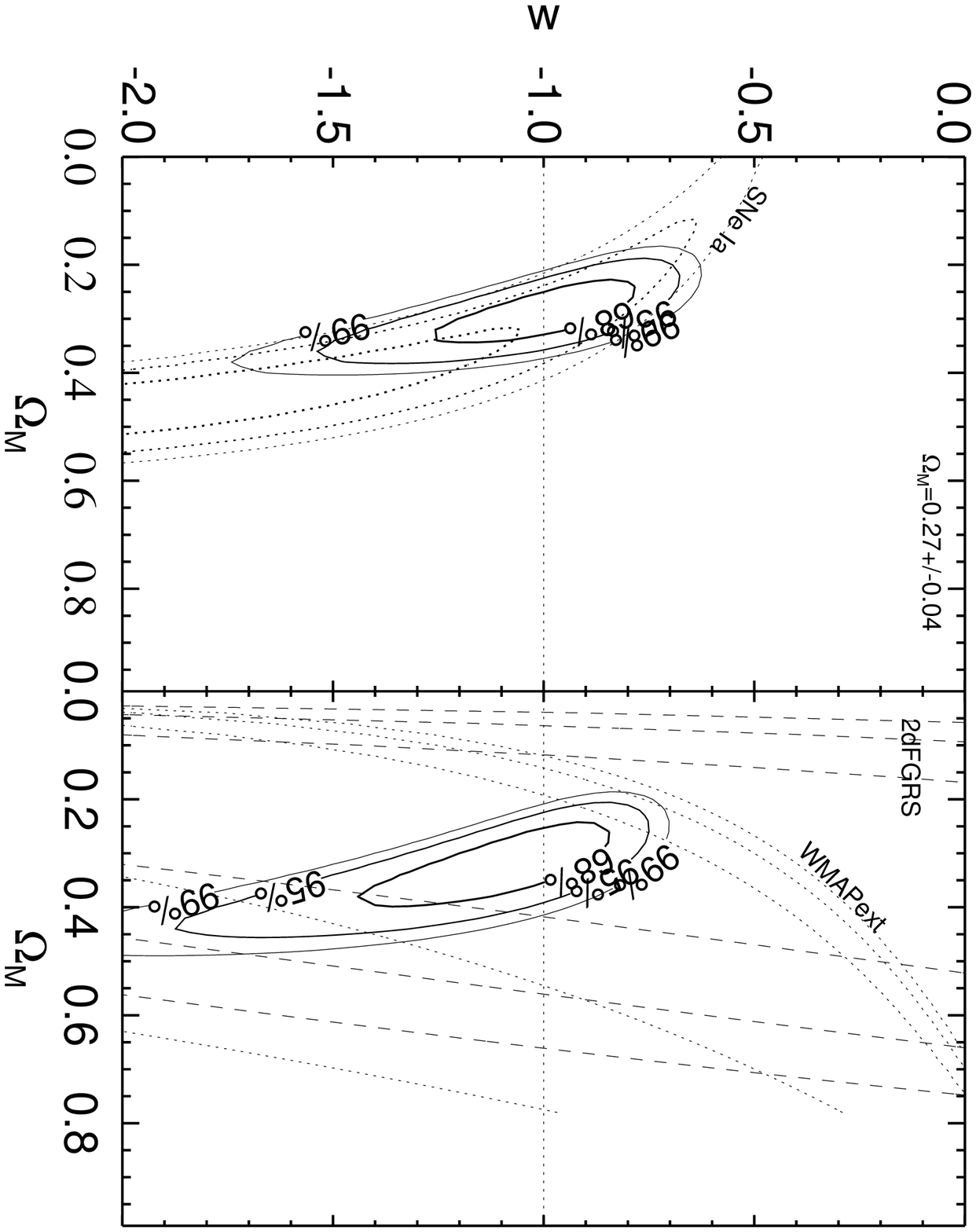}
\end{figure}

\vfill \eject

\begin{figure}[h]
\vspace*{150mm}
\includegraphics{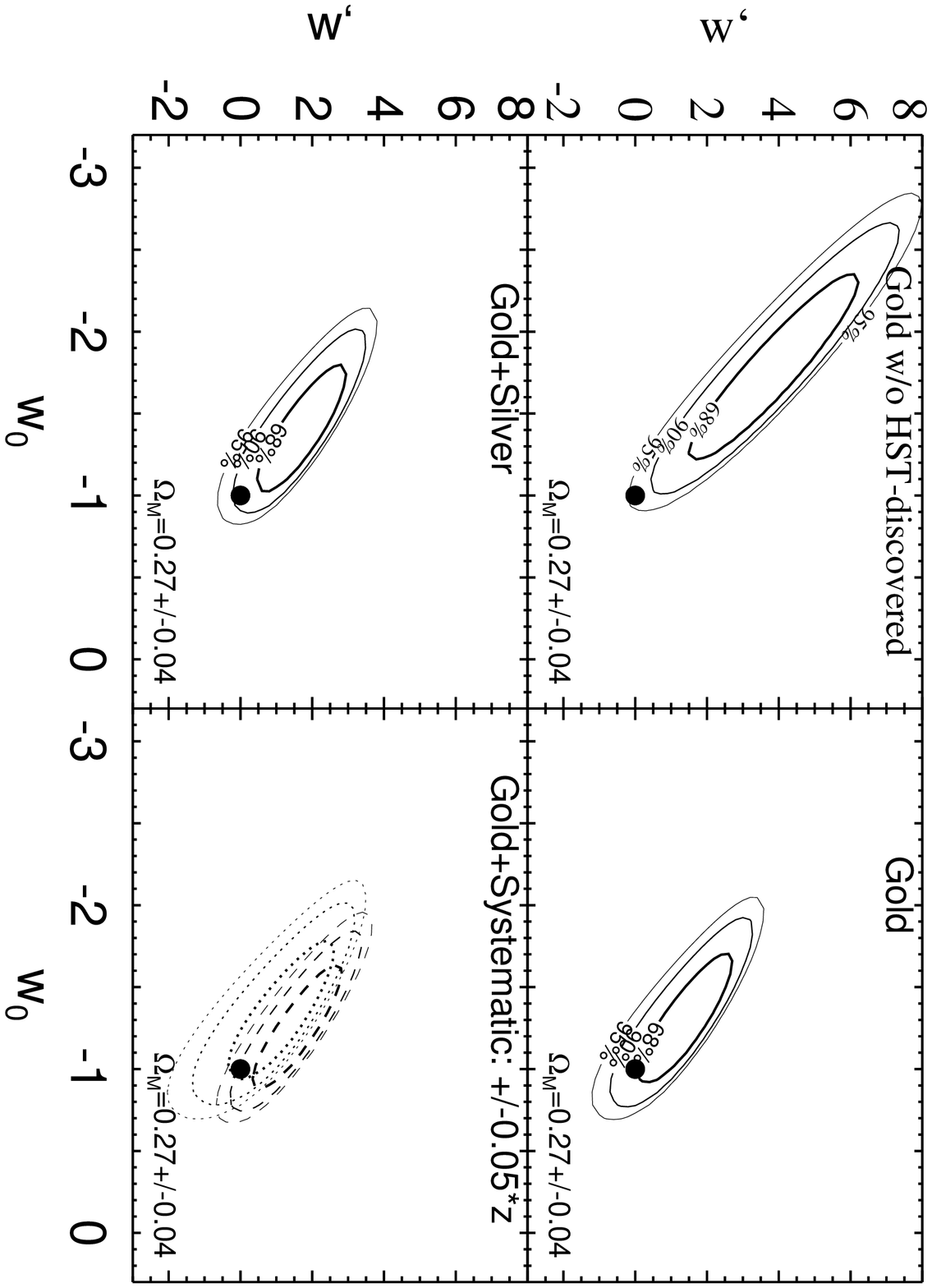}
\end{figure}

\vfill \eject

\begin{figure}[h]
\vspace*{150mm}
\includegraphics{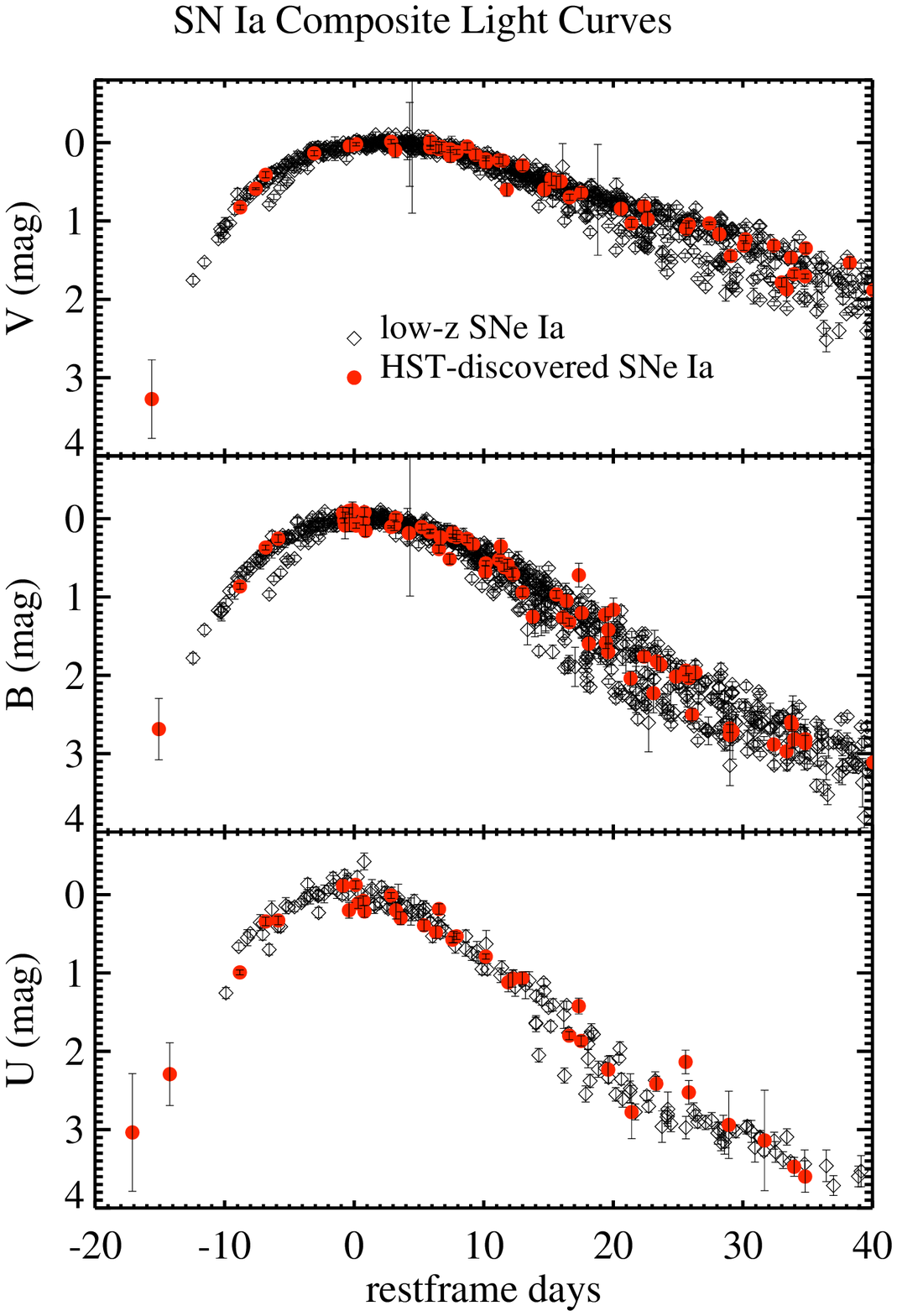}
\end{figure}

\vfill \eject

\begin{figure}[h]
\vspace*{150mm}
\includegraphics{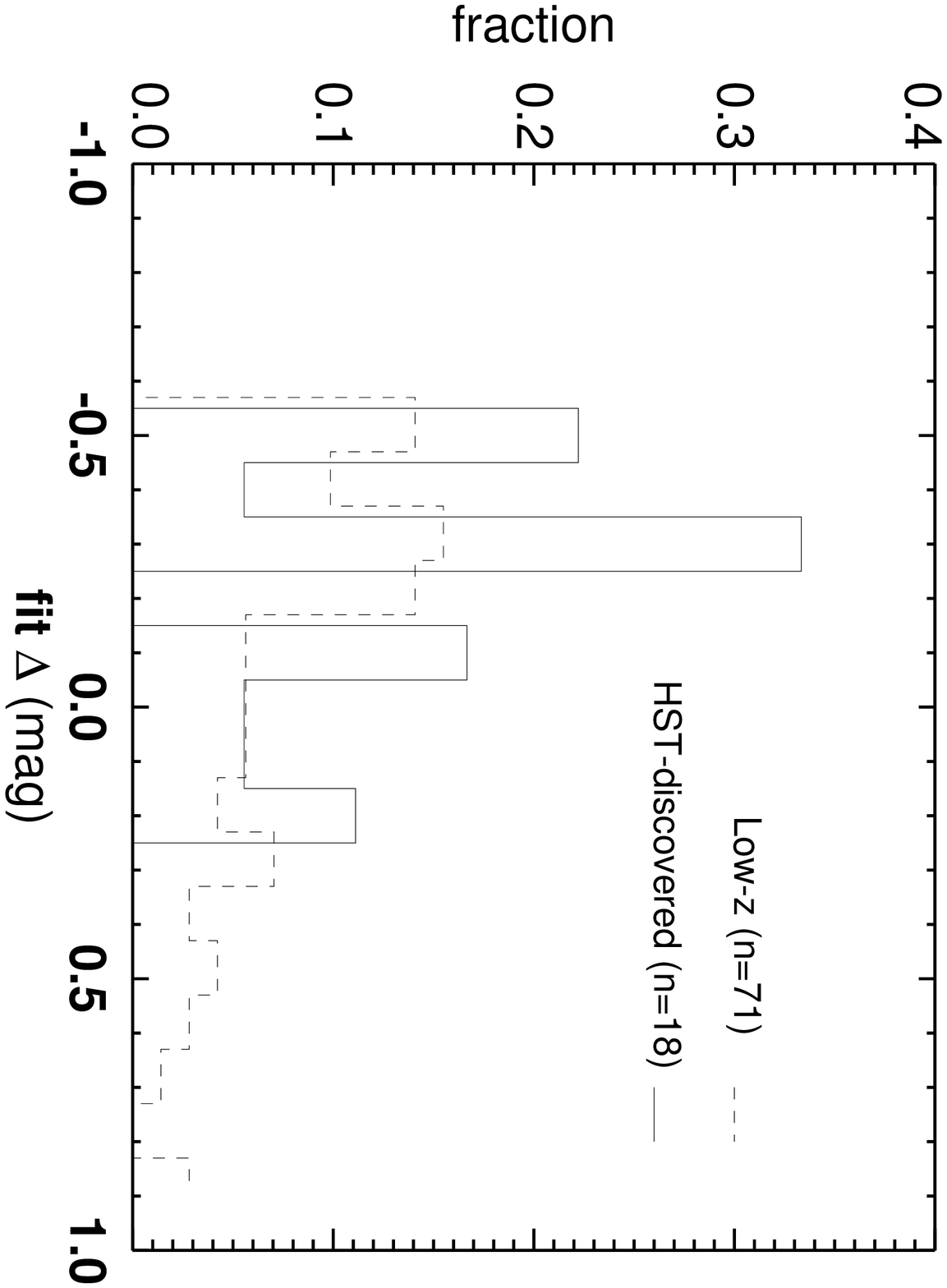}
\end{figure}

\vfill \eject

\begin{figure}[h]
\vspace*{150mm}
\includegraphics{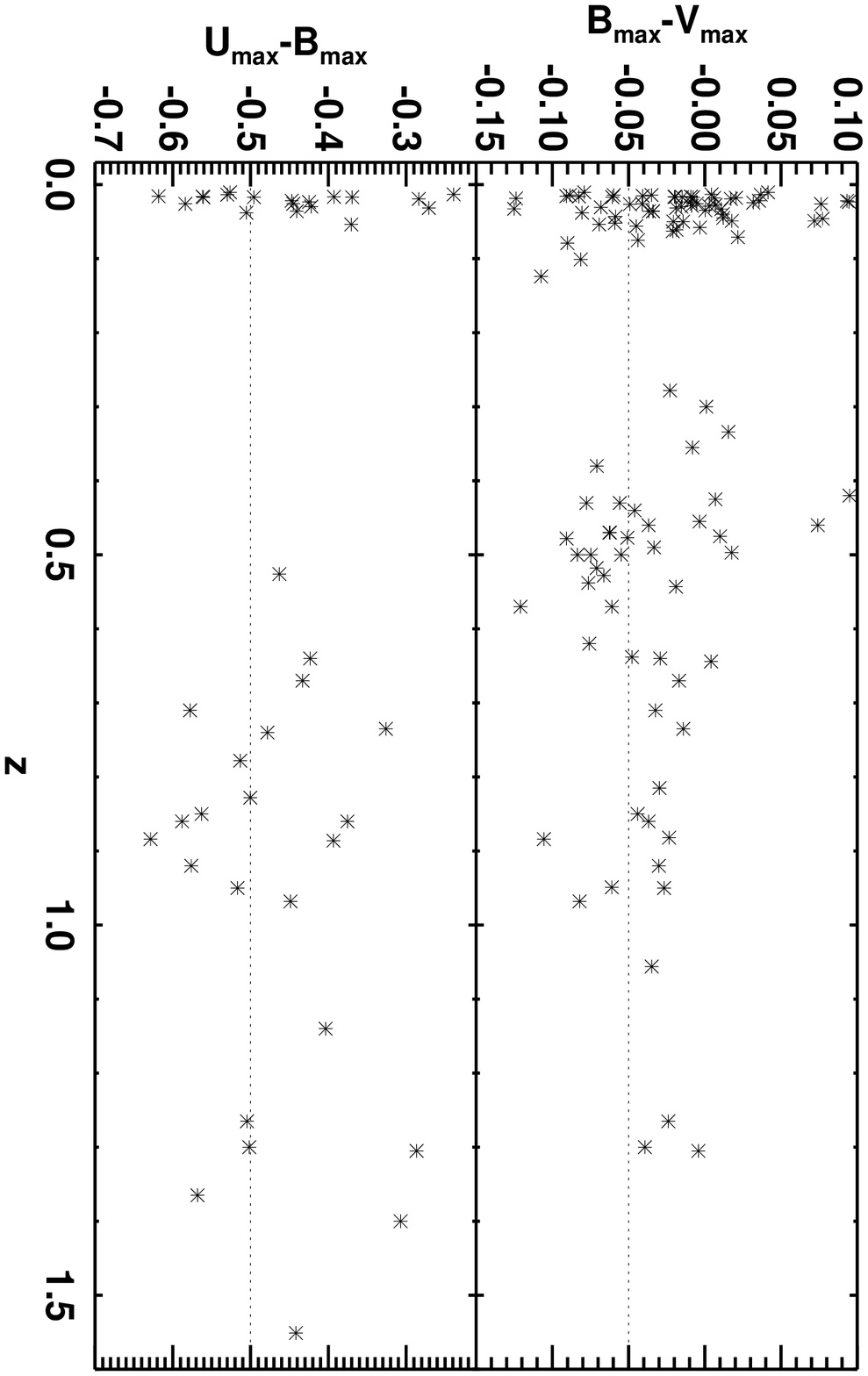}
\end{figure}

\vfill \eject

\begin{figure}[h]
\vspace*{150mm}
\includegraphics{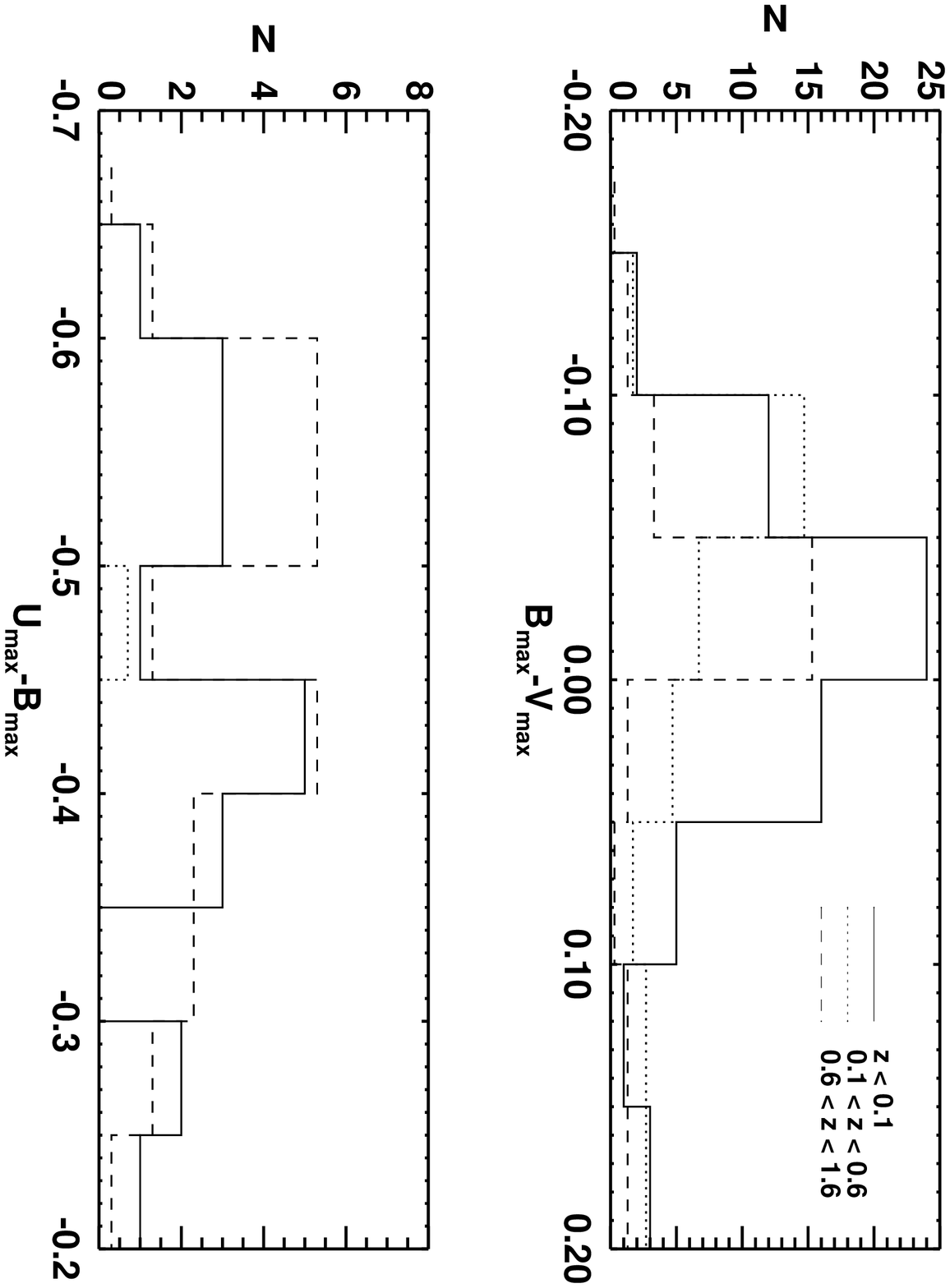}
\end{figure}

\vfill \eject

\begin{figure}[h]
\vspace*{150mm}
\includegraphics{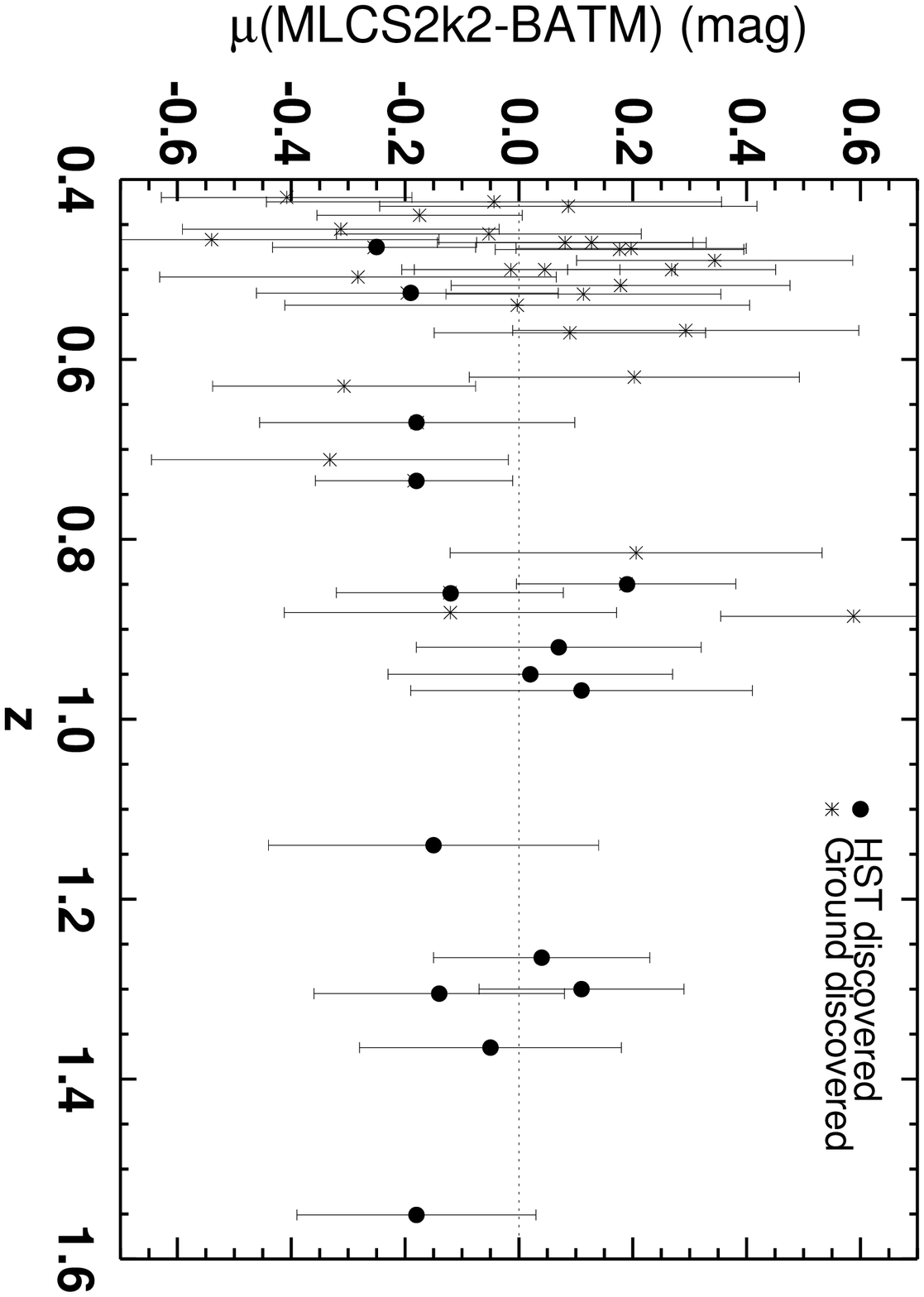}
\end{figure}

\vfill \eject

\begin{figure}[h]
\vspace*{150mm}
\includegraphics{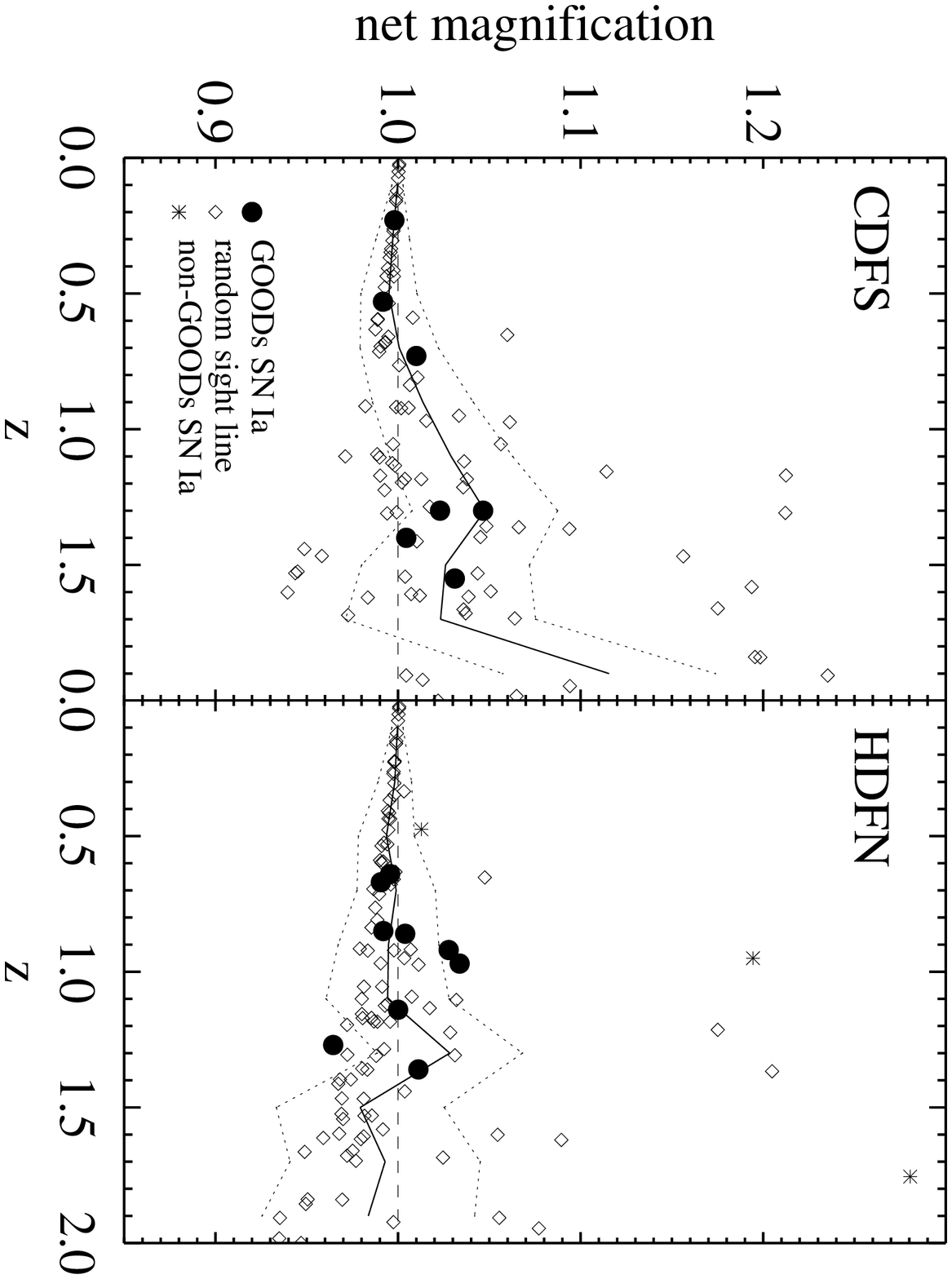}
\end{figure}

\vfill \eject

\begin{figure}[h]
\vspace*{150mm}
\includegraphics{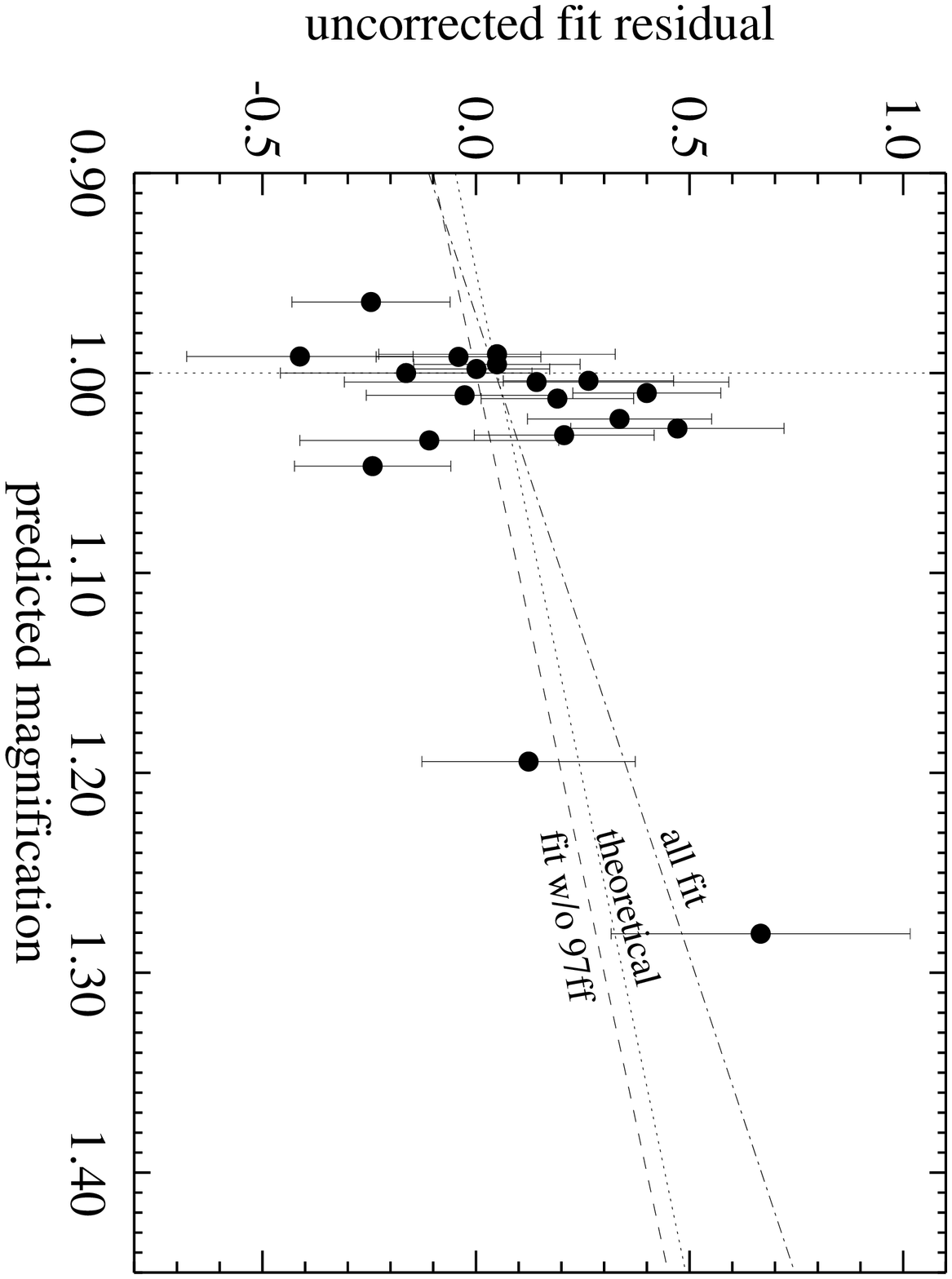}
\end{figure}

\vfill
\eject
 
\end{document}